\documentclass[aps, prd, twocolumn, superscriptaddress, showpacs]{revtex4}

\usepackage{epsf}
\usepackage{amssymb}
\usepackage{dcolumn}

\newlength{\zero}
\newcommand{\wz}{\hspace{\zero}}
\newcommand{\wwz}{\hspace{\zero}\hspace{\zero}}

\newlength{\szero}
\newcommand{\swz}{\hspace{\szero}}

\newcommand{\mb}[1]{\mbox{\boldmath $#1$}}

\newcommand{\dotI}[1]{\dot{I}_{#1}}

\newcommand{\ddotI}{\hbox{$ I $}
  \raise1.92ex\hbox{\kern-1.0ex..}}
\newcommand{\ddotIij}{\hbox{$ I_{ij} $}%
  \raise1.92ex\hbox{\kern-2.45ex..}\kern1.2ex}
\newcommand{\ddotIsub}[1]{\hbox{$ I_{#1} $}%
  \raise1.92ex\hbox{\kern-2.8ex..}\kern1.6ex}
\newcommand{\ddotIhat}{\hbox{$ I $}%
  \raise0.6ex\hbox{\kern-0.7ex$ \hat{\dot{} \kern0.7ex \dot{}} $}}
\newcommand{\ddotIhatij}{\hbox{$ I_{ij} $}%
  \raise0.6ex\hbox{\kern-2.2ex$ \hat{\dot{} \kern0.7ex \dot{}} $}\kern1.5ex}
\newcommand{\ddotIhatsub}[1]{\hbox{$ I_{#1} $}%
  \raise0.6ex\hbox{\kern-2.5ex$ \hat{\dot{} \kern0.7ex \dot{}} $}\kern1.8ex}
\newcommand{\ddotIsquared}{\hbox{$ I $}
  \raise1.92ex\hbox{\kern-1.05ex..}^2\kern-0.3ex}
\newcommand{\ddotIijsquared}{\hbox{$ I_{ij} $}%
  \raise1.92ex\hbox{\kern-2.5ex..}^2\kern0.3ex}
\newcommand{\ddotIsubsquared}[1]{\hbox{$ I_{#1} $}%
  \raise1.92ex\hbox{\kern-2.8ex..}^2\kern0.7ex}
\newcommand{\ddotIhatsquared}{\hbox{$ I $}%
  \raise0.6ex\hbox{\kern-0.7ex$ \hat{\dot{} \kern0.7ex \dot{}} $}^{\,2}\kern-0.5ex}
\newcommand{\ddotIhatijsquared}{\hbox{$ I_{ij} $}%
  \raise0.6ex\hbox{\kern-2.2ex$ \hat{\dot{} \kern0.7ex \dot{}} $}^{\,2}\kern0.3ex}
\newcommand{\ddotIhatsubsquared}[1]{\hbox{$ I_{#1} $}%
  \raise0.6ex\hbox{\kern-2.5ex$ \hat{\dot{} \kern0.7ex \dot{}} $}^{\,2}\kern0.5ex}

\newcommand{\dddotI}{\hbox{$ I $}%
  \raise1.92ex\hbox{\kern-1.35ex...}}
\newcommand{\dddotIij}{\hbox{$ I_{ij} $}%
  \raise1.92ex\hbox{\kern-2.8ex...}\kern0.9ex}
\newcommand{\dddotIsub}[1]{\hbox{$ I_{#1} $}%
  \raise1.92ex\hbox{\kern-3.15ex...}\kern1.2ex}
\newcommand{\dddotIsquared}{\hbox{$ I $}%
  \raise1.92ex\hbox{\kern-1.35ex...}^2}
\newcommand{\dddotIijsquared}{\hbox{$ I_{ij} $}%
  \raise1.92ex\hbox{\kern-2.8ex...}^2}
\newcommand{\dddotIsubsquared}[1]{\hbox{$ I_{#1} $}%
  \raise1.92ex\hbox{\kern-3.15ex...}^2\kern0.2ex}
\newcommand{\dddotIhatsquared}{\hbox{$ I $}%
  \raise0.6ex\hbox{\kern-1.05ex$ \hat{\dot{} \kern0.7ex \dot{} \kern0.7ex \dot{}} $}^{\,2}%
  \kern-0.3ex}

\newcommand{\dash}{\multicolumn{1}{c}{--}}

\newenvironment{equationarray}
{\arraycolsep 0.14 em
\begin {eqnarray}}
{\end {eqnarray}}

\newenvironment{equationarray*}
{\arraycolsep 0.14 em
\begin {eqnarray*}}
{\end {eqnarray*}}

\begin{document}

\title{``Mariage des Maillages'': \\
  A new numerical approach for 3D relativistic core collapse
  simulations}

\author{Harald Dimmelmeier}
\email{harrydee@mpa-garching.mpg.de}
\affiliation{Max-Planck-Institut f\"ur Astrophysik,
  Karl-Schwarzschild-Strasse 1, D-85741 Garching, Germany}

\author{J\'er\^ome Novak}
\email{Jerome.Novak@obspm.fr}
\affiliation{Laboratoire de l'Univers et de ses Th\'eories,
  Observatoire de Paris, F-92195 Meudon Cedex, France}

\author{Jos\'e A.\ Font}
\email{J.Antonio.Font@uv.es}
\affiliation{Departamento de Astronom\'{\i}a y Astrof\'{\i}sica, 
  Universidad de Valencia, Dr.\ Moliner 50, E-46100 Burjassot (Valencia), Spain}

\author{Jos\'e M.\ Ib\'a\~nez}
\email{Jose.M.Ibanez@uv.es}
\affiliation{Departamento de Astronom\'{\i}a y Astrof\'{\i}sica, 
  Universidad de Valencia, Dr.\ Moliner 50, E-46100 Burjassot (Valencia), Spain}

\author{Ewald M\"uller}
\email{emueller@mpa-garching.mpg.de}
\affiliation{Max-Planck-Institut f\"ur Astrophysik,
  Karl-Schwarzschild-Strasse 1, D-85741 Garching, Germany}

\date{\today}

\begin{abstract}
  We present a new three-dimensional general relativistic
  hydrodynamics code which is intended for simulations of stellar
  core collapse to a neutron star, as well as pulsations and
  instabilities of rotating relativistic stars. Contrary
  to the common approach followed in most existing three-dimensional 
  numerical relativity codes which are based in Cartesian coordinates, 
  in this code both the metric and the hydrodynamics equations are formulated 
  and solved numerically using spherical polar coordinates. A
  distinctive feature of this new code is the combination of two types
  of accurate numerical schemes specifically designed to solve each
  system of equations. More precisely, the code uses spectral methods
  for solving the gravitational field equations, which are
  formulated under the assumption of the conformal flatness condition
  (CFC) for the three-metric. Correspondingly, the hydrodynamics
  equations are solved by a class of finite difference methods called
  high-resolution shock-capturing schemes, based upon state-of-the-art
  Riemann solvers and third-order cell-reconstruction procedures. We
  demonstrate that the combination of a finite difference grid and a
  spectral grid, on which the hydrodynamics and metric equations are
  respectively solved, can be successfully accomplished. This
  approach, which we call {\em Mariage des Maillages\/} (French
  for grid wedding), results in high accuracy of the metric solver
  and, in practice, allows for fully three-dimensional applications
  using computationally affordable resources, along with ensuring long
  term numerical stability of the evolution. We compare our new
  approach to two other, finite difference based, methods to solve the
  metric equations which we already employed in earlier axisymmetric
  simulations of core collapse. A variety of tests in two and three
  dimensions is presented, involving highly perturbed neutron star
  spacetimes and (axisymmetric) stellar core collapse, which
  demonstrate the ability of the code to handle spacetimes with and
  without symmetries in strong gravity. These tests are also employed
  to assess the gravitational waveform extraction capabilities of the
  code, which is based on the Newtonian quadrupole formula. The code
  presented here is not limited to approximations of the Einstein
  equations such as CFC, but it is also well suited, in principle, to
  recent constrained formulations of the metric equations where elliptic 
  equations have a preeminence over hyperbolic equations.
\end{abstract}


\maketitle

\section{Introduction}
\label{section:introduction}

\subsection{Relativistic core collapse simulations}
\label{subsection:core_collapse_simulations}

Improving our understanding of the formation of neutron stars as a
result of the gravitational collapse of the core of massive stars is a
difficult endeavour involving many aspects of extreme and not
very well understood physics of the supernova explosion
mechanism~\cite{buras_03_a}. Numerical simulations of core collapse
supernova are driving progress in the field despite the limited
knowledge on issues such as realistic precollapse stellar models
(including rotation) or realistic equation of state, as well as
numerical limitations due to Boltzmann neutrino transport,
multidimensional hydrodynamics, and relativistic gravity. Axisymmetric
and three-dimensional approaches based on Newtonian gravity are
available since a few decades now (see e.g.~\cite{mueller_97_a} and
references therein). These approaches, which are constantly improving
over time, have provided valuable information on important issues such
as the dynamics of the collapse of a stellar core to nuclear density,
the formation of a proto-neutron star, and the propagation of the shock
front which ultimately is believed to eject the outer layers of the
stellar progenitor. Currently, however, even the most realistic
simulations of both nonrotating and rotating progenitor models do not
succeed in producing explosions (see~\cite{buras_03_a} and references
therein).

In addition, the incorporation of full relativistic gravity in the
simulations is likely to bring in well-known difficulties of numerical
relativity, where the attempts are traditionally hampered by
challenging mathematical, computational, and algorithmic issues as
diverse as the formulation of the field equations, 
robustness, efficiency, and long-term stability (particularly if
curvature singularities are either initially present or develop during
black hole formation). As high densities and velocities are involved
in combination with strong gravitational fields, gravitational
collapse and neutron star formation constitute a challenging
problem for general relativistic hydrodynamic simulations. The pace of
the progress is, no wonder, slow; for instance, in the
three-dimensional case, there is still no description of core collapse
in full general relativity today, even for the simplest matter models
one can conceive, where all microphysics is neglected.

In recent years, the interest in performing core collapse simulations
has been further motivated by the necessity of obtaining reliable
gravitational waveforms from (rotating) core collapse, one of the main
targets of gravitational radiation for the present and planned
interferometer detectors such as LIGO, GEO600, and VIRGO
(see~\cite{new_03_a} for a review). As a result of the complexities
listed above, it is not surprising that most previous studies aimed at
computing the gravitational wave signature of core collapse supernovae
have considered greatly simplified parameterized
models~\cite{mueller_82_a, finn_90_a, moenchmeyer_91_a, yamada_95_a,
  zwerger_97_a, rampp_98_a, dimmelmeier_01_a, dimmelmeier_02_a,
  dimmelmeier_02_b, fryer_02_a, fryer_03_a, imamura_03_a, kotake_03_a,
  shibata_04_a, ott_04_a}. In addition to the burst signal of
gravitational waves emitted during core bounce, multidimensional
simulations have also provided the signals produced by
convection~\cite{mueller_97_b} (see also~\cite{mueller_04_a} for the
most realistic simulations available at present), as well as those
from the resulting neutrino emission~\cite{burrows_96_a,
  mueller_97_b}.

From the above references it becomes apparent that our understanding
of core collapse and neutron star formation has advanced mainly by
studies carried out employing Newtonian dynamics. The situation is now
slowly changing, at least for simplified matter models where
microphysics and radiation transport are not yet included, with
new formulations of the Einstein field equations and of the general
relativistic hydrodynamics equations. Unfortunately, the $ 3 + 1 $
Einstein equations describing the dynamics of spacetime are a
complicated set of coupled, highly nonlinear hyperbolic-elliptic
equations with plenty of terms. Their formulation in a form suitable
for accurate and stable numerical calculations is not unique, and
constitutes one of the major fields of current research in numerical
relativity (see~\cite{lehner_01_a, lindblom_03_a} and references
therein). Not surprisingly, approximations of those equations have
been suggested, such as the conformal flatness condition of
Isenberg--Wilson--Mathews~\cite{isenberg_78_a, wilson_96_a} (CFC hereafter),
who proposed to approximate the 3-metric of the $ 3 + 1 $
decomposition by a conformally flat metric.

Using this approximation, Dimmelmeier et al.~\cite{dimmelmeier_01_a,
  dimmelmeier_02_a, dimmelmeier_02_b}
presented the first relativistic simulations of the core collapse of
rotating polytropes and neutron star formation in axisymmetry,
providing an in-depth analysis of the dynamics of the process as
well as of the gravitational wave emission. The results showed that
relativistic effects may qualitatively change in some cases the
dynamics of the collapse obtained in previous Newtonian
simulations~\cite{moenchmeyer_91_a, mueller_97_a}. In particular, core
collapse with multiple bounces was found to be strongly suppressed when
employing relativistic gravity. In most cases, compared to Newtonian
simulations, the gravitational wave signals are weaker and their
spectra exhibit higher average frequencies, as the newly born
proto-neutron stars have stronger compactness in the deeper
relativistic gravitational potential. Therefore, telling from
simulations based on rotating polytropes, the prospects for
detection of gravitational wave signals from supernovae are most
likely not enhanced by taking into account relativistic gravity. The
gravitational wave signals computed by Dimmelmeier et
al.~\cite{dimmelmeier_01_a,dimmelmeier_02_a,dimmelmeier_02_b} are
within the sensitivity range of the planned laser interferometer
detectors if the source is located within our Galaxy or in its local
neighbourhood. A catalogue of the core collapse
waveforms presented in~\cite{dimmelmeier_02_b} is available
electronically~\cite{garching_results}. This catalogue is currently
being employed by gravitational wave data analysis groups to
calibrate their search algorithms (see e.g.~\cite{pradier_01_a} for
results concerning the VIRGO group).

More recently, Shibata and Sekiguchi~\cite{shibata_04_a} have
presented simulations of axisymmetric core collapse of rotating
polytropes to neutron stars in {\it full} general relativity. These 
authors used a conformal-traceless reformulation of the $ 3 + 1 $
gravitational field equations commonly referred to in the literature 
by the acronym BSSN after the works of~\cite{shibata_95_a,baumgarte_99_a} 
(but note that many of the new features of the BSSN formulation were 
anticipated as early as 1987 by Nakamura, Oohara, and 
Kojima~\cite{nakamura_87_a}). The results obtained for initial models 
similar to those of~\cite{dimmelmeier_02_b} agree to high precision 
in both the dynamics of the collapse and the gravitational waveforms. 
This conclusion, in turn, implies that, at least for core collapse 
simulations to neutron stars, CFC is a very precise approximation 
of general relativity.

We note that in the relativistic core collapse simulations mentioned
thus far~\cite{dimmelmeier_02_b, shibata_04_a}, the gravitational
radiation is computed using the (Newtonian) quadrupole formalism. To
the best of our knowledge the only exception to this is the work of
Siebel et al.~\cite{siebel_03_a}, where, owing to the use of the
characteristic (light-cone) formulation of the Einstein equations, the
gravitational radiation from axisymmetric core collapse simulations was
unambiguously extracted at future null infinity without any
approximation. 

\subsection{Einstein equations and spectral methods}
\label{subsection:einstein_equations}

The most common approach to numerically solve the Einstein equations
is by means of finite differences (see~\cite{lehner_01_a} and
references therein). However, it is well known that spectral
methods~\cite{gottlieb_77_a, canuto_88_a} are far more accurate than
finite differences for smooth solutions (e.g.\ best for initial data
without discontinuities), being particularly well suited to solve
elliptic and parabolic equations. Good results can be obtained for
hyperbolic equations as well, as long as no discontinuities appear in
the solution. The basic principle underlying spectral methods is the
representation of a given function $ f (x) $ by its coefficients in a
complete basis of orthonormal functions: sines and cosines (Fourier
expansion) or a family of orthogonal polynomials (e.g.\ Chebyshev
polynomials $ T_i (x) $ or Legendre polynomials). In practice, of
course, only a {\em finite\/} set of coefficients is used and one
approximates $ f $ by the truncated series
$ f (x) \simeq \sum_{i = 0}^n c_i T_i (x) $ of such functions. The use
of spectral methods results in a very high accuracy, since the error
made by this truncation decreases like $ e^{-n} $ for smooth functions
(exponential convergence).

In an astrophysical context spectral methods have allowed to study
subtle phenomena such as the development of physical instabilities
leading to gravitational collapse~\cite{bonazzola_90_a}. In the last
few years, spectral methods have been successfully employed by the
{\em Meudon group\/}~\cite{meudon_group} in a number of relativistic
astrophysics scenarios~\cite{bonazzola_99_a}, among them the
gravitational collapse of a neutron star to a black hole, the infall
phase of a tri-axial stellar core in a core collapse supernova
(extracting the gravitational waves emitted in such process), the
construction of equilibrium configurations of rapidly rotating neutron
stars endowed with magnetic fields, or the tidal interaction of a star
with a massive black hole. Their most recent work concerns the
computation of the inertial modes of rotating
stars~\cite{villain_02_a}, of quasi-equilibrium configurations of
co-rotating binary black holes in general
relativity~\cite{grandclement_02_a}, as well as the evolution of pure
gravitational wave spacetimes~\cite{bonazzola_03_a}. To carry out
these numerical simulations the group has developed a fully
object-oriented library called {\sc Lorene}~\cite{lorene_code} (based
on the C++ computer language) to implement spectral methods in
spherical coordinates. Spectral methods are now employed in numerical
relativity by other groups as well~\cite{frauendiener_99_a,
pfeiffer_03_a}.

\subsection{Hydrodynamics equations and HRSC schemes}
\label{subsection:hydro_equations}

On the other hand, robust finite difference schemes to solve
hyperbolic systems of conservation (and balance) laws, such as the
Euler equations of fluid dynamics, are known for a long time and have
been employed successfully in computational fluid dynamics (see
e.g.~\cite{toro_97_a} and references therein). In particular, the
so-called upwind high-resolution shock-capturing schemes (HRSC schemes
hereafter) have shown their advantages over other type of methods even
when dealing with relativistic flows with highly ultrarelativistic
fluid speeds (see e.g.~\cite{marti_03_a, font_03_a} and references
therein). HRSC schemes are based on the mathematical information
contained in the characteristic speeds and fields (eigenvalues and
eigenvectors) of the Jacobian matrices of the system of partial
differential equations. This information is used in a fundamental
way to build up either exact or approximate Riemann solvers to
propagate forward in time the collection of local Riemann problems
contained in the initial data, once these data are discretized on a
numerical grid. These schemes have a number of interesting properties:
(1) The convergence to the physical solution (i.e.\ the unique weak
solution satisfying the so-called entropy condition) is guaranteed by
simply writing the scheme in conservation form, (2) the
discontinuities in the solution are sharply and stably resolved, and
(3) these methods attain a high order of accuracy in smooth parts of
the solution.

\subsection{Mariage des Maillages}
\label{subsection:mdm}

From the above considerations, it seems a promising strategy, in
the case of relativistic problems where coupled systems of elliptic
(for the spacetime) and hyperbolic (for the hydrodynamics) equations
must be solved, to use spectral methods for the former and HRSC
schemes for the latter (where discontinuous solutions may
arise). Showing the feasibility of such an approach is, in fact, the
main motivation and aim of this paper. Therefore, we present and
assess here the capabilities of a new, fully three-dimensional code
whose distinctive features are that it combines both types of
numerical schemes and implements the field equations and the
hydrodynamic equations using spherical coordinates. It should be
emphasized that our {\em Mariage des Maillages\/} approach is hence best
suited for formulations of the Einstein equations which favor the
appearance of elliptic equations against hyperbolic equations,
i.e.\ either approximations such as CFC~\cite{isenberg_78_a, wilson_96_a}
(the formulation we adopt in the simulations reported in this paper),
higher-order post-Newtonian extensions~\cite{cerda_04_a}, or {\em
exact\/} formulations as recently proposed by~\cite{bonazzola_03_a,
schaefer_04_a}. The hybrid approach put forward here has a successful 
precedent in the literature; using such combined methods, first results 
were obtained in one-dimensional core collapse in the framework of a 
tensor-scalar theory of gravitation~\cite{novak_00_a}.

We note that one of the main limitations of the previous axisymmetric
core collapse simulations presented
in~\cite{dimmelmeier_01_a,dimmelmeier_02_a,dimmelmeier_02_b} was the
CPU time spent when solving the elliptic equations describing the
gravitational field in CFC. The restriction was severe enough to
prevent the practical extension of the investigation to the
three-dimensional case. In that sense, spectral methods are again
particularly appropriate as they provide accurate results with
reasonable sampling, as compared with finite difference methods.

The three-dimensional code we present in this paper has been designed
with the aim of studying general relativistic astrophysical scenarios
such as rotational core collapse to neutron stars (and, eventually, to
black holes), as well as pulsations and instabilities of the formed
compact objects. Core collapse may involve, obviously, matter fields
which are not rotationally symmetric. While during the infall phase of
the collapse the deviations from axisymmetry should be rather small,
for rapidly rotating neutron stars which form as a result of the
collapse, or which may be spun up by accretion at later times,
rotational (nonaxisymmetric) bar mode instabilities may develop,
particularly in relativistic gravity and for differential rotation. In
this regard, in the previous axisymmetric simulations of
Dimmelmeier et al.~\cite{dimmelmeier_02_b}, some of the most extremely
rotating initial models yielded compact remnants which are above the
thresholds for the development of such bar mode instabilities on
secular or even dynamic time scales for Maclaurin spheroids in
Newtonian gravity (which are $ \beta_{\rm s} \sim 0.14 $ and
$ \beta_{\rm d} \sim 0.27 $, respectively, with
$ \beta = E_{\rm r} / |E_{\rm b}| $ being the ratio of rotational
energy and gravitational binding energy).

Presently, only a few groups worldwide have developed finite difference,  
three-dimensional (Cartesian) codes capable of performing the kind of 
simulations we aim at, where the joint integration of the Einstein and 
hydrodynamics equations is required~\cite{shibata_99_a, font_99_a, font_02_a}. 
Further 3D codes are currently being developed by a group in the
U.S.~\cite{duez_02_a} and by a E.U.\ Research Training Network
collaboration~\cite{baiotti_04_a, whisky_code}.

\subsection{Organization of the paper}
\label{subsection:organization}

The paper is organized as follows: In 
Section~\ref{section:model_and_equations} we introduce the assumptions
of the adopted physical model and the equations governing the dynamics
of a general relativistic fluid and the gravitational
field. Section~\ref{section:numerical_methods} is devoted to
describing algorithmic and numerical features of the code, such as the
setup of both the spectral and the finite difference grids, as
well as the basic ideas behind the HRSC schemes we have implemented to
solve the hydrodynamics equations. In addition, a detailed comparison 
of the three different solvers for the metric equations and their practical 
applicability is given. In Section~\ref{section:tests} we present a
variety of tests of the numerical code, comparing the metric solver
based on spectral methods to two other alternative methods using
finite differences. We conclude the paper with a summary and an
outlook to future applications of the code in
Section~\ref{section:summary}. We use a spacelike signature
$ (-, +, +, +)$ and units in which $ c = G = 1 $ (unless explicitly
stated otherwise). Greek indices run from 0 to 3, Latin indices from 1
to 3, and we adopt the standard convention for the summation over
repeated indices.

\section{Physical model and equations}
\label{section:model_and_equations}

\subsection{General relativistic hydrodynamics}
\label{subsection:gr_hydrodynamics}

\subsubsection{Flux-conservative hyperbolic formulation}
\label{subsubsection:hyperbolic_formulation}

Let $ \rho $ denote the rest-mass density of the fluid, $ u^\mu $ its
four-velocity, and $ P $ its pressure. The hydrodynamic evolution of a
relativistic perfect fluid with rest-mass current
$ J^{\mu} = \rho u^{\mu} $ and energy-momentum tensor
$ T^{\mu\nu} = \rho h u^\mu u^\nu + P g^{\mu\nu} $ in a (dynamic)
spacetime $ g^{\mu\nu}$ is determined by a system of local
conservation equations, which read
\begin{equation}
  \nabla_{\mu} J^{\mu} = 0, \qquad \nabla_{\mu} T^{\mu \nu} = 0,
  \label{eq:gr_equations_of_motion}
\end{equation}
where $\nabla_{\mu}$ denotes the covariant
derivative. The quantity $ h $ appearing in the energy-momentum tensor
is the specific enthalpy, defined as $ h = 1 + \epsilon + P / \rho $,
where $\epsilon$ is the specific internal energy. The three-velocity
of the fluid, as measured by an Eulerian observer at rest in a
spacelike hypersurface $ \Sigma_{t} $ is given by
\begin{equation}
  v^i = \frac{u^i}{\alpha u^0} + \frac{\beta^i}{\alpha},
  \label{eq:three_velocity}
\end{equation}
where $ \alpha $ is the lapse function and $ \beta^i $ is the shift
vector (see Section~\ref{subsection:metric_equations}).

Following the work laid out in~\cite{banyuls_97_a} we now introduce the
following set of conserved variables in terms of the primitive
(physical) hydrodynamic variables $ (\rho, v_i, \epsilon) $:
\begin{displaymath}
  \setlength{\arraycolsep}{0.14 em}
  \begin{array}{rcl}
    D & \equiv & \rho W, \\ [0.5 em]
    S_i  & \equiv  & \rho h W^2 v_i, \\ [0.5 em]
    \tau & \equiv & \rho h W^2 - P - D.
  \end{array}
\end{displaymath}
In the above expressions $ W $ is the Lorentz factor defined as
$ W = \alpha u^0 $, which satisfies the relation
$ W = 1 / \sqrt{1 - v_i v^i} $ and $ v_i = \gamma_{ij} v^j $, where
$ \gamma_{ij} $ is the 3-metric.

Using the above variables, the local conservation
laws~(\ref{eq:gr_equations_of_motion}) can be written as a
first-order, flux-conservative hyperbolic system of equations,
\begin{equation}
  \frac{1}{\sqrt{- g}} \left[
  \frac{\partial \sqrt{\gamma} \mb{U}}{\partial t} +
  \frac{\partial \sqrt{- g} \mb{F}^i}{\partial x^i} \right] = \mb{Q},
  \label{eq:hydro_conservation_equation}
\end{equation}
with the state vector, flux vector, and source vector given by
\begin{equation}
  \setlength{\arraycolsep}{0.14 em}
  \begin{array}{rcl}
  \mb{U} & = & [D, S_j, \tau], \\ [1.0 em]
  \mb{F}^i & = & \displaystyle
  \left[ D \hat{v}^i, S_j \hat{v}^i + \delta^i_j P,
  \tau \hat{v}^i + P v^i \right], \\ [1.0 em]
  \mb{Q} & = & \displaystyle
  \left[ 0, T^{\mu \nu} \left(
  \frac{\partial g_{\nu j}}{\partial x^\mu} - 
  {\it \Gamma}^\lambda_{\mu \nu} g_{\lambda j} \right),
  \right. \\ [1.0 em]
  & & \displaystyle \quad \left. \alpha \left( T^{\mu 0}
  \frac{\partial \ln \alpha}{\partial x^\mu} -
  T^{\mu \nu} {\it \Gamma}^0_{\mu \nu} \right) \right]  .
  \end{array}
  \label{eq:hydro_conservation_equation_constituents}
\end{equation}
Here $ \hat{v}^i = v^i - \beta^i / \alpha $, and
$ \sqrt{-g} = \alpha \sqrt{\gamma} $, with
$ g = \det (g_{\mu \nu}) $ and $ \gamma = \det (\gamma_{ij}) $ being
the determinant of the 4-metric and 3-metric, respectively (see
Section~\ref{subsection:adm_metric_equations}). In
addition, $ {\it \Gamma}^\lambda_{\mu \nu} $ are the Christoffel
symbols associated with $ g_{\mu \nu} $.

\subsubsection{Equation of state}
\label{subsubsection:eos}

The system of hydrodynamic
equations~(\ref{eq:hydro_conservation_equation}) is closed by an
equation of state (EoS) which relates the pressure to some 
thermodynamically independent quantities, e.g.\
$ P = P (\rho, \epsilon) $. As in~\cite{dimmelmeier_02_a,
  dimmelmeier_02_b, siebel_03_a} we have implemented in the code a
hybrid ideal gas EoS~\cite{janka_93_a}, which consists of a polytropic
pressure contribution and a thermal pressure contribution,
$ P = P_{\rm p} + P_{\rm th} $. This EoS, which despite its simplicity
is particularly suitable for stellar core collapse simulations, is
intended to model the degeneracy pressure of the electrons and (at
supranuclear densities) the pressure due to nuclear forces in the
polytropic part, and the heating of the matter by shock waves in the
thermal part. The hybrid EoS is constructed as follows.

For a rotating stellar core before collapse the polytropic relation
between the pressure and the rest mass density,
\begin{equation}
  P_{\rm p} = K \rho^{\gamma},
  \label{eq:polytropic_relation}
\end{equation}
with $ \gamma = \gamma_{\rm ini} = 4 / 3 $ and
$ K = 4.897 \times 10^{14} $ (in cgs units) is a fair approximation
of the density and pressure stratification~\cite{mueller_97_a}.

In order to start the gravitational collapse of a configuration
initially in equilibrium, the effective adiabatic index $ \gamma $ is
reduced from $ \gamma_{\rm ini} $ to $ \gamma_1 $ on the initial time
slice. During the infall phase of core collapse the matter is assumed
to obey a polytropic EoS~(\ref{eq:polytropic_relation}), which is
consistent with the ideal gas EoS for a compressible inviscid fluid,
$ P = (\gamma - 1) \rho \epsilon $.

To approximate the stiffening of the EoS for densities larger than
nuclear matter density $ \rho_{\rm nuc} $, we assume that the
adiabatic index $ \gamma $ jumps from $ \gamma_1 $ to $ \gamma_2 $ at
$ \rho = \rho_{\rm nuc} $. At core bounce a shock forms and propagates
out, and the matter accreted through the shock is heated,
i.e.\ its kinetic energy is dissipated into internal energy. This is
reflected by a nonzero
$ P_{\rm th} = \rho \epsilon_{\rm th} (\gamma_{\rm th} - 1) $, where
$ \epsilon_{\rm th} = \epsilon - \epsilon_{\rm p} $ with
$ \epsilon_{\rm p} = P_{\rm p} / [\rho (\gamma - 1)] $, in the
post-shock region. We choose $ \gamma_{\rm th} = 1.5 $. This choice
describes a mixture of relativistic ($ \gamma = 4/3 $) and
nonrelativistic ($ \gamma = 5/3 $) components of an ideal fluid.

Requiring that $ P $ and $ \epsilon $ are continuous at the transition
density $ \rho_{\rm nuc} $, one can construct an EoS for which both the
total pressure $ P $ and the individual contributions $ P_{\rm p} $
and $ P_{\rm th} $ are continuous at $ \rho_{\rm nuc} $, and which
holds during all stages of the collapse:
\begin{equationarray}
  P & = & \frac{\gamma - \gamma_{\rm th}}{\gamma - 1}
  K \rho_{\rm nuc}^{\gamma_1 - \gamma}
  \rho^{\gamma} - \frac{(\gamma_{\rm th} - 1) (\gamma - \gamma_1)}
  {(\gamma_1 - 1) (\gamma_2 - 1)}
  K \rho_{\rm nuc}^{\gamma_1 - 1} \rho \nonumber \\
  & & + (\gamma_{\rm th} - 1) \rho \epsilon.
  \label{eq:hybrid_eos}
\end{equationarray}%
For more details about this EoS, we refer to~\cite{dimmelmeier_02_a,
  janka_93_a}.

Our implementation of the hybrid EoS allows us to suppress the
contribution of the thermal pressure $ P_{\rm th} $. In this case the
EoS~(\ref{eq:hybrid_eos}) analytically reduces to the polytropic
relation~(\ref{eq:polytropic_relation}). We use this EoS, with
different values for $ \gamma $ and $ K $, in the simulations of 
polytropic neutron star models presented below.

\subsection{Metric equations}
\label{subsection:metric_equations}

\subsubsection{ADM metric equations}
\label{subsection:adm_metric_equations}

We adopt the ADM $ 3 + 1 $ formalism~\cite{arnowitt_62_a} to
foliate the spacetime into a set of non-intersecting spacelike
hypersurfaces. The line element reads
\begin{equation}
  ds^2 = - \alpha^2 dt^2 + \gamma_{ij} (dx^i + \beta^i dt)
  (dx^j + \beta^j dt),
  \label{eq:line_element}
\end{equation}
where $ \alpha $ is the lapse function which describes the rate
of advance of time along a timelike unit vector $ n^{\mu} $ normal
to a hypersurface, $ \beta^i $ is the spacelike shift three-vector
which describes the motion of coordinates within a surface, and
$ \gamma_{ij} $ is the spatial three-metric.

In the $ 3 + 1 $ formalism, the Einstein equations are split into
evolution equations for the three-metric $ \gamma_{ij} $ and the
extrinsic curvature $ K_{ij} $, and constraint equations (the
Hamiltonian and momentum constraints) which must be fulfilled at every
spacelike hypersurface:
\begin{equation}
  \setlength{\arraycolsep}{0.14 em}
  \begin{array}{rcl}
    \partial_t \gamma_{ij} & = & - 2 \alpha K_{ij} +
    \nabla_i \beta_j + \nabla_j \beta_i, \\ [0.8 em]
    \partial_t K_{ij} & = & - \nabla_i \nabla_j \alpha +
    \alpha (R_{ij} + K K_{ij} - 2 K_{ik} K_j^k) \\ [0.5 em]
    & & + \beta^k \nabla_k K_{ij} + K_{ik} \nabla_j \beta^k + K_{jk}
    \nabla_i \beta^k \\ [0.5 em]
    & & - \displaystyle
    8 \pi \alpha \! \left( \! S_{ij} - \frac{\gamma_{ij}}{2}
    (S_k^k - \rho_{\rm H}) \right), \\ [0.8 em]
    0 & = & R + K^2 - K_{ij} K^{ij} - 16 \pi \rho_{\rm H}, \\ [0.8 em]
    0 & = & \nabla_i (K^{ij} - \gamma^{ij} K) - 8 \pi S^j.
  \end{array}
  \label{eq:adm_metric_equations}
\end{equation}
In these equations $ \nabla_i $ is the covariant derivative with
respect to the three-metric $ \gamma_{ij} $, $ R_{ij} $ is the
corresponding Ricci tensor, $ R $ is the scalar curvature, and $ K $
is the trace of the extrinsic curvature $ K_{ij} $. The matter fields
appearing in the above equations, $ S_{ij} $, $ S^j $, and $ \rho_{\rm
H} = \rho h W^2 - P $, are the spatial components of the stress-energy
tensor, the three momenta, and the total energy, respectively.

The ADM equations have been repeatedly shown over the years to be 
intrinsically numerically unstable. Recently, there have been numerous 
attempts to reformulate above equations into forms better suited 
for numerical investigations (see~\cite{shibata_95_a, baumgarte_99_a,
lehner_01_a, lindblom_03_a} and references therein). These
approaches to delay or entirely suppress the excitation of constraint
violating unstable modes include the BSSN reformulation of the ADM 
system~\cite{nakamura_87_a, shibata_95_a, baumgarte_99_a}
(see Section~\ref{subsection:einstein_equations}), hyperbolic reformulations
(see~\cite{reula_98_a} and references therein), or a new form with
maximally constrained evolution~\cite{bonazzola_03_a}. In our opinion
a consensus seems to be emerging currently in numerical relativity, which in
general establishes that the more constraints are used in the formulation of
the equations the more numerically stable the evolution is.

\subsubsection{Conformal flatness approximation for the spatial metric}
\label{subsection:cfc_approximation}

Based on the ideas of Isenberg~\cite{isenberg_78_a} and Wilson et
al.~\cite{wilson_96_a}, and as it was done in the work
of Dimmelmeier et al.~\cite{dimmelmeier_02_b}, we
approximate the general metric $ g_{\mu\nu} $ by replacing the spatial
three-metric $ \gamma_{ij} $ with the conformally flat three-metric,
$ \gamma_{ij} = \phi^4 \hat{\gamma}_{ij} $, where $ \hat{\gamma}_{ij} $
is the flat metric ($ \hat{\gamma}_{ij} = \delta_{ij} $ in Cartesian
coordinates). In general, the conformal factor $ \phi $ depends on the
time and space coordinates. Therefore, at all times during a numerical
simulation we assume that all off-diagonal components of the
three-metric are zero, and the diagonal elements have the common
factor $ \phi^4 $.

In CFC the following relation between the time derivative of the
conformal factor and the shift vector holds:
\begin{equation}
  \partial_t \phi = \frac{\phi}{6} \nabla_k \beta^k.
  \label{eq:phi_time_evolution}
\end{equation}
With this the expression for the extrinsic curvature becomes
time-independent and reads
\begin{equation}
  K_{ij} =
  \frac {1}{2 \alpha} \left( \nabla_i \beta_j + \nabla_j
  \beta_i - \frac{2}{3} \gamma_{ij} \nabla_k \beta^k \right).
  \label{eq:extrinsic_curvature_cfc}
\end{equation}
If we employ the maximal slicing condition, $ K = 0 $, then 
in the CFC approximation the ADM
equations~(\ref{eq:adm_metric_equations}) reduce to a set of five
coupled elliptic (Poisson-like) nonlinear equations for the metric
components,
\begin{equation}
  \setlength{\arraycolsep}{0.14 em}
  \begin{array}{rcl}
    \hat{\Delta} \phi & = & \displaystyle - 2 \pi \phi^5 \left( \rho h
    W^2 - P + \frac{K_{ij} K^{ij}}{16 \pi} \right), \\ [1.0 em]
    \hat{\Delta} (\alpha \phi) & = & \displaystyle 2 \pi \alpha \phi^5
    \! \left( \! \rho h (3 W^2 - 2) \! + \! 5 P \! + \! \frac{7 K_{ij}
    K^{ij}}{16 \pi} \! \right), \!\!\!\!\!\!\!\!\!\!\!\! \\ [1.0 em]
    \hat{\Delta} \beta^i & = & \displaystyle 16 \pi \alpha \phi^4 S^i +
    2 \phi^{10} K^{ij} \hat{\nabla}_j
    \left( \frac{\alpha}{\phi^6} \right) -
    \frac{1}{3} \hat{\nabla}^i \hat{\nabla}_k \beta^k,
    \!\!\!\!\!\!\!\!\!\!\!\!
  \end{array}
  \label{eq:metric_equations}
\end{equation}
where $ \hat{\nabla}_i $ and $ \hat{\Delta} $ are the flat space Nabla
and Laplace operators, respectively. We note that the way of writing the 
metric equations with a Laplace operator on the left hand side can be 
exploited by numerical methods specifically designed to solve such kind 
of equations (see Sections~\ref{subsubsection:conventional_poisson_solver}
and~\ref{subsubsection:spectral_metric_solver} below).

These elliptic metric equations couple to each other via their right
hand sides, and in case of the three equations for the components of
$ \beta^i $ also via the operator $ \hat{\Delta} $ acting on the
vector $ \beta^i $. They do not contain explicit time derivatives, and
thus the metric is calculated by a fully constrained approach, at the
cost of neglecting some evolutionary degrees of freedom in the spacetime
metric. In the astrophysical situations we plan to explore (e.g.\
evolution of neutron stars or core collapse of massive stars), the
equations are entirely dominated by the source terms involving the
hydrodynamic quantities $ \rho $, $ P $, and $ v^i $, whereas the
nonlinear coupling through the remaining, purely metric, source terms
becomes only important for strong gravity. On each time slice the
metric is hence solely determined by the instantaneous hydrodynamic
state, i.e.\ the distribution of matter in space. 

Recently, Cerd\'a-Dur\'an et al.~\cite{cerda_04_a} have extended the
above CFC system of equations (and the corresponding core collapse
simulations in CFC reported in~\cite{dimmelmeier_02_b}) by the
incorporation of additional degrees of freedom in the approximation,
which render the spacetime metric exact up to the second
post-Newtonian order. Despite the extension of the five original
elliptic CFC metric equations for the lapse, the shift vector, and the
conformal factor by additional equations, the final system of equations
in the new formulation is still elliptic. Hence, the same code and
numerical schemes employed in~\cite{dimmelmeier_02_b} and in the
present work can be used. The results obtained 
by Cerd\'a-Dur\'an et al.~\cite{cerda_04_a} for a representative subset 
of the core collapse models in~\cite{dimmelmeier_02_b} show only minute
differences with respect to the CFC results, regarding both the
collapse dynamics and the gravitational waveforms. We point out that
Shibata and Sekiguchi~\cite{shibata_04_a}
have recently considered axisymmetric core collapse of rotating
polytropes to neutron stars in full general relativity (i.e.\ no
approximations) using the $ 3 + 1 $ BSSN formulation of the Einstein
equations. Interestingly, the results obtained for initial models
similar to those of~\cite{dimmelmeier_02_b} agree to high precision in
the dynamics of the collapse and on the gravitational waveforms, which
supports the suitability and accuracy of the CFC approximation for
simulations of relativistic core collapse to neutron stars (see also
Section~\ref{subsubsection:comparison_with_full_gr}).

In addition, there has been a direct comparison between the CFC
approximation and perturbative analytical approaches (post-Newtonian
and effective-one-body), which shows a very good agreement in the
determination of the innermost stable circular orbit of a system of 
two black holes~\cite{damour_02_a}.

\subsubsection{Metric equation terms with noncompact support}
\label{subsection:noncompact_support}

In general, the right hand sides of the metric
equations~(\ref{eq:metric_equations}) contain nonlinear source terms
of noncompact support. For a system with an isolated matter
distribution bounded by some stellar radius $ r_{\rm s} $, the source
term of each of the metric equations for a metric quantity $ u $ can
be split into a ``hydrodynamic'' term with compact support
$ S_{\rm h} $ and a purely ``metric'' term with noncompact support
$ S_{\rm m} $. Where no matter is present, only the metric term
remains:
\begin{equation}
  \hat{\Delta} u = \left\{
    \begin{array}{ll}
      S_{\rm h} (u) + S_{\rm m} (u) & \qquad \mbox{for } r \le r_{\rm s}, \\
      S_{\rm m} (u) & \qquad \mbox{for } r > r_{\rm s}.
    \end{array}
  \right.
  \label{eq:metric_equations_rhs_split}
\end{equation}
The source term $ S_{\rm m} $ vanishes only for $ K_{ij} = 0 $ and thus
$ \beta^i = 0 $, i.e.\ if the three-velocity vanishes and the matter
is static. As a consequence of this, only a spherically symmetric
static matter distribution will yield a time-independent solution
to Eq.~(\ref{eq:metric_equations_rhs_split}), which is equivalent to the 
spherically symmetric Tolman--Oppenheimer--Volkoff (TOV) solution of
hydrostatic equilibrium. In this case the vacuum metric is given by
the solution of a homogeneous Poisson equation, $ u = k_1 + k_2 / r $,
the constants $ k_1 $ and $ k_2 $ being determined by boundary values
e.g.\ at $ r_{\rm s} $.

A time-dependent spherically symmetric matter interior suffices to
yield a nonstatic vacuum metric ($ u = u (t) $ everywhere). However,
this is not a contradiction to Birkhoff's theorem, as it is purely a gauge
effect. A transformation of the vacuum part of the metric from an
isotropic to a Schwarzschild-like radial coordinate leads to the
static (and not conformally flat) standard Schwarzschild vacuum
spacetime.

Thus, in general, the vacuum metric solution to
Eqs.~(\ref{eq:metric_equations}) cannot be obtained analytically, and
therefore (except for TOV stars) no exact boundary values can be
imposed for $ \phi $, $ \alpha $, and $ \beta^i $ at some finite
radius $ r $. We note that this property of the metric equations is no
consequence of the {\em approximative\/} character of conformal
flatness, as in spherical symmetry the CFC renders the exact ADM
equations~(\ref{eq:adm_metric_equations}), but rather results from the
choice of the (isotropic) radial coordinate.

\section{Numerical methods}
\label{section:numerical_methods}

\subsection{Finite difference grid}
\label{subsection:finite_difference_grid}

The expressions for the hydrodynamic and metric quantities outlined in
Section~\ref{section:model_and_equations} are in covariant form. For a
numerical implementation of these equations, however, we have to
choose a suitable coordinate system adapted to the geometry of the
astrophysical situations intended to be simulated with the code.

As we plan to investigate isolated systems with matter configurations
not too strongly departing from spherical symmetry with a spacetime
obeying asymptotic flatness, the formulation of the hydrodynamic and
metric equations, Eqs.~(\ref{eq:hydro_conservation_equation}) and
(\ref{eq:metric_equations}), and their numerical implementation are
based on spherical polar coordinates $ (t, r, \theta, \varphi) $. This 
coordinate choice facilitates the use of fixed grid refinement in form
of nonequidistant radial grid spacing. Additionally, in spherical
coordinates the boundary conditions for the system of partial
differential metric equations~(\ref{eq:metric_equations}) are simpler
to impose (at finite or infinite distance) on a spherical surface than
on a cubic surface if Cartesian coordinates were used. We have found
no evidence of numerical instabilities arising at the coordinate
singularities at the origin ($ r = 0 $) or at the axis
($ \theta = 0, \pi $) in all simulations performed thus far with the
code (see~\cite{evans_86_a, stark_89_a} for related discussions on
instabilities in codes based upon spherical coordinates).

Both the discretized hydrodynamic and metric quantities are located on
the Eulerian {\em finite difference grid\/} at cell centers
$ (r_i, \theta_j, \varphi_k) $, where $ i, j, k $ run from 1 to
$ n_r, n_\theta, n_\varphi $, respectively. The angular grid zones in
the $ \theta $- and $ \varphi $-direction are each equally spaced, while
the radial grid, which extends out to a finite radius $ r_{\rm fd} $
larger than the stellar radius $ r_{\rm s} $, can be chosen to be
equally or logarithmically spaced. Each cell is bounded by two
interfaces in each coordinate direction. Values on ghost zone cell
centers, needed to impose boundary conditions, are obtained with the
symmetry conditions described in~\cite{dimmelmeier_02_a}. We further
assume equatorial plane symmetry in all simulations presented below (the 
code, however, is not restricted to this symmetry condition).
Expressions containing finite differences in space on this grid are
calculated with second order accuracy.

Note that the space between the surface of the star, the radius of
which in general is angular dependent, and the outer boundary of the
finite difference grid is filled with an artificial atmosphere (as
done in codes similar to ours, see~\cite{font_02_a, duez_02_a,
  baiotti_04_a}). This atmosphere obeys the polytropic
EoS~(\ref{eq:polytropic_relation}), and has a very low density such
that its presence does not affect the dynamics of the
star~\cite{dimmelmeier_02_a}. As an example, we observe a slight
violation of conservation of rest mass and angular momentum in
simulations of axisymmetric rotational core collapse of the order of
$ 10^{-4} $. This small violation can be entirely attributed to the
interaction of the stellar matter with the artificial atmosphere (see
Appendix~\ref{subsection:exact_conservation}).

\subsection{Spectral methods and grid}
\label{subsection:sm_and_grids}

\subsubsection{Spectral methods}
\label{subsubsection:spectral_methods}

Our most general metric solver is based on spectral methods (see
Section~\ref{subsubsection:spectral_metric_solver}). The basic
principle of these methods has been given in
Section~\ref{subsection:einstein_equations}. Let us now describe some
details of our implementation in the case of 3D functions in spherical
coordinates. The interested reader can refer to~\cite{bonazzola_99_a} for
details. A function $ f $ can be decomposed as follows
($ \xi $ is linked with the radial coordinate $ r $, as given below):
\begin{equation}
  f (\xi, \theta, \varphi) =
  \sum_{k=0}^{\hat{n}_\varphi}
  \sum_{j=0}^{\hat{n}_\theta}
  \sum_{i=0}^{\hat{n}_r}
  c_{ijk} T_i(\xi) Y_j^k (\theta, \varphi),
  \label{eq:spec_decomp}
\end{equation}
where $ Y_j^k (\theta, \varphi) $ are spherical harmonics. The angular
part of the function can also be decomposed into a Fourier series, to
compute angular derivatives more easily. If $ f $ is represented by
its coefficients $ c_{ijk} $, it is easy to obtain the coefficients of
e.g.\ $\partial f / \partial r $, $ \Delta f $ (or the result of any
linear differential operator applied to $ f $) thanks to the
properties of Chebyshev polynomials or spherical harmonics. For
instance, to compute the coefficients of the radial derivative of $ f
$, we make use of the following recursion formula on Chebyshev
polynomials:
\begin{equation}
  \frac{dT_{n+1}(x)}{dx} = 2 (n + 1) T_n (x) + \frac{n + 1}{n - 1}
  \frac{dT_{n-1}(x)}{dx}
  \quad \forall n > 1.
  \label{eq:recdcheb}
\end{equation}
A grid is still needed for two reasons: firstly, to calculate these
coefficients through the computation of integrals, and secondly to
evaluate non-linear operators (e.g.\ $ \nabla f \times \nabla f $),
using the values of the functions at grid points (in physical
space). The spectral grid points, called collocation points
are situated at $ (\hat{r}_i, \hat{\theta}_j, \hat{\varphi}_k) $, where
$ i, j, k $ run from 1 to $ \hat{n}_r, \hat{n}_\theta, \hat{n}_\varphi $,
respectively. They are the nodes of a Gauss--Lobato quadrature used to
compute the integrals giving the spectral coefficients. The use of Fast
Fourier Transforms (FFT) for the angular part requires equally spaced
points in the angular directions, whereas a fast Chebyshev transform
(also making use of FFT) requires that the radial grid points
correspond, in $ \xi $, to the zeros of $ T_{\hat{n}_r} $.
Note that in our simulations each of the domains contains the same
number of radial and angular collocation points.

In order to be able to cover the entire space ($ r \in [0, +\infty] $)
and to handle coordinate singularities at the origin ($ r = 0 $), we
use several grid {\em domains\/}:
\begin{itemize}
\item a nucleus spanning from $ r = 0 $ to $ r_{\rm d} $, where we set
  $ r = \alpha \xi $, with $ \xi \in [0,1] $ and $ \alpha $ being a
  constant (we use either only even Chebyshev polynomials
  $ T_{2i} (\xi) $, or only odd polynomials $ T_{2 i + 1} (\xi) $);
\item an arbitrary number (including zero) of shells bounded by the
  inner radius $ r_{{\rm d\,}i} $ and outer radius
  $ r_{{\rm d\,}i + 1} $, where we set $ r = \alpha_i \xi + \beta_i $
  with $ \xi \in [-1, 1] $ and $ \alpha_i $ and $ \beta_i $ being
  constants depending on the shell number $ i $;
\item a compactified external domain extending from the outer boundary
  of the finite difference grid at $ r_{\rm fd} $ to radial infinity,
  where we set $ r = 1 / [\alpha_{\rm c} (\xi + 1)] $, with
  $ \xi \in [-1, 1] $ and $ \alpha_{\rm c} $ being a constant.
\end{itemize}
Furthermore, we assume that the ratio $ f_{\rm d} $ between the outer
boundary radii of two consecutive domains is constant, which yields
the relation
\begin{equation}
  f_{\rm d} =
  \left( \frac{r_{\rm fd}}{r_{\rm d}} \right)^{1 / (n_{\rm d} - 2)}
  \!\!\!\!\!\!\!\!\!\!\!\!\!\!\!\!\!\!\!\!\!,
  \label{eq:domain_boundary_factor}
\end{equation}
where $ n_{\rm d} $ is the number of domains (including the nucleus
and the external compactified domain). Thus a particular choice of
$ n_{\rm d} $ and fixing the radius of the nucleus $ r_{\rm d} $
completely specifies the setup of the spectral grid:
\begin{equation}
  \setlength{\arraycolsep}{0.14 em}
  \begin{array}{rcl}
    r_{{\rm d\,}1} & = & r_{\rm d}, \\ [-0.4 em]
    & \vdots & \\ [-0.2 em]
    r_{{\rm d\,}i} & = & f_{\rm d} \times r_{{\rm d\,}i - 1}, \\ [-0.4 em]
    & \vdots & \\ [-0.2 em]
    r_{{\rm d\,}n_{\rm d} - 1} & = & r_{\rm fd}, \\
    r_{{\rm d\,}n_{\rm d}} & = & \infty.
  \end{array}
  \label{eq:domain_radius_calculation}
\end{equation}

\begin{figure}[t]
  \epsfxsize = 8.6 cm
  \centerline{\epsfbox{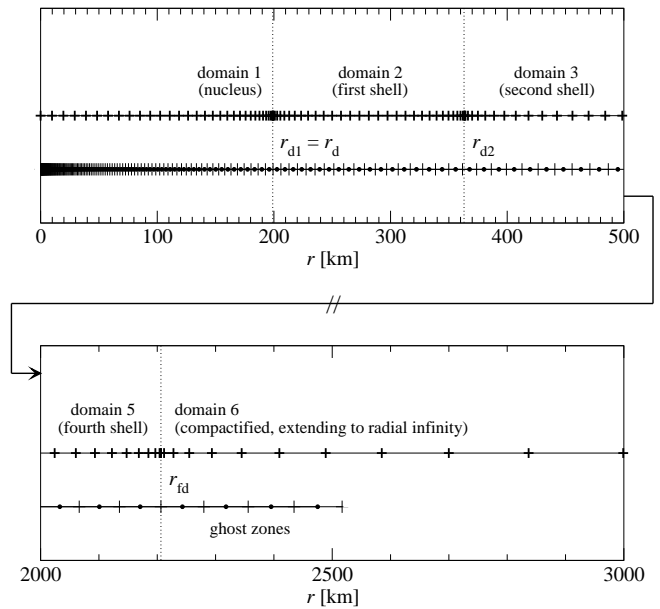}}
  \caption{Radial setup of the initial spectral grid (collocation
    points are marked by plus symbols) and the time-independent finite
    difference grid (cell centers are marked by filled circles,
    separated by cell interfaces symbolized by vertical dashes) for a
    typical core collapse simulation. The upper panel shows the
    innermost $ 500 {\rm\ km} $ containing the nucleus (ending at
    $ r_{\rm d} \approx 200 {\rm\ km} $), the first shell, and a part
    of the second shell of the spectral grid. In the lower panel a
    part of the last regular shell (which is confined by the outer
    boundary of the finite difference grid at
    $ r_{\rm fd} \approx 2200 {\rm\ km} $) and the beginning of the
    compactified domain of the spectral grid are plotted. The domain
    boundaries are indicated by vertical dotted lines.}
  \label{fig:grid_setup}
\end{figure}

The setup of the spectral grid and the associated finite difference
grid for a typical stellar core collapse model is exemplified in
Fig.~\ref{fig:grid_setup} for $ \hat{n}_r = 33 $ grid points per
spectral radial domain and $ n_r = 200 $ finite difference grid
points. Particularly in the central parts of the star (upper panel)
the logarithmic radial spacing of the finite difference grid is
obvious. While the finite difference grid ends at the finite radius $
r_{\rm fd} $ (with the exception of four ghost zones, which are needed
for the hydrodynamic reconstruction scheme; see
Section~\ref{subsection:hrsc}), the radially compactified outermost
6th domain of the spectral grid covers the entire space to radial
infinity (lower panel). The finite difference grid is fixed in time,
while the boundaries $ r_{{\rm d}i} $ of the spectral radial domains
(and thus the radial collocation points) change adaptively during the
evolution (for details, we refer to
Section~\ref{subsubsection:axisymmetric_core_collapse}). Note that the
radial collocation points of the spectral grid, which correspond to
the roots of the Chebyshev polynomials (for the Gauss--Lobato
quadrature), are concentrated towards the domain boundaries.

Generally speaking, in order to achieve a comparable accuracy in the
representation of functions and their derivatives, the finite
difference grid needs much more points than the spectral one. For
example, when considering the representation of some function like
$ \exp(-x^2) $ on the interval $ [0, 1] $, spectral methods using
Chebyshev polynomials need $ \sim 30 $ coefficients (and grid points)
to reach machine double precision ($ 10^{-16} $) for the representation
of the function and $ 10^{-13} $ for the representation of its first
derivative. For comparison, a third order scheme based on finite
differences needs $ \sim 10^5 $ points to achieve the same accuracy.

\subsubsection{Communication between grids}
\label{subsubsection:grid_communication}

Passing information from the spectral grid to the finite difference
grid is technically very easy. Knowing the spectral coefficients of a
function, this step simply requires the evaluation of the sum
(\ref{eq:spec_decomp}) at the finite difference grid points. The
drawback of this method, as it will be discussed in
Section~\ref{subsection:interpolation_tests}, is the computational
time spent. In 3D this time can even be larger than the time spent by
the spectral elliptic solver. Going from the finite difference grid to
the spectral grid requires an actual interpolation, taking special
care to avoid Gibbs phenomena that can appear in the spectral
representation of discontinuous functions. The matter terms entering
in the sources of the gravitational field equations can be
discontinuous when a shock forms. Thus, it is necessary to smooth or
filter out high frequencies that would otherwise spoil the spectral
representation. This introduces a numerical error in the fields that
should remain within the overall error of the code. The important
point to notice is that an accurate description needs not be achieved
in the spectral representation of the sources (the hydrodynamic
quantities are well described on the finite difference grid), but in
that of the gravitational field, which is always continuous, as well
as its first derivatives.

Technically, we interpolate from the finite difference grid to the
spectral grid using a one-dimensional algorithm and intermediate
grids. We first perform an interpolation in the $ r $-direction, then
in the $ \theta $-direction and finally in the
$ \varphi $-direction. We can choose between piecewise linear or
parabolic interpolations, and a scheme that globally minimizes the
norm of the second derivative of the interpolated
function~\cite{novak_00_a}. The filtering of spectral coefficients is
performed {\em a posteriori\/} by removing the coefficients 
corresponding to higher frequencies. For example, in the radial
direction, this is done by canceling the $ c_{ijk} $ in
Eq.~(\ref{eq:spec_decomp}) for $ i $ larger than a given threshold. In
practice, best results were found when cancelling the last third of
radial coefficients. This can be linked with the so-called ``two-thirds
rule'' used for spectral computations of quadratically nonlinear
equations~\cite{boyd_01_a}. Nevertheless, a different (higher)
threshold would also give good results, in the sense that there are no
high-frequency terms rising during the metric iteration.

\subsection{High-resolution shock-capturing schemes}
\label{subsection:hrsc}

As in our previous axisymmetric
code~\cite{dimmelmeier_02_a,dimmelmeier_02_b}, in the present code the
numerical integration of the system of hydrodynamic equations is
performed using a Godunov-type scheme. Such schemes are
specifically designed to solve nonlinear hyperbolic systems of
conservation laws (see, e.g.~\cite{toro_97_a} for general definitions
and~\cite{marti_03_a,font_03_a} for specific details regarding their
use in special and general relativistic hydrodynamics). In a
Godunov-type method the knowledge of the characteristic structure of
the equations is crucial to design a solution procedure based upon
either exact or approximate Riemann solvers. These solvers, which
compute at every cell-interface of the numerical grid the solution of
local Riemann problems, guarantee the proper capturing of all
discontinuities which may appear in the flow.

The time update of the hydrodynamic
equations~(\ref{eq:hydro_conservation_equation}) from $ t^n $ to
$ t^{n + 1} $ is performed using a method of lines in combination with
a second-order (in time) conservative Runge--Kutta scheme. The basic
conservative algorithm reads:
\begin{equationarray}
    \mb{U}_{i,j,k}^{n + 1} = \mb{U}_{i,j,k}^n
    & - & \frac{\Delta t}{\Delta r_i}
    \left( \,\widehat{\!\mb{F}}{}^{r}_{\!i + 1/2,j,k} -
    \,\widehat{\!\mb{F}}{}^{r}_{\!i - 1/2,j,k} \right)
    \nonumber \\
    & - & \frac{\Delta t}{\Delta \theta}
    \left( \,\widehat{\!\mb{F}}{}^{\theta}_{\!i,j + 1/2,k} -
    \,\widehat{\!\mb{F}}{}^{\theta}_{\!i,j - 1/2,k} \right)
    \nonumber \\
    & - & \frac{\Delta t}{\Delta \varphi}
    \left( \,\widehat{\!\mb{F}}{}^{\varphi}_{\!i,j,k + 1/2} -
    \,\widehat{\!\mb{F}}{}^{\varphi}_{\!i,j,k - 1/2} \right)
    \nonumber \\ [0.5 em]
    & + & \Delta t \, \mb{Q}_{i,j,k}.
\end{equationarray}%
The index $ n $ represents the time level, and the time and space
discretization intervals are indicated by $ \Delta t $ and
$ \Delta r_i $, $ \Delta \theta $, and $ \Delta \varphi $ for the
$ r $-, $ \theta $-, and $ \varphi $-direction, respectively. The
numerical fluxes along the three coordinate directions,
$ \widehat{\!\mb{F}}{}^{r} $, $ \widehat{\!\mb{F}}{}^{\theta} $,
and $ \widehat{\!\mb{F}}{}^{\varphi} $, are computed by means of
Marquina's flux formula~\cite{donat_98_a}. A family of local Riemann 
problems is set up at every cell-interface, whose jumps are minimized 
with the use of a piecewise parabolic reconstruction procedure (PPM) 
which provides third-order accuracy in space.

We note that Godunov-type schemes have also been implemented recently
in 2D and 3D Cartesian codes designed to solve the coupled system of
the Einstein and hydrodynamic equations, as reported
in~\cite{font_99_a, font_02_a, shibata_03_a, baiotti_04_a}.

\subsection{Elliptic solvers}
\label{subsection:fd_elliptic_solvers}

In the following we present the three different approaches we have
implemented in our code to numerically solve the system of metric
equations~(\ref{eq:metric_equations}). We compare the properties of
these solvers with special focus on issues like \\
- radius and order of convergence, \\
- scaling with resolution in various coordinate directions, \\
- imposition of boundary conditions, \\
- assumptions about the radial extension of the grid, \\
- computational performance, \\
- parallelization issues, and \\
- extensibility from two to three spatial dimensions.

In order to formalize the metric equations we define a vector of unknowns
\begin{equation}
  \hat{u} = u^p = (\phi, \alpha \phi, \beta^1, \beta^2, \beta^3).
  \label{eq:metric_solution_vector}
\end{equation}
Then the metric equations~(\ref{eq:metric_equations}) can be written as
\begin{equation}
  \hat{f} (\hat{u}) = f^q (u^p) = 0,
  \label{eq:metric_equations_vector}
\end{equation}
with $ \hat{f} = f^q $ denoting the vector of the five metric
equations for $ \hat{u} $ ($ p, q = 1, \dots, 5 $). For metric
solvers~1 and~2 the metric equations are discretized at cell centers
$ (r_i, \theta_j, \varphi_k) $ on the finite difference
grid. Correspondingly, for metric solver~3 the metric equations are
evaluated at collocation points
$ (\hat{r}_i, \hat{\theta}_j, \hat{\varphi}_k) $ on the spectral
grid. Thus, when discretized, Eq.~(\ref{eq:metric_equations_vector})
transforms into the following coupled nonlinear system of equations of
dimension $ 5 \times n_r \times n_\theta \times n_\varphi $ or
$ 5 \times \hat{n}_r \times \hat{n}_\theta \times \hat{n}_\varphi $,
respectively:
\begin{equation}
  \mb{\hat{f}} (\mb{\hat{u}}) = \hat{f}_{i,j,k} (\hat{u}_{l,m,n}) =
  f_{i,j,k}^q (u_{l,m,n}^p) = 0,
  \label{eq:discretized_metric_vector_equations}
\end{equation}
with the vector of discretized equations
$ \mb{\hat{f}} = \hat{f}_{i,j,k} = f_{i,j,k}^q $ for the
unknowns $ \mb{\hat{u}} = \hat{u}_{l,m,n} = u_{l,m,n}^p $. For this
system we have to find the roots. Note that, in
general, each discretized metric equation $ f_{i,j,k}^q $ couples both
to the other metric equations through the five unknowns
(indices $ p $), and to other (neighboring) cell locations on the
grid (indices $ l, m, n $).

All three metric solvers are based on iterative methods, where the new
value for the metric $ \mb{\hat{u}}^{s + 1} $ is computed from the
value at the current iteration $ s $ by adding an increment
$ \Delta \mb{\hat{u}}^s $ which is weighted with a relaxation factor
$ f_{\rm r} $. The tolerance measure we use to control convergence of
the iteration is the maximum increment of the solution vector on the
grid the iteration is executed on, i.e.\
\begin{equation}
  \Delta \hat{u}^s_{\rm max} =
  \max\, (\Delta \mb{\hat{u}}^s) =
  \max\, (\Delta u_{i,j,k}^{p\,s}).
  \label{eq:maximum_solution_vector}
\end{equation}

\subsubsection{Multidimensional Newton--Raphson solver (Solver 1)}
\label{subsubsection:newton_raphson_solver}

Solver~1, which was already introduced in the core collapse
simulations reported in~\cite{dimmelmeier_02_a,dimmelmeier_02_b}, uses
a multi-dimensional Newton--Raphson iteration method to find the roots
of Eq.~(\ref{eq:discretized_metric_vector_equations}). Thus, solving
the nonlinear system is reduced to finding the solution of a linear
problem of the same dimension during each iteration. The matrix
$ \mb{A} $ defining the linear problem consists of the Jacobi matrix
of $ \mb{\hat{f}} $ and additional contributions originating from
boundary and symmetry conditions (see~\cite{dimmelmeier_02_a} for
further details). As the spatial derivatives in the metric equations
(which also contain mixed derivatives of second order) are
approximated by second-order central differences with a three-point
stencil, $ \mb{A} $ has a band structure with $ 1 + 2 d^2 $ bands of
blocks of size $ 5 \times 5 $, where $ d $ is the number of spatial
dimensions of the finite difference grid. Furthermore, matrix
$ \mb{A} $ is sparse and usually diagonally dominated.

A simple estimate already shows that the size $ n \times n $ of the
linear problem grows impractically large in 3D. A resolution of 100
grid points in each coordinate direction results in a square
$ (5 \times 10^6) \times (5 \times 10^6) $ matrix $ \mb{A} $. Thus,
direct (exact) inversion methods, like Gauss--Jordan elimination or
exact LU decomposition, are beyond practical applicability, as these
are roughly $ n^3 $ processes, where $ n $ is the dimension of the
matrix. Even when exploiting the sparsity and band structure of
$ \mb{A} $ the linear problem remains too large to be solved on
present-day computers in a reasonable time by using iterative methods
like successive over-relaxation (SOR) or conjugate gradient (CG)
methods with appropriate preconditioning.

Because of these computational restrictions, the use of solver~1 is
restricted to 2D axisymmetric configurations, where the matrix $
\mb{A} $ has nine bands of blocks. Even in this case, for coupled
spacetime and hydrodynamic evolutions, the choice of linear solver
methods is limited: The computational time spent by the metric solver
should not exceed the time needed for one hydrodynamical time step by
an excessive amount. We have found that a recursive block tridiagonal
sweeping method~\cite{potter_73_a} (for the actual numerical
implementation, see~\cite{dimmelmeier_02_a}) yields the best
performance for the linear problem. Here the three leftmost, middle,
and rightmost bands are combined into three new bands of $ n_r $
blocks of size $ (5 \times n_\theta) \times (5 \times n_\theta) $ and which are
inverted in a forward-backward recursion along the bands using a
standard LU decomposition scheme for dense matrices. Actual execution
times for this method and the scaling with grid resolution are given
in Section~\ref{subsubsection:convergence_2d}.

We point out that the recursion method provides us with a
non-iterative linear solver, and the Newton--Raphson method exhibits
in general very rapid and robust convergence. Therefore, solver~1
converges rapidly to an accurate solution of the metric
equations~(\ref{eq:metric_equations_vector}) even for strongly
gravitating, distorted configurations, irrespective of the relative
strength of the ``hydrodynamics'' term $ S_{\rm h} $ and ``metric''
term $ S_{\rm m} $ in the metric equations (see
Eq.~(\ref{eq:metric_equations_rhs_split})). Its convergence radius is
sufficiently large, so that even the flat Minkowski metric can be used
as an initial guess for the iteration, and the relaxation factor
$ f_{\rm r} $ can be set equal to 1. Note that in solver~1 every
metric function is treated numerically in an equal way; in particular,
the equations for each of the three vector components of the shift
vector $ \beta^i $ are solved separately.

In its current implementation, solver~1 exhibits a particular
disadvantage, which will be discussed in more detail in
Section~\ref{subsubsection:metric_fall_off}. As its spatial grid, on
which the metric equations are discretized, is not radially
compactified, there is a need for explicit boundary conditions of the
metric functions $ \hat{u} $ at the outer radial boundary of the
finite difference grid. This poses a severe problem, as there exists
no general analytic solution for the vacuum spacetime surrounding an
arbitrary rotating fluid configuration in any coordinate system. Even
in spherical symmetry, our choice of isotropic coordinates yields
equations with noncompact support terms, which leads to imprecise
boundary conditions, as demonstrated in
Section~\ref{subsection:noncompact_support}. Therefore, as an
approximate boundary condition for an arbitrary matter configuration
with gravitational mass $ M_{\rm g} $, we use the monopole field for a
static TOV solution,
\begin{equation}
  \phi = 1 + \frac{M_{\rm g}}{2 r},
  \qquad
  \alpha = \frac{1 - \frac{M_{\rm g}}{2 r}}{1 + \frac{M_{\rm g}}{2 r}},
  \qquad
  \beta^i = 0,
  \label{eq:vacuum_matching}
\end{equation}
evaluated at $ r_{\rm fd} $. The influence of this approximation on
the accuracy of the solution for typical compact stars is discussed in
Section~\ref{subsubsection:metric_fall_off}. We emphasize that the use
of a noncompactified finite radial grid is not an inherent restriction
of this solver method. However in the case of metric solver~1, for
practical reasons we have chosen to keep the original grid setup as
presented in~\cite{dimmelmeier_02_a}, where both the metric and
hydrodynamic equations are solved on the same finite difference grid.

Finally, a further drawback of solver~1 is its inefficiency regarding
scalability on parallel or vector computer architectures. The
recursive nature of the linear solver part of this method prevents
efficient distribution of the numerical load onto multiple processors
or a vector pipeline. In combination with the disadvantageous scaling
behavior of the linear solver with resolution (see also
Table~\ref{tab:runtime_metric_solver_2d} below), these practical
constraints render any extension of solver~1 to 3D beyond
feasibility.

\subsubsection{Conventional iterative integral nonlinear Poisson solver (Solver 2)}
\label{subsubsection:conventional_poisson_solver}

While solver~1 makes no particular assumption about the form of the
(elliptic) equations to be solved, solver~2 exploits the fact that the
metric equations~(\ref{eq:metric_equations}) can be written in the
form of a system of nonlinear coupled equations with a Laplace
operator on the left hand
side~(\ref{eq:metric_equations_rhs_split}). A common method to solve
such kind of equations is to keep the right hand side $ S (\hat{u}) $
fixed, solve each of the resulting decoupled linear Poisson equations,
$ \hat{\Delta} \hat{u}^{s + 1} = S (\hat{u}^s) $, and iterate until
the convergence criterion~(\ref{eq:maximum_solution_vector}) is
fulfilled.

The linear Poisson equations are transformed into
integral form by using a three-dimensional Green's function,
\begin{equationarray}
  & & \hat{u}^{s + 1} (r, \theta, \varphi) = \nonumber \\
  & & \qquad - \frac{1}{4 \pi}
  \! \int \! r'^2 dr'
  \!\! \int \! \sin \theta' d\theta'
  \!\! \int \! d \varphi' \,
  \frac{S (\hat{u}^s (r', \theta', \varphi'))}{|\mb{x} - \mb{x}'|},
  \qquad
  \label{eq:integral_poisson_equation}
\end{equationarray}%
where the spatial derivatives in $ S $ are approximated by central
finite differences. The volume integral on the right hand side of
Eq.~(\ref{eq:integral_poisson_equation}) is numerically evaluated by
expanding the denominator into a series of radial
functions $ f_l (r, r') $ and associated Legendre polynomials
$ P_l^m (\cos \theta) $, which we cut at $ l = 10 $. The integration
in Eq.~(\ref{eq:integral_poisson_equation}), which has to be performed
at every grid point, yields a problem of numerical size
$ (n_r \times n_\theta \times n_\varphi)^2 $. However, the problem
size can be reduced to $ n_r \times n_\theta \times n_\varphi $ by
recursion. Thus, solver~2 scales linearly
with the grid resolution in all spatial dimensions (see
Section~\ref{subsubsection:convergence_2d}). However, while the
numerical solution of an integral equation like
Eq.~(\ref{eq:integral_poisson_equation}) is well
parallelizable, the recursive method which we employ to improve the
resolution scaling performance poses a severe obstacle. In practice
only the parallelization across the expansion series index $ l $ (or
possibly cyclic reduction) can be used to distribute the computational
workload over several processors.

An advantage of solver~2 is that it does not require the
imposition of explicit boundary conditions at a finite radius due to
the integral form of the equations. Demanding asymptotic flatness at
spatial infinity fixes the integration constants in
Eq.~(\ref{eq:integral_poisson_equation}). However, as the metric
equations contain in general source terms with noncompact support (see
Section~\ref{subsection:noncompact_support}), the radial integration
must be performed up to infinity to account for the source term
contributions. As the discretization scheme used in solver~2 limits the
radial integration to some finite radius $ r_{\rm fd} $, the metric
equations are solved only approximately if the source terms with
noncompact support are nonzero. The consequences of this fact are
discussed in Section~\ref{subsubsection:metric_fall_off}. As in the
case of metric solver~1, the metric solver~2 could be used with a
compactified radial coordinate as well.

One major disadvantage of solver~2 is its slow convergence rate and a
small convergence radius. For simplicity, we decompose the metric
vector equation for the shift vector $ \beta^i $
into three scalar equations for its components. If the
$ \theta $-component of the shift vector does not vanish,
$ \beta^2 \ne 0 $, and if the spacetime is nonaxisymmetric, solver~2
does not converge at all (probably due to diverging terms like
$ \beta^\theta / \sin^2 \theta $ in the vector Laplace operator). Even
when using a known solution obtained with another metric solver as
initial guess, solver~2 fails to converge. Thus, the use of
solver~2 is limited to axisymmetry. Even so, when $ \beta^2 \ne 0 $,
a quite small relaxation factor $ f_{\rm r} \approx 0.05 $ is
required. Furthermore, as the iteration scheme is of fix-point
type, it already has a much lower convergence rate than e.g.\ a
Newton--Raphson scheme. Both factors result in typically
several hundred iterations until convergence is reached (see
Section~\ref{subsubsection:convergence_2d}). For strong gravity, the
small convergence radius restricts the initial guess to a metric close
to the actual solution of the discretized equations.

\subsubsection{Iterative spectral nonlinear Poisson solver (Solver 3)}
\label{subsubsection:spectral_metric_solver}

The basic principles of this iterative solver are similar to the ones
used for solver~2: A numerical solution of the nonlinear elliptic
system of the metric differential equations is obtained by solving the
associated linear Poisson equations with a fix-point iteration
procedure until convergence.  However, instead of using finite
difference scalar Poisson solvers, solver~3 is built from routines of
the publicly available {\sc Lorene} library~\cite{lorene_code} and
uses spectral methods to solve scalar and vector Poisson
equations~\cite{grandclement_01_a}.

Before every computation of the spacetime metric, the hydrodynamic and
metric fields are interpolated from the finite difference to the
spectral grid by the methods detailed in
Section~\ref{subsubsection:grid_communication}. All three-dimensional
functions are decomposed into Chebyshev polynomials $ T_n(r) $
and spherical harmonics $ Y_l^m (\theta, \varphi) $ in each
domain. When using solver~3 the metric
equations~(\ref{eq:adm_metric_equations}) are rewritten in order to
gain accuracy according to the following transformations. The scalar
metric functions $ \phi $ and $ \alpha $ have the same type of
asymptotic behavior near spatial infinity,
$ \phi|_{r \to \infty} \sim 1 + \Delta \phi (r) $, 
$ \alpha|_{r \to \infty} \sim 1 + \Delta \alpha (r) $,
with $ \Delta \phi (r) $ and $ \Delta \alpha (r) $ approaching 0 as
$ r \to \infty $. Therefore, to obtain a more precise numerical
description of the (usually small) deviations of $ \phi $ and
$ \alpha $ from unity, we solve the equations for the logarithm of
$ \phi $ and $ \alpha \phi $, imposing that $ \ln \phi $ and
$ \ln (\alpha \phi) $ approach zero at spatial infinity. Another
important difference to the other two solvers is that the
vector Poisson equation for the shift vector $ \beta^i $ is not
decomposed into single scalar components, but instead the entire
linear vector Poisson equation is solved, including the
$ \frac{1}{3} \hat{\nabla}^i \hat{\nabla}_k $ operator on the left
hand side. Therefore, the system of metric equation to be solved reads
\begin{equation}
  \setlength{\arraycolsep}{0.14 em}
  \begin{array}{rcl}
    \hat{\Delta} \ln \phi & = & \displaystyle
    - 4 \pi \phi^4 \left(\rho h W^2 - P +
    \frac{K_{ij} K^{ij}}{16 \pi} \right) \\ [0.7 em]
    & & \displaystyle
    - \hat{\nabla}^i \ln \phi \; \hat{\nabla}_i \ln \phi, \\ [0.7 em]
    \hat{\Delta} \ln \alpha \phi & = & \displaystyle
    2 \pi \phi^4 \left( \rho h (3 W^2 - 2) + 5 P +
    \frac{7 K_{ij} K^{ij}}{16 \pi} \right) 
    \!\!\!\!\!\!\!\!\!\!\! \\ [0.7 em]
    & & \displaystyle
    - \hat{\nabla}^i \ln \alpha \phi \;
    \hat{\nabla}_i \ln \alpha \phi, \\ [0.7 em]
    \hat{\Delta} \beta^i & + & \displaystyle 
    \frac{1}{3} \hat{\nabla}^i \hat{\nabla}_k \beta^k =
    \displaystyle 16 \pi \alpha \phi^4 S^i +
    2 \phi^{10} K^{ij} \hat{\nabla}_j
    \left( \frac{\alpha}{\phi^6} \right)\!.
    \!\!\!\!\!\!\!\!\!\!\!\!\!\!\!\!\!
  \end{array}
  \label{eq:ln_phi_alpha_metric_equation}
\end{equation}
During each iteration a spectral representation of the solution of the
linear scalar and vector Poisson equations associated with the above
system is obtained. The Laplace operator is inverted (i.e.\ the
linear Poisson equation is solved) in the following way: For a given pair of
indices $ l $ and $ m $ of $ Y_l^m (\theta, \varphi) $, the linear
scalar Poisson equation reduces to an ordinary differential equation
in $ r $. The action of the differential operator 
\begin{equation}
  \frac{\partial^2}{\partial r^2} +
  \frac{2}{r} \frac{\partial}{\partial r} -
  \frac{l (l + 1)}{r^2}
  \label{eq:poisson_operator_lm}
\end{equation}
acting thus on each multipolar component ($ l $ and $ m $) of a scalar
function corresponds to a matrix multiplication in the Chebyshev
coefficient space. The corresponding matrix is inverted to obtain a
particular solution in each domain, which is then combined with
homogeneous solutions ($ r^l $ and $ 1 / r^l $, for a given $ l $) to
satisfy regularity and boundary conditions. The matrix has a small
size (about $ 30 \times 30 $) and can be put into a banded form, owing
to the properties of the Chebyshev polynomials, which facilitates its
fast inversion. For more details about this procedure, and how the
vector Poisson equation is treated, the interested reader is addressed
to~\cite{grandclement_01_a}. Note also that when solving the shift
vector equation, $ \beta^i $ is decomposed into Cartesian components
defined on the spherical polar grid (see~\cite{grandclement_01_a}).

The spatial differentials in the source terms on the right hand sides
of the metric equations are approximated by second-order central
differences in solvers~1 and~2, while they are obtained by spectral
methods in solver~3 (see
Section~\ref{subsubsection:spectral_methods}). When using $ \sim 30 $
collocation points, very high precision ($ \sim 10^{-13} $) can be
achieved in the evaluation of these derivatives. Another advantage of
metric solver~3 is that a compactified radial coordinate $ u = 1 / r $
enables us to solve for the entire space, and to impose exact boundary
conditions at spatial infinity, $ u = 0 $. This ensures both
asymptotic flatness and fully accounts for the effects of the source
terms in the metric equations with noncompact support. Solver~3 uses
the same fix-point iteration method as solver~2, but does not suffer
from the convergence problem encountered with that solver. Due to the
direct solution of the vector Poisson equation for the shift vector
$ \beta^i $, it converges to the correct solution in all investigated
models (including highly distorted 3D matter configurations with
velocity perturbations, see
Section~\ref{subsubsection:convergence_2d}). Furthermore, this can be
achieved with the maximum possible relaxation factor,
$ f_{\rm r} = 1 $, starting from the flat metric as initial guess.

However, the strongest reason in favor of solver~3 is its
straightforward extension to 3D. As mentioned previously, both
metric solvers~1 and~2 are limited to axisymmetric situations. The
spectral elliptic solvers provided by the {\sc Lorene} library are
already intrinsically three-dimensional. Indeed, even in
axisymmetry the spectral grid of solver~3 requires
$ \hat{n}_\varphi = 4 $ grid points in the $ \varphi $-direction
order to correctly represent the Cartesian components of the shift
vector.

There is an additional computational overhead due to the communication
between the finite difference and the spectral grids. These
computational costs may actually become a dominant part when
calculating the metric (as will be shown in
Section~\ref{subsection:interpolation_tests}). The interpolation
methods also have to be chosen carefully to obtain the desired
accuracy. Furthermore, spectral methods may suffer from Gibbs
phenomena if the source terms of the Poisson-like equations contain
discontinuities. For the particular type of simulations we are aiming
at, discontinuities are present (supernova shock front, discontinuity
at the transition from the stellar matter distribution to the
artificial atmosphere at the boundary of the star). This can result in
high-frequency spurious oscillations of the metric solution, if too
few radial domains are used, or if the boundaries of the spectral
domains are not chosen properly. As mentioned before, a simple way to
reduce the oscillations is to filter out part of the high-frequency
spectral coefficients.

As the C++ routines of the {\sc Lorene} library in the current release
are optimized for neither vector nor parallel computers, solver~3
cannot yet exploit these architectures. However, we were able to
improve the computational performance by coarse-grain parallelizing
the routines which interpolate the metric solution in the spectral
representation to the finite difference grid.

\subsection{Extraction of gravitational waves}
\label{subsection:wave_extraction}

In a conformally flat spacetime the dynamical gravitational wave
degrees of freedom are not present~\cite{dimmelmeier_02_a}. Therefore,
in order to extract information regarding the gravitational radiation
emitted in core collapse events and in rotating neutron star
evolutions, we have implemented in the code the 3D generalization of
the axisymmetric Newtonian quadrupole formula used
in~\cite{dimmelmeier_01_a, dimmelmeier_02_a, dimmelmeier_02_b}. Note
that we use spherical polar components for the tensors of
the radiation field.

Whereas in axisymmetry there exists only one independent component of
the quadrupole gravitational radiation field $ h_{ij}^{\rm TT} $ in
the transverse traceless gauge,
\begin{equation}
  h_{ij}^{\rm TT} (r, \theta) = \frac{1}{r}
  A_+ (\theta) e_+,
  \label{eq:h_TT_2d}
\end{equation}
in three dimensions we have
\begin{equation}
  h_{ij}^{\rm TT} (r, \theta, \varphi) = \frac{1}{r}
  \left[ A_+ (\theta, \varphi) \mb{e}_+ +
  A_\times (\theta, \varphi) \mb{e}_\times  \right],
  \label{eq:h_TT_3d}
\end{equation}
with the unit vectors $ \mb{e}_+ $ and $ \mb{e}_\times $ defined as
\begin{equationarray}
  \mb{e}_+ & = & \mb{e}_\theta \otimes \mb{e}_\theta -
  \mb{e}_\varphi \otimes e_\varphi, \\
  \mb{e}_\times & = & \mb{e}_\theta \otimes \mb{e}_\varphi +
  \mb{e}_\varphi \otimes e_\theta.
  \label{eq:e_plus_e_cross}
\end{equationarray}%

The amplitudes $ A_+ $ and $ A_\times $ are linear combinations of
the second time derivative of some components of the quadrupole moment
tensor $ I_{ij} $, which for simplicity we evaluate at
$ \varphi = 0 $ on the polar axis and in the equatorial plane,
respectively:
\begin{equationarray}
  \begin{array}{rcl}
    A_+^{\rm p} & = & \ddotIsub{11} - \ddotIsub{22}, \\ [0.25 em]
    A_\times^{\rm p} & = & 2 \ddotIsub{12},
  \end{array} & & \qquad \mbox{at } \theta = 0 \mbox{ (pole)}, \\ [0.5 em]
  \begin{array}{rcl}
    A_+^{\rm e} & = & \ddotIsub{33} - \ddotIsub{22}, \\ [0.25 em]
    A_\times^{\rm e} & = & - 2 \ddotIsub{13},
  \end{array} & & \qquad \mbox{at } \theta = \pi / 2 \mbox{ (equator)}.
  \label{eq:A_plus_A_cross_pole_equator_3d}
\end{equationarray}%

A direct numerical calculation of the quadrupole moment in the {\em
standard quadrupole formulation\/},
\begin{equation}
  I_{ij} = \!\! \int \! dV \rho^* \!
  \left[ x_i x_j - \frac{1}{3} \delta_{ij}
  \left( x_1^2 + x_2^2 + x_3^2 \right) \right]\!,
  \label{eq:I_ij_sqf_3d}
\end{equation}
results in high frequency noise completely dominating the wave signal
due to the presence of the second time derivatives in
Eq.~(\ref{eq:A_plus_A_cross_pole_equator_3d}). Therefore, we make use
of the time-differentiated quadrupole moment in the {\em first moment
of momentum density formulation\/},
\begin{equation}
  \dotI{ij} = \!\! \int \! dV \rho^* \!
  \left[ v_i x_j + v_j x_i - \frac{2}{3} \delta_{ij}
  \left( v_1 x_1 + v_2 x_2 + v_3 x_3 \right) \right]\!,
  \label{eq:I_ij_fmomdf_3d}
\end{equation}
and {\em stress formulation\/},
\begin{equation}
  \ddotIij = \!\! \int \! dV \rho^*
  \left[ 2 v_i v_j - x_i \partial_j \Phi - x_j \partial_i \Phi \right]\!,
  \label{eq:I_ij_sf_3d}
\end{equation}
of the quadrupole formula~\cite{finn_89_a, blanchet_90_a}.

In the above equations, $ x_i $ and $ v_i $ are the coordinates and
velocities in Cartesian coordinates, respectively. When evaluating
Eq.~(\ref{eq:I_ij_sf_3d}) numerically, we transform $ v_i $
to spherical polar coordinates. In the quadrupole moment, we use
$ \rho^* = \rho W \phi^6 $ instead of $ \rho $ as
in~\cite{dimmelmeier_01_a, dimmelmeier_02_a, dimmelmeier_02_b}, as
this quantity is evolved by the continuity equation (note that
both quantities have the same Newtonian limit). This also allows a
direct comparison with the results presented in~\cite{shibata_03_b},
which we show in
Section~\ref{subsubsection:comparison_with_full_gr}. For a discussion
about the ambiguities arising from the spatial derivatives of the
Newtonian potential $ \Phi $ in Eq.~(\ref{eq:I_ij_sf_3d}) in a general
relativistic framework and their solution (which we also employ in
this work), we refer to~\cite{dimmelmeier_02_b}.

The total energy emitted by gravitational waves can be expressed
either as a time integral,
\begin{equationarray}
  E_{\rm gw} & = & \frac{2}{15} \int dt
  \biggl[ - \dddotIsub{11} \dddotIsub{22} -
  \dddotIsub{11} \dddotIsub{33} - \dddotIsub{22} \dddotIsub{33} 
  \nonumber \\
  & & + \dddotIsubsquared{11} + \dddotIsubsquared{22} +
  \dddotIsubsquared{33} + 3 \left( \dddotIsubsquared{12} +
  \dddotIsubsquared{13} + \dddotIsubsquared{23} \right) \biggr],
  \quad
  \label{eq:E_grav_time_3d}
\end{equationarray}%
or, equivalently, as a frequency integral,
\begin{equationarray}
  E_{\rm gw} & = & \frac{1}{15} \int \nu^2 d\nu
  \biggl[ - \ddotIhatsub{11} \ddotIhatsub{22} -
  \ddotIhatsub{11} \ddotIhatsub{33} -
  \ddotIhatsub{22} \ddotIhatsub{33} 
  \nonumber \\
  & & + \ddotIhatsubsquared{11} + \ddotIhatsubsquared{22} +
  \ddotIhatsubsquared{33} + 3 \left( \ddotIhatsubsquared{12} +
  \ddotIhatsubsquared{13} + \ddotIhatsubsquared{23} \right) \biggr],
  \qquad
  \label{eq:E_grav_frequency_3d}
\end{equationarray}%
where $ \ddotIhatij (\nu) $ is the Fourier transform of
$ \ddotIij (t) $. We point out that the above general expressions
reduce to the following ones in axisymmetry:
\begin{equationarray}
  &
  A_+^{\rm p} = 0,
  \qquad
  A_\times^{\rm p} = 0,
  \qquad
  A_+^{\rm e} = \ddotI,
  \qquad
  A_\times^{\rm e} = 0,
  &
  \qquad
  \\
  & \displaystyle
  E_{\rm gw} =
  \frac{2}{15} \int dt \, \dddotIsquared =
  \frac{1}{15} \int \nu^2 d\nu \, \ddotIhatsquared,
  &
  \label{eq:E_grav_2d}
\end{equationarray}%
with $ I = I_{33} - I_{22} $ being the only nonzero independent
component of the quadrupole tensor, and $ \ddotIhatsquared $ being the
Fourier transform of $ \ddotIsquared $. The quadrupole wave amplitude
$ A^{\rm E2}_{20} $ used in~\cite{zwerger_97_a, dimmelmeier_01_a,
  dimmelmeier_02_b} is related to $ I $ according to
$ A^{\rm E2}_{20} = 8 \sqrt{\pi / 15} \, \ddotI $.

We have tested the equivalence between the waveforms obtained by the
axisymmetric code presented
in~\cite{dimmelmeier_01_a, dimmelmeier_02_a, dimmelmeier_02_b} and
those by the current three-dimensional code using the corresponding
axisymmetric model. In all investigated cases, they agree with
excellent precision.

\section{Code tests and applications}
\label{section:tests}

We turn now to an assessment of the numerical code with a variety of
tests and applications. We recall that we do not attempt in the
present paper to investigate any realistic astrophysical scenario,
which is deferred to subsequent publications. Instead, we
focus here on discussing standard tests for general relativistic
three-dimensional hydrodynamics code, which were all passed by our
code. In particular, we show that the code exhibits long-term
stability when evolving strongly gravitating systems like rotational
core collapse and equilibrium configurations of (highly perturbed)
rotating relativistic stars. Each separate constituent methods of the
code (HRSC schemes for the hydrodynamics equations and elliptic
solvers based on spectral methods for the gravitational field
equations) has already been thoroughly tested and successfully applied
in the past (see e.g.~\cite{font_03_a, marti_03_a, grandclement_01_a}
and references therein). Therefore, we mainly demonstrate here that
the coupled numerical schemes work together as desired.

\subsection{Interpolation efficiency and accuracy}
\label{subsection:interpolation_tests}

\begin{table}[t]
  \caption{Execution time $ t_{\rm fd \rightarrow sp} $ and accuracy
    $ \Delta f_{\rm fd \rightarrow sp} $ for the interpolation of a
    test function $ f_{\rm t} (r, \theta, \varphi) $ (see text) from
    the finite difference grid to the spectral grid, listed for
    different finite difference grid resolutions
    $ n_r \times n_\theta \times n_\varphi $ and interpolation
    types. The interpolation methods are piecewise linear (type 1),
    piecewise parabolic (type 2), and globally minimizing the norm of
    the second derivative of the interpolated
    function~\cite{novak_00_a} (type 0). The spectral grid has a
    resolution of $ \hat{n}_r = 17 $, $ \hat{n}_\theta = 17 $, and
    $ \hat{n}_\varphi = 16 $ grid points.}
  \label{tab:runtime_interpolation_fd_to_sp}
  \begin{ruledtabular}
    \begin{tabular}{c@{~~}|c@{~~}|c@{~~}|c@{~~}}
      $ n_r \times n_\theta \times n_\varphi $ &
      Type &
      $ t_{\rm fd\rightarrow sp} $ [s] & 
      $ \Delta f_{\rm fd\rightarrow sp} $ [$ L_0 $ norm] \\
      \hline \rule{0 em}{1.0 em}%
      $    400 \times    200 \times    800 $ & 2 & 5.13 & $ 5.0 \times 10^{-8} $ \\
      $    400 \times    200 \times    800 $ & 1 & 5.12 & $ 7.0 \times 10^{-6} $ \\
      $    400 \times    200 \times    800 $ & 0 & 9.44 & $ 1.8 \times 10^{-6} $ \\ [0.7 em]
      $    400 \times    200 \times    400 $ & 2 & 2.92 & $ 3.1 \times 10^{-7} $ \\
      $    400 \times    200 \times    200 $ & 2 & 1.43 & $ 1.6 \times 10^{-6} $ \\
      $    400 \times    200 \times    100 $ & 2 & 0.77 & $ 1.7 \times 10^{-5} $ \\
      $    400 \times    200 \times \wz 10 $ & 2 & 0.09 & $ 1.3 \times 10^{-2} $ \\ [0.7 em]
      $    400 \times    100 \times    800 $ & 2 & 2.55 & $ 3.1 \times 10^{-7} $ \\
      $    400 \times \wz 50 \times    800 $ & 2 & 1.60 & $ 1.8 \times 10^{-6} $ \\
      $    400 \times \wwz 5 \times    800 $ & 2 & 0.32 & $ 2.0 \times 10^{-3} $ \\ [0.7 em]
      $    200 \times    200 \times    800 $ & 2 & 3.61 & $ 2.7 \times 10^{-7} $ \\
      $    100 \times    200 \times    800 $ & 2 & 1.81 & $ 2.1 \times 10^{-6} $ \\
      $ \wz 50 \times    200 \times    800 $ & 2 & 1.40 & $ 1.6 \times 10^{-5} $ \\
      $ \wwz 5 \times    200 \times    800 $ & 2 & 0.99 & $ 1.4 \times 10^{-2} $ \\
    \end{tabular}
  \end{ruledtabular}
\end{table}

The interpolation procedure from the finite difference grid to the
spectral grid has been described in
Section~\ref{subsubsection:grid_communication}. Among the three
possible algorithms we have implemented in the code, the most
efficient turned out to be the one based on a piecewise parabolic
interpolation (see Table~\ref{tab:runtime_interpolation_fd_to_sp}). It
is as fast as the piecewise linear interpolation, and more accurate than
the algorithm based on the minimization of the second derivative of
the interpolated function. Table~\ref{tab:runtime_interpolation_fd_to_sp}
shows, for a particular example of an interpolated test function
$ f_{\rm t} (r, \theta, \varphi) =
\exp \left[ - r^2 (1 + \sin^2 \theta \cos^2 \varphi) \right] $,
the relative accuracy $ \Delta f_{\rm int} $ (in the $ L_0 $ norm)
achieved by this interpolation, as well as the CPU time spent on a
Pentium IV Xeon processor at 2.2 GHz. The spectral grid consists of two
domains (nucleus + shell) with $ \hat{n}_r = 17 $,
$ \hat{n}_\theta = 17 $, and $ \hat{n}_\varphi = 16 $. The outer
radius of the nucleus is located at 0.5, and the outer boundary of the
shell is at 1.5 (corresponding to the radius of the finite difference
grid $ r_{\rm fd} $).

This test demonstrates that the piecewise parabolic interpolation is
indeed third-order accurate, and that the time spent scales roughly
linearly with the number of points of the finite difference grid in
any direction. We have made other tests which show that the
interpolation accuracy is independent of $ \hat{n} $, and that it
scales in time like $ {\cal O} \left( \hat{n}^3 \right) + {\cal O}
\left( n^3 \right) $, where $ \hat{n} $ and $ n $ are the number of
points used in each dimension by the spectral and the finite
difference grid, respectively. The interpolation is exact, up to
machine precision, for functions which can be expressed as polynomials
of degree $ \leq 2 $ with respect to all three coordinates.

\begin{table}[t]
  \caption{Execution time $ t_{\rm sp\rightarrow fd} $ and accuracy
    $ \Delta f_{\rm sp \rightarrow fd} $ for the evaluation of a test
    function $ f_{\rm t} (r, \theta, \varphi) $ (see text) on the
    finite difference grid from its representation in spectral
    coefficients, listed for different numbers of spectral
    grid points $ \hat{n}_r \times \hat{n}_\theta \times \hat{n}_\varphi $.
    The finite difference grid has a resolution of $ n_r = 100 $,
    $ n_\theta = 50 $, and $ n_\varphi = 30 $ grid points.}
  \label{tab:runtime_interpolation_sp_to_fd}
  \begin{ruledtabular}
    \begin{tabular}{@{~~~~}c@{~~~~}|c@{~~~~}|c@{~~~~}}
      $ \hat{n}_r \times \hat{n}_\theta \times \hat{n}_\varphi $ &
      $ t_{\rm sp\rightarrow fd} $ [s] & 
      $ \Delta f_{\rm sp\rightarrow fd} $ [$ L_0 $ norm] \\
      \hline \rule{0 em}{1.0 em}%
      $ 33    \times    17 \times    64 $ & 75.8 & $ 1.5 \times 10^{-15}     $ \\
      $ 33    \times    17 \times    32 $ & 38.4 & $ 5.5 \times 10^{-9 \swz} $ \\
      $ 33    \times    17 \times    16 $ & 19.6 & $ 2.6 \times 10^{-4 \swz} $ \\
      $ 33    \times    17 \times \wz 8 $ & 10.3 & $ 2.8 \times 10^{-2 \swz} $ \\ [0.7 em]
      $ 33    \times \wz 9 \times    64 $ & 40.8 & $ 6.4 \times 10^{-9 \swz} $ \\
      $ 33    \times \wz 5 \times    64 $ & 23.4 & $ 3.2 \times 10^{-4 \swz} $ \\ [0.7 em]
      $ 17    \times    17 \times    64 $ & 41.2 & $ 1.9 \times 10^{-13}     $ \\
      $ \wz 9 \times    17 \times    64 $ & 24.6 & $ 9.2 \times 10^{-7 \swz} $ \\
      $ \wz 5 \times    17 \times    64 $ & 16.7 & $ 1.9 \times 10^{-3 \swz} $ \\
    \end{tabular}
  \end{ruledtabular}
\end{table}

The direct spectral summation from the spectral to the finite
difference grid is a very precise way of evaluating a function: For
smooth functions, the relative error decreases like
$ \exp({- \hat{n}}) $ (infinite order scheme). This property is
fulfilled in our code, as shown in
Table~\ref{tab:runtime_interpolation_sp_to_fd} for the same test function
$ f_{\rm t} (r, \theta, \varphi) $ and the same domain setup as for
Table~\ref{tab:runtime_interpolation_fd_to_sp} (again the timings are
for a Pentium IV Xeon processor at 2.2 GHz). Double precision
accuracy is reached with a reasonable number of points ($ \hat{n}_r = 33 $,
$ \hat{n}_\theta = 17 $, and $ \hat{n}_\varphi = 64 $). According to
Table~\ref{tab:runtime_interpolation_sp_to_fd}
the CPU cost scales linearly with the number of coefficients
$ \hat{n} $ in any direction. We have also confirmed that it scales
linearly with the number of finite difference grid points $ n $ in any
direction. The drawback of this most straightforward procedure
is that it requires $ {\cal O} \left(\hat{n}^3 n^3 \right) $
operations, which is much more expensive than the interpolation from
the finite difference grid to the spectral one, and even more
expensive than the iterative procedure providing the solution of
system~(\ref{eq:ln_phi_alpha_metric_equation}). Nevertheless, it is
computationally not prohibitive since the overall accuracy of the code
does not depend on $ \hat{n} $ (which can thus remain small). A way to
reduce the execution time is to use a partial summation
algorithm (see e.g.~\cite{boyd_01_a}), which needs
only $ {\cal O} \left( \hat{n} n^3 \right) + {\cal O} \left( \hat{n}^2
n^2 \right) + {\cal O} \left( \hat{n}^3 n \right) $ operations, at the
additional cost of increased central memory requirement. Another
alternative is to truncate the spectral sum, staying at an accuracy
level comparable to that of finite difference differential operators.

\subsection{Solver comparison in 2D}
\label{subsection:2d_comparison_tests}

\subsubsection{Convergence properties}
\label{subsubsection:convergence_2d}

The theoretical considerations about the convergence properties of the
three implemented metric solvers (as outlined in
Section~\ref{subsection:fd_elliptic_solvers}) are checked by solving
the spacetime metric for a 2D axisymmetric rotating
neutron star model in equilibrium (labeled model RNS), which we have
constructed with the method described in Komatsu et al.~\cite{komatsu_89_a}. 
This model has a central density
$ \rho_{\rm c} = 7.905 \times 10^{14} {\rm\ g\ cm}^{-3} $, obeys a
polytropic EoS with $ \gamma = 2 $ and
$ K = 1.455 \times 10^5 $ (in cgs units), and rotates
rigidly at the mass shedding limit, which corresponds to a
polar-to-equatorial axis ratio of 0.65. These model parameters are
equivalent to those used for neutron star models in~\cite{font_00_a,
  font_02_a}.

\begin{figure}[t]
  \epsfxsize = 8.6 cm
  \centerline{\epsfbox{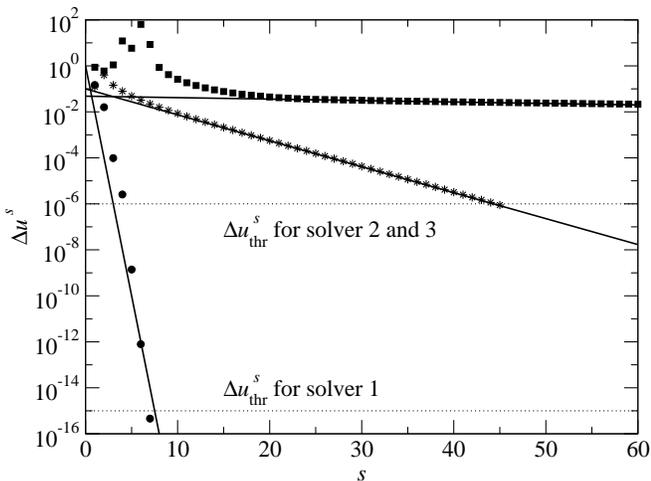}}
  \caption{Comparison of the convergence behavior for the three metric
    solvers in 2D. For solver~1 (filled circles), the maximum
    increment $ \Delta \hat{u}^s_{\rm max} $ per iteration $ s $
    decreases to the threshold $ \Delta \hat{u}^s_{\rm thr}=10^{-15} $
    (lower horizontal dotted line) within less than 10 iterations,
    while solver~3 (asterisks) needs more than 40 iterations to reach
    its (less restrictive) threshold (upper horizontal dotted line) of
    $ 10^{-6} $. The very low relaxation factor needed for solver~2
    (filled squares) results in a remarkably slow convergence,
    requiring more than 700 iterations. The solid lines mark the
    approximate linear decrease of
    $ \log \Delta \hat{u}^s_{\rm max} $.}
  \label{fig:convergence_metric_solver_2d}
\end{figure}

To the initial equilibrium model we add an $ r $- and
$ \theta $-dependent density and velocity perturbation,
\begin{equation}
  \setlength{\arraycolsep}{0.14 em}
  \begin{array}{rcl}
    \rho & = &
    \displaystyle \rho_{\rm ini} \left[ 1 +
    0.02 \sin^2 \! \left( \! \pi \frac{r}{r_{\rm s}} \right)
    \left( 1 + \sin^2 (2 \theta) \right) \right], \\ [1 em]
    v_r & = & \displaystyle
    0.05 \sin^2 \! \left( \! \pi \frac{r}{r_{\rm s}} \right)
    \left( 1 + \sin^2 (2 \theta) \right), \\ [1 em]
    v_\theta & = & \displaystyle
    0.05 \sin^2 \! \left( \! \pi \frac{r}{r_{\rm s}} \right)
    \sin^2 (2 \theta), \\ [1 em]
    v_\varphi & = & \displaystyle v_{\varphi\,\rm ini} +
    0.05 \sin^2 \! \left( \! \pi \frac{r}{r_{\rm s}} \right)
    \left( 1 + \sin^2 (2 \theta) \right),
  \end{array}
  \label{eq:perturbation_metric_test_2d}
\end{equation}
where $ r_{\rm s} $ is the ($ \theta $-dependent) stellar radius, and
$ v_r = \sqrt{v_1 v^1} $, $ v_\theta = \sqrt{v_2 v^2} $, and
$ v_\varphi = \sqrt{v_3 v^3} $. The metric
equations (Eqs.~(\ref{eq:metric_equations}) for solvers~1 and~2, and 
Eqs.~(\ref{eq:ln_phi_alpha_metric_equation}) in the case of
solver~3) are then solved using the three implemented metric solvers.
The perturbation of $ v_r $ and $ v_\theta $ ensures that the metric
equations yield the general case of a shift vector with three nonzero
components, which cannot be obtained with an initial model in
equilibrium.

We point out that by adding the perturbations specified in
Eq.~(\ref{eq:perturbation_metric_test_2d}) and calculating the
metric for these perturbed initial data, we add a small inconsistency
to the initial value problem. As the Lorentz factor $ W $ in the
right hand sides of the metric equations contains metric contributions
(which are needed for computing the covariant velocity components), it
would have to be iterated with the metric solution until
convergence. However, as the perturbation amplitude is small, and as
we do not evolve the perturbed initial data, we neglect this small
inconsistency.

The most relevant quantity related to convergence properties of the
metric solver is the maximum increment $ \Delta \hat{u}^s_{\rm max} $
of all metric components on the grid (see
Fig.~\ref{fig:convergence_metric_solver_2d}). As expected solver~1
exhibits the typical quadratic decline of a Newton--Raphson solver to
its threshold value $ \Delta \hat{u}^s_{\rm thr} = 10^{-15} $. As the
methods implemented in solvers~2 and~3 correspond to a fix-point
iteration, the decline of their metric increment is significantly
slower. Therefore, for the Poisson-based solvers, we typically use a
less restrictive threshold $ \Delta \hat{u}^s_{\rm thr} = 10^{-6} $.
While the spectral Poisson solver~3 allows for a relaxation factor of
1 and thus for a still quite rapid convergence, the conventional
Poisson solver~2 requires more than 700 iterations due to its much
smaller relaxation factor imposed by the $ \beta^2 $-equation.

It is worth stressing that all three solvers show rather robust
convergence, if one keeps in mind that the initial guess is the flat
spacetime metric. If the metric is changing dynamically during an
evolution, the metric values from the previous computation can be used
as new starting values, which reduces the number of iterations by
about a factor of two with respect to those reported in
Fig.~\ref{fig:convergence_metric_solver_2d}.

Besides the convergence rate, the execution time $ t_{\rm m} $
required for a single metric computation and its dependence on the
grid resolution is also of paramount relevance for the practical
usefulness of a solver. These times for one metric computation of the
perturbed RNS stellar model on a finite difference grid with various $
r $- and $ \theta $-resolutions on an IBM RS/6000 Power4 processor are
summarized in Table~\ref{tab:runtime_metric_solver_2d}. As
theoretically expected, both solver~1 and 2 show a linear scaling of
$ t_{\rm m} $ with the number of radial grid points $ n_r $, i.e.\ the
ratio $ r_{n_r} = t_{\rm m} (n_r) / t_{\rm m} (n_{r/2}) $ is
approximately 2. While the integration method of solver~2 shows linear
dependence also for the number of meridional grid zones $ n_\theta $,
the inversion of the dense $ n_\theta \times n_\theta $ matrices
during the radial sweeps in solver~1 is roughly a $ n_\theta^3 $
process. Thus, the theoretical value of $ r_{n_\theta} = 8 $ for that
solver is well met by the results shown in
Table~\ref{tab:runtime_metric_solver_2d}. We note that for even larger
values of $ n_\theta $, specific processor properties like cache-miss
problems can even worsen the already cubic scaling of solver~1, while
for $ n_\theta \gtrsim 64 $ solver~2 fails to converge altogether. On
the other hand for solver~3 $ t_{\rm m} $ is approximately
independent of the number of finite difference grid points in either
coordinate direction, as the number of spectral collocation points is
fixed. A dependence on $ n_r $ and $ n_\theta $ can only enter via the
interpolation procedure between the two grids, the time for which is,
however, entirely negligible in 2D.

\begin{table}[t]
  \caption{Metric solver execution time $ t_{\rm m} $ for different
    finite difference grid resolutions $ n_r \times n_\theta $ for the
    three metric solvers in 2D applied to the perturbed rotating
    neutron star model RNS. The ratios $ a_{n_r} $ ($ a_{n_{\theta}} $)
    between execution times for a given $ n_r $ ($ n_\theta $) and for
    half that resolution exhibit the behavior expected from
    theoretical considerations. The spectral grid has a resolution of
    $ \hat{n}_r = 33 $, $ \hat{n}_\theta = 17 $, and
    $ \hat{n}_\varphi = 4 $ grid points.}
  \label{tab:runtime_metric_solver_2d}
  \begin{ruledtabular}
    \begin{tabular}{c@{~}|rrr@{~}|rrr@{~}|rrr@{~}}
      &
      \multicolumn{3}{c@{~}|}{Solver 1} &
      \multicolumn{3}{c@{~}|}{Solver 2} &
      \multicolumn{3}{c@{~}}{Solver 3} \\
      $ n_r \times n_\theta $ &
      \multicolumn{1}{c}{$ t_{\rm m} $ [s]} &
      \multicolumn{1}{c}{$ a_{n_r} $} &
      \multicolumn{1}{c@{~}|}{$ a_{n_\theta} $} &
      \multicolumn{1}{c}{$ t_{\rm m} $ [s]} &
      \multicolumn{1}{c}{$ a_{n_r} $} &
      \multicolumn{1}{c@{~}|}{$ a_{n_\theta} $} &
      \multicolumn{1}{c}{$ t_{\rm m} $ [s]} &
      \multicolumn{1}{c}{$ a_{n_r} $} &
      \multicolumn{1}{c@{~}}{$ a_{n_\theta} $} \\
      \hline \rule{0 em}{1.0 em}%
      $ \wz 50 \times 16 $ &   1.8 &     &      &  2.8 &     &     & 20.7 &     &     \\
      $    100 \times 16 $ &   3.7 & 2.0 &      &  5.9 & 2.1 &     & 20.6 & 1.0 &     \\
      $    200 \times 16 $ &   7.4 & 2.0 &      & 12.9 & 2.2 &     & 20.8 & 1.0 &     \\ [0.7 em]
      $ \wz 50 \times 32 $ &  12.5 &     & 6.9 &   5.9 &     & 2.1 & 20.8 &     & 1.0 \\
      $    100 \times 32 $ &  25.4 & 2.0 & 6.9 &  12.3 & 2.1 & 2.1 & 20.5 & 1.0 & 1.0 \\
      $    200 \times 32 $ &  50.8 & 2.0 & 6.9 &  27.1 & 2.2 & 2.1 & 21.7 & 1.1 & 1.0 \\ [0.7 em]
      $ \wz 50 \times 64 $ & 109.7 &     & 8.8 &  12.4 &     & 2.1 & 20.9 &     & 1.0 \\
      $    100 \times 64 $ & 224.2 & 2.0 & 8.8 & \dash &     &     & 21.5 & 1.0 & 1.1 \\
      $    200 \times 64 $ & 445.2 & 2.0 & 8.8 & \dash &     &     & 21.7 & 1.0 & 1.1 \\
    \end{tabular}
  \end{ruledtabular}
\end{table}

The break even point for the three solvers corresponds roughly to a
resolution of $ 100 \times 32 $ grid points at
$ t_{\rm m} \sim 20 {\rm\ s} $. We emphasize that this value of
$ t_{\rm m} $ is {\em much larger\/} than the time needed for one
hydrodynamic step at the same resolution, which is roughly
$ t_{\rm h} \sim 0.1 {\rm\ s} $. From the results reported in
Table~\ref{tab:runtime_metric_solver_2d} it becomes evident that due
to the independence of $ t_{\rm m} $ on the finite difference grid
resolution in the spectral metric solver~3, this method is far
superior to the other two solvers for simulations requiring a large
number of grid points in general, and particularly in
$ \theta $-direction.

\subsubsection{Radial fall-off of the metric components}
\label{subsubsection:metric_fall_off}

When comparing in Section~\ref{subsection:fd_elliptic_solvers} the
theoretical foundations of the three alternative metric solvers
implemented in the code, we already raised the issue of the existence
of source terms with noncompact support in the metric
equations~(\ref{eq:metric_equations}) (see
Section~\ref{subsection:noncompact_support}). Neither the
Newton--Raphson-based solver~1, which requires explicit boundary
conditions at the finite radius $ r_{\rm fd} $ (which are in general
not exactly known and possibly time-dependent), nor the conventional
iterative Poisson solver~2, which integrates the Poisson-like
metric equations only up to
the same finite radius $ r_{\rm fd} $, are able to fully account for
the nonlinear source terms, even if the radial boundary of the
finite difference grid is in the vacuum region outside the star,
$ r_{\rm fd} > r_{\rm s} $.

Hence, both solvers yield a numerical solution of the exact metric equations
only in very few trivial cases, like e.g.\ the solution for the metric
of a spherically symmetric static matter distribution (TOV solution), 
when the metric equations reduce to Poisson-like equations with
compact support. However, due to the radial compactification of the
spectral grid, which allows for the Poisson equations to be
numerically integrated out to spacelike infinity, the spectral
solver~3 can consistently handle all noncompact support source terms
in the metric equations in a non-approximative way. This property holds
even when the metric quantities are mapped from the spectral grid onto
the finite difference grid, the latter extending only to
$ r_{\rm fd} $. Thus, we expect that only solver~3 captures the
correct radial fall-off behavior of the metric quantities outside the
matter distribution.

\begin{figure}[t]
  \epsfxsize = 8.6 cm
  \centerline{\epsfbox{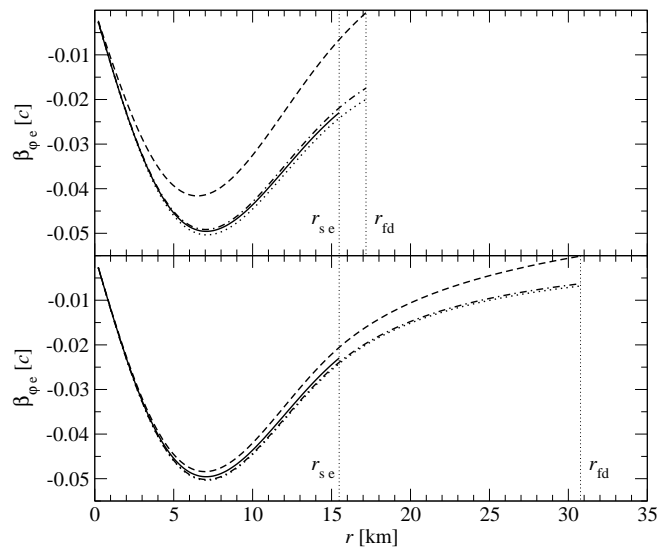}}
  \caption{Equatorial profile of the shift vector component
    $ \beta_{\varphi \rm\,e} $ obtained by different metric solvers
    compared with the correct profile from the initial data solver
    (solid line) for the rotating neutron star model RNS. Due to its
    approximate boundary value, the profile from solver~1 (dashed
    line) shows large deviation from the correct solution,
    particularly for a grid boundary $ r_{\rm fd} $ close to the
    stellar equatorial radius $ r_{\rm s\,e} $ (upper panel). As
    solver~2 (dashed-dotted line) needs no explicit boundary
    conditions, its solution matches well with the correct solution,
    with improving agreement as $ r_{\rm fd} $ is at larger distance
    from $ r_{\rm s\,e} $ (lower panel). The
    compactified radial grid of solver~3 (dotted line) fully accounts
    for non-compact support terms, and thus agrees very well
    with the correct solution, independent of the location of
    $ r_{\rm fd} $. The radii $ r_{\rm s\,e} $ and $ r_{\rm fd} $ are
    indicated by vertical dotted lines.}
  \label{fig:rns_shift_vector_comparison_2d}
\end{figure}

In the following we illustrate the effects of noncompact support terms
in the metric equations on the numerical solution using the three
different solvers. Fig.~\ref{fig:rns_shift_vector_comparison_2d} shows
the radial equatorial profiles of the rotational shift vector
component $ \beta_\varphi = \sqrt{\gamma_{33}} \beta^3 $ for the
rapidly rotating neutron star initial model (RNS) specified in
Section~\ref{subsubsection:convergence_2d}, obtained with the three
alternative metric solvers. While we restrict our discussion to the
particular metric quantity $ \beta_{\varphi {\rm\,e}} $ we notice that
the radial fall-off behavior and the dependence on the solver method
is equivalent for all other metric components.

In the upper panel of Fig.~\ref{fig:rns_shift_vector_comparison_2d}
the equatorial stellar boundary $ r_{\rm s\,e} $ is very close to the
radial outer boundary of the finite difference grid,
$ r_{\rm s\,e} = 0.9 \, r_{\rm fd} $ (both indicated by vertical
dotted lines). The star and the exterior atmosphere are resolved using
$ n_{r {\rm\,s}} = 90 $ radial grid points for the star and
$ n_{r {\rm\,a}} = 10 $ radial grid points for the atmosphere (along
the equator), respectively, and $ n_\theta = 30 $ meridional
points. The spectral solver~3 uses $ \hat{n}_r = 33 $ radial and
$ \hat{n}_\theta = 17 $ meridional grid points.

If the boundary value for the metric at $ r_{\rm fd} $ is exact, solver~1
always yields the correct solution, irrespective of the source terms
not having compact support. For stationary solutions like rotating
neutron stars these exact values can in principle be provided by the
initial data solver. However, for instance in a dynamical situation,
exact values cannot be provided, and we are forced to use approximate
boundary conditions, which we choose according to
Eq.~(\ref{eq:vacuum_matching}). As the approximate boundary value for
solver~1, $ \beta_\varphi (r_{\rm fd}) = 0 $, is far from the exact
value, the corresponding profile of the shift vector (dashed line)
strongly deviates from the correct $ \beta_{\varphi {\rm\, e}} $
obtained by the initial data solver (solid line). Note that the exact
solution is given only for $ r \le r_{\rm s\,e} $, due to limitations
of the initial solver method~\cite{komatsu_89_a}. As shown in the
lower panel of the figure, with increasing distance of the finite
difference grid boundary from the stellar boundary
($ r_{\rm fd} = 2.0 \, r_{\rm s\,e} $ with
$ n_{r {\rm\,s}} = n_{r {\rm\,a}} = 90 $), the approximation for
$ \beta_{\varphi {\rm\,e}} (r_{\rm fd}) $ improves noticeably, and so
does the matching of $ \beta_{\varphi {\rm\,e}} $ with the correct
solution.

On the other hand, as the integral approach of solvers~2 and~3
requires no specific boundary conditions at a finite radius (contrary to
solver~1), the numerical solution for $ \beta_{\varphi {\rm\,e}} $
agrees well with the correct solution even for an integration boundary
$ r_{\rm fd} $ close to the stellar boundary $ r_{\rm s\,e} $
(dashed-dotted and dotted lines in
Fig.~\ref{fig:rns_shift_vector_comparison_2d}, respectively). For
$ r_{\rm fd} \gg r_{\rm s\,e} $, when the influence of the source
terms with noncompact support is increasingly picked up by the radial
integral, the solutions supplied by solver~2 rapidly approach the
correct one. The terms with noncompact support usually do not
contribute strongly to the solution of the metric equations (except in
cases of very strong gravity and extremely rapid contraction or
rotation). Thus, solver~2 is superior to solver~1 when approximate
boundary values must be used,
Eq.~(\ref{eq:vacuum_matching}). Solver~3, on the other hand, has the
key advantage over solver~2 of using very accurate
spectral methods for solving the Poisson equation over the
{\em entire\/} spatial volume due to its compactified radial
coordinate. Hence, irrespective of the distance of $ r_{\rm fd} $ from
$ r_{\rm s\,e} $, it yields the same results on the finite difference
grid, onto which the results are mapped from the spectral grid.

The (small) difference between the results for
$ \beta_{\varphi {\rm\,e}} $ from solver~3 and from the initial data
solver is partly due to the accuracy of the numerical schemes and the
mapping between different grids, and particularly a result of the CFC
approximation of the field equations employed by the evolution code
(note that the initial data are generated from a numerical solution of
the exact Einstein metric equations). In the case of rapidly rotating
neutron star models we have found that the truncation error and the
error arising from the mapping of the initial data to the evolution
code is typically more than one order of magnitude smaller than the
error which can be attributed to the CFC approximation, if a grid with
a resolution $ n_r \sim 100 $, $ n_\theta \sim 30 $ and
$ \hat{n}_r = 33 $, $ \hat{n}_\theta = 17 $ is used. For estimates of
the quality of the CFC approximation in such cases,
see~\cite{dimmelmeier_02_a} and references therein.

We again note that, in principle, the use of a compactified radial
grid is not confined to the spectral solver~3. A finite difference grid
extending to spatial infinity could be used for solvers~1 and~2 as well.
However, in that case either the exterior atmosphere would also have to be
extended to the entire grid too (generating unnecessary
computations), or only the relevant portion of the grid containing the
star would have to be evolved in time (creating an additional
boundary). When using solver~3, there is a clearcut split between the
finite difference grid and the spectral grid. Thus, the hydrodynamic
quantities can be defined on a grid with an atmosphere of only small
size, while the metric in the compactified domain can be computed very
accurately with only few radial collocation points due to the
exponential convergence of spectral methods in this smooth
region. Additionally, the {\sc Lorene} library provides the use of a
compactified radial domain as an already implemented option at no
extra cost.

\subsubsection{Axisymmetric core collapse to a neutron star -- \\
  Construction of the spectral grid domains}
\label{subsubsection:axisymmetric_core_collapse}

\begin{figure}[t]
  \epsfxsize = 8.6 cm
  \centerline{\epsfbox{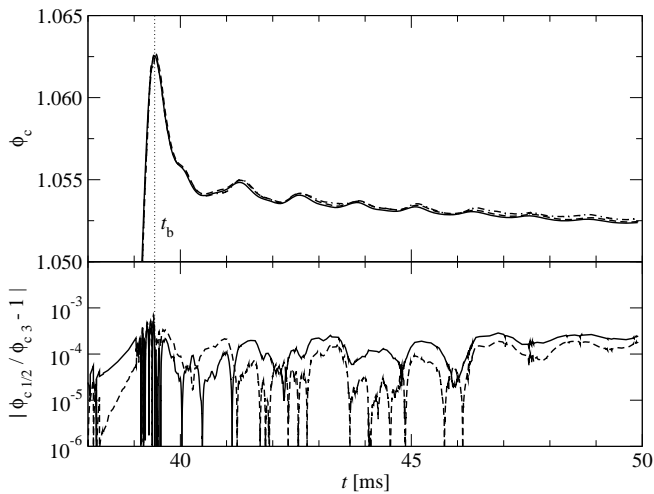}}
  \caption{Time evolution of the central conformal factor
    $ \phi_{\rm c} $ (upper panel) for the core collapse model SCC,
    using metric solver~1 (solid line), 2 (dashed line), and~3
    (dashed-dotted line), respectively. All three solvers yield
    similar results. The small relative differences of less than
    $ 10^{-3} $ in $ \phi_{\rm c} $ (lower panel) obtained with
    solvers~1 and~3 (solid line) and solver~2 and~3 (dashed line)
    prove that numerical variations of the metric from each solver
    are of the order of the small overall discretization error of the
    entire evolution code. The time of bounce $ t_{\rm b} $ is
    indicated by the vertical dotted line.}
  \label{fig:metric_solver_comparison_2d}
\end{figure}

As all three metric solvers yield equally precise numerical solutions
of the spacetime metric in 2D, they give nearly identical results when
applied to simulations of rotational core collapse, as shown in
Fig.~\ref{fig:metric_solver_comparison_2d}. For the results presented
in this figure we have chosen the stellar core collapse model labeled
A3B2G4 in~\cite{dimmelmeier_02_b} (model SCC in the following), which
rotates differentially and moderately fast, and has an initial central
density $ \rho_{\rm c} = 10^{10} {\rm\ g\ cm}^{-3} $. The initial
adiabatic index is reduced from $ \gamma_{\rm i} = 4/3 $ to
$ \gamma_1 = 1.3 $ during contraction, and is increased to
$ \gamma_2 = 2.5 $ beyond supranuclear matter densities,
$ \rho > \rho_{\rm nuc} = 2.0 \times 10^{14} {\rm\ g\ cm}^{-3} $. The
details of the EoS for this model are given by
Eq.~(\ref{eq:hybrid_eos}). As the metric calculation is
computationally very expensive, it is done only every 100/10/50
hydrodynamic time steps before/during/after core bounce, and
extrapolated in between (for details on the satisfactory accuracy of
this procedure see~\cite{dimmelmeier_02_a}). The number of zones used
in the finite difference grid is $ n_r = 200 $ and $ n_\theta = 30 $,
with logarithmic spacing in the $ r $-direction and a central
resolution of 500~m, and an equidistant spacing in the
$ \theta $-direction. Again, the grid resolution of the spectral
solver~3 is $ \hat{n}_r = 33 $ and $ \hat{n}_\theta = 17 $.

In the upper panel of Fig.~\ref{fig:metric_solver_comparison_2d} we
plot the time evolution of the central conformal factor
$ \phi_{\rm c} $, which rises steeply when the central density
increases to supranuclear densities, reaches a maximum at the time of
core bounce $ t_{\rm b} $ (vertical dotted line), and subsequently
approaches a new equilibrium value with decreasing ringdown
oscillations. This new state, which is reached asymptotically, signals
the formation of a pulsating compact remnant which can be identified
with the nascent proto-neutron star. Each of the three curves in this
upper panel is the result of using one of the three available metric
solver (see caption for details). The lower panel of the figure
demonstrates that the relative differences found in the dynamical
evolution of our representative core collapse model are negligibly
small when using either metric solver, which proves the applicability
of any of the metric solvers in 2D.

\begin{figure}[t]
  \epsfxsize = 8.6 cm
  \centerline{\epsfbox{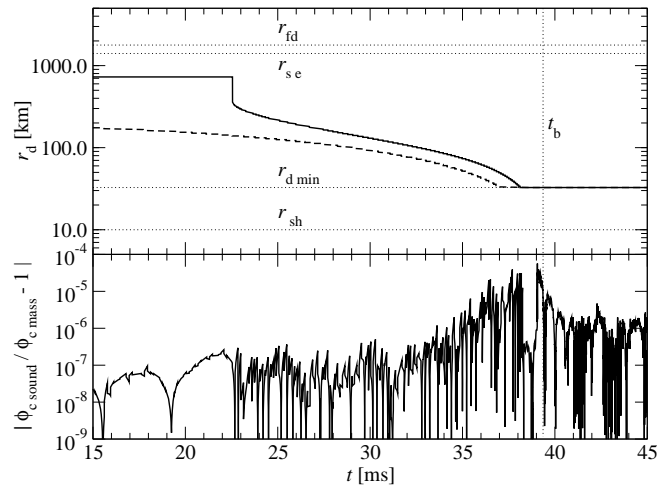}}
  \caption{Two different methods for determining the domain radii of
    the spectral grid boundary. The upper panel shows the time
    evolution of the domain radius parameter $ r_{\rm d} $ for the
    core collapse model SCC, where $ r_{\rm d} $ is either set by the
    sonic point method (solid line; sonic point first detected at
    $ t \sim 23 {\rm\ ms} $) or by the rest mass fraction
    method (dashed line). The boundary of the finite difference grid
    $ r_{\rm fd} $, the stellar equatorial radius $ r_{\rm s\,e} $,
    the minimal domain radius $ r_{\rm d\,min} $ (set to $ r_{40} $),
    and the approximate location of shock formation $ r_{\rm sh} $ are
    indicated by horizontal dotted lines. The relative difference
    between the values of $ \phi_{\rm c} $ from simulations using the
    two methods (lower panel) is less than $ 10^{- 4} $ throughout the
    evolution. The time of bounce $ t_{\rm b} $ is indicated by the
    vertical dotted line.}
  \label{fig:domain_radius_method_comparison_2d}
\end{figure}

However, in such a highly dynamical situation, where the relevant
radial scales vary by a factor of about 100, solver~3 requires a
special treatment of the radial domain setup of the spectral grid
defined in Section~\ref{subsubsection:spectral_methods}. During the
infall phase of a core collapse simulation the contracting core must
be suffiently resolved by the radial grid, and thus we adjust the
radius of the nucleus $ r_{\rm d} $ dynamically before core bounce.
(Note that this is no contradiction to the assumption\
$ f_{\rm d} = {\rm const.} $ in Eq.~(\ref{eq:domain_boundary_factor}),
as $ f_{\rm d} $ may change between {\em subsequent} metric
calculations during the evolution.) Initially the value of
$ r_{\rm d} $ is given by half the stellar radius. As the evolution
proceeds it is set equal to the radial location of the sonic point in
the equatorial plane (once unambiguously detected). Alternatively
$ r_{\rm d} $ can be determined by the radius enclosing a shell of a
fixed fraction of the total rest mass of the star (typically 10\%),
whereby $ r_{\rm d} $ moves inward during the collapse, too. In either
case $ r_{\rm d} $ is held fixed when some minimal radial threshold
$ r_{\rm d\,min} $ is crossed, which we set equal to the radius of
some given radial grid point (e.g.\ the 40th grid point at
$ r_{40} $). This ensures that there is always a sufficient number of grid
points on the finite difference grid, such that the interpolation to
the spectral grid is well behaved. For $ n_{\rm d} = 6 $ domains, both
approaches yield equally accurate results, the relative difference
between the values of $ \phi_{\rm c} $ being less than $ 10^{- 4} $
throughout the evolution of the collapse model SCC (see lower panel of
Fig.~\ref{fig:domain_radius_method_comparison_2d}).

\begin{figure}[t]
  \epsfxsize = 8.6 cm
  \centerline{\epsfbox{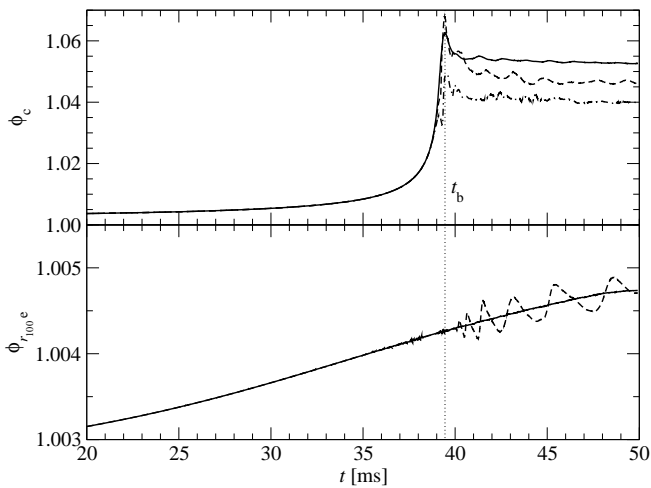}}
  \caption{Importance of the correct spectral domain setup for highly
    dynamic simulations, shown for the core collapse model SCC. If the
    domain radius parameter is not reasonably
    adjusted (upper panel), e.g.\ $ r_{\rm d} $ is held fixed at 10\%
    of the initial stellar equatorial radius (dashed line), or if the
    minimal domain radius is too large, $ r_{\rm d\,min} = r_{100} $
    (dashed-dotted line), the central conformal factor
    $ \phi_{\rm c} $ deviates strongly from the correct value (solid
    line; c.f.\ Fig.~\ref{fig:metric_solver_comparison_2d}). If the
    number of domains is too small (lower panel), e.g.\
    $ n_{\rm d} = 3 $ (dashed line) instead of $ n_{\rm d} = 6 $
    (solid line), the metric inside the star (here the equatorial
    conformal factor $ \phi_{r_{100}{\rm\,e}} $ at the 100th radial
    grid point) shows strong oscillations after core
    bounce. The time of bounce $ t_{\rm b} $ is indicated by the
    vertical dotted line.}
  \label{fig:domain_number_and_radius_effects_2d}
\end{figure}

At least for core collapse simulations, the appropriate choice of the
radial spectral domain setup parameters $ n_{\rm d} $ and
$ r_{\rm d} (t) $ is crucial, as exemplified in
Fig.~\ref{fig:domain_number_and_radius_effects_2d}. The reduction of
$ r_{\rm d} $ with time must follow the contraction of the core to a
sufficiently small radius, while $ r_{\rm d\,min} $ must retain enough
grid points for the nucleus. Furthermore, when splitting the spectral
grid into several radial domains, well-behaved differential operators
(in particular, the Poisson operator) are only obtained if, for a {\em
  shell\/}-type domain, the criterion of thin shell-type domains,
$ f_{\rm d} \lesssim 2 $, is fulfilled. This restriction for the ratio
$ f_{\rm d} $ between the outer and the inner radii originates
from the requirement to keep the condition number of the matrix
representing (for a given multipolar momentum $ l $) the radial Poisson
operator~(\ref{eq:poisson_operator_lm}), which is a very fast growing
function of $ f_{\rm d} $, lower than $ \sim 10^3 $. 

In particular Fig.~\ref{fig:domain_number_and_radius_effects_2d} shows
that if $ r_{\rm d} $ is not properly adjusted or if
$ r_{\rm d\,min} $ is too large, the central conformal factor deviates
strongly from the correct value (upper panel). In addition, if the
number of domains is too small while keeping the radial resolution
$ \hat{n}_r = 33 $ fixed, the conformal factor inside the core shows
large amplitude oscillations after core bounce, due to a too large
value of $ f_{\rm d} $ (lower panel). If $ f_{\rm d} \lesssim 2 $ is
violated because of too few domains in a collapse situation, such
oscillations are even present if the radial resolution $ \hat{n}_r $
is increased.

On the other hand, in quasi-stationary situations with no large
dynamical radial range (e.g.\ oscillations of neutron stars), one can
safely reduce $ n_{\rm d} $ from 6 to 3 and keep $ r_{\rm d} $ fixed
throughout the evolution. The optimal number of domains $ n_{\rm d} $
is thus determined by balancing radial resolution and the requirement
of thin shell-type domains against computational costs.

\subsubsection{Axisymmetric core collapse to a neutron star -- \\
  Comparison with fully general relativistic simulations}
\label{subsubsection:comparison_with_full_gr}

Only recently, fully general relativistic simulations of axisymmetric
rotational core collapse have become available~\cite{shibata_04_a}. We
now estimate the quality of the CFC approximation adopted in our code
by simulating one of the core collapse models presented
in~\cite{shibata_04_a} and comparing the results. 

In their simulations, Shibata and Sekiguchi~\cite{shibata_04_a} make
use of the {\sc Cartoon} method~\cite{alcubierre_01_a} which reduces
the dimensionality of a code based on 3D Cartesian coordinates to 2D
in the case of axisymmetric configurations. Using this approach, and
solving the full set of BSSN metric equations, these authors present a
series of rotational core collapse models with parameters close (but
not exactly equal) to the ones simulated by Dimmelmeier et
al.~\cite{dimmelmeier_02_b}. As an additional difference,
$ \rho^* = \rho W \phi^6 $ is employed by~\cite{shibata_04_a} in the
gravitational wave extraction with the first moment of momentum
density formula, while in~\cite{dimmelmeier_02_b} the wave extraction
is performed with the stress formula using the density $ \rho $ (see
Section~\ref{subsection:wave_extraction} for details). Furthermore, in
the simulations reported in~\cite{shibata_04_a}, the equidistant
Cartesian finite difference grid is repeatedly remapped during the collapse,
so that the grid spacing in the center increases from initially $
\sim 3 {\rm\ km} $ to $ \sim 300 {\rm\ m} $ during core bounce. As the
outer boundary moves in accordingly, matter leaves the computational
grid, resulting in a mass loss of about 3\%.

In their paper, Shibata and Sekiguchi investigated a core collapse
model which is identical to our model SCC (A3B2G4
in~\cite{dimmelmeier_02_b}) with the exception of a slightly smaller
rotation length parameter $ \hat{A} = A / r_{\rm s\,e} = 0.25 $
(compared to $ \hat{A} = 0.32 $ in~\cite{dimmelmeier_02_b}) in the
initial equilibrium model. They found that the evolution of this model
(labeled $ {\rm SCC_{SS}} $ hereafter) computed with their fully
general relativisitc code agrees qualitatively well with the evolution
of our model SCC simulated with our CFC code. However, it produces an
increased gravitational wave amplitude of about 20\% at the peak
during core bounce, and up to a factor 2 in the ringdown. Furthermore,
the damping time of the ringdown signal of model $ {\rm SCC_{SS}} $ as
shown in~\cite{shibata_04_a} is significantly longer compared to that
of model SCC presented in~\cite{dimmelmeier_02_b}.

\begin{figure}[t]
  \epsfxsize = 8.6 cm
  \centerline{\epsfbox{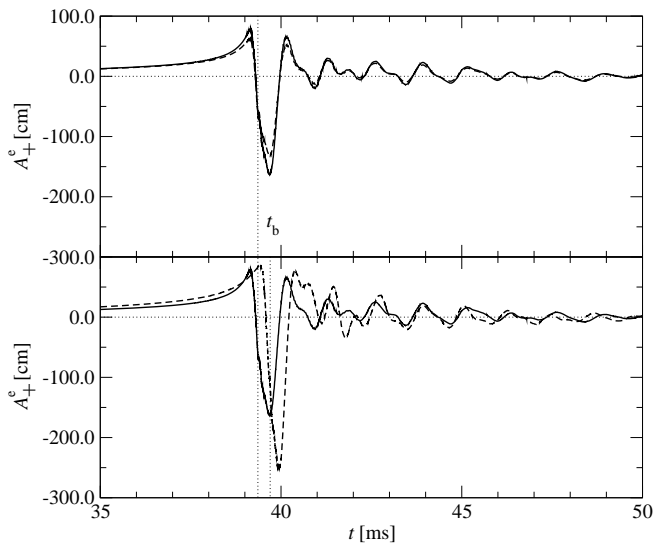}}
  \caption{Influence of the density used in the wave extraction equations
    (upper panel) and of small differences in the initial model (lower
    panel) on the gravitational waveforms from rotational core collapse.
    If $ \rho^* = \rho h W^2 $ is used in the quadrupole formula (solid
    line) instead of $ \rho $ (dashed line), the wave amplitude
    $ A_+^{\rm e} $ increases by about 20\% at core bounce (upper
    panel). A change from model SCC (solid line)
    to model $ {\rm SCC_{SS}} $ (dashed line), which corresponds
    solely to a difference in the initial configuration, results in a
    qualitatively different waveform, in particular during the
    ringdown phase (lower panel). The times of bounce $ t_{\rm b} $
    are indicated by the vertical dotted lines.}
  \label{fig:comparison_with_shibata_waveform_2d}
\end{figure}

Shibata and Sekiguchi offer several possible explanations for this
noticeable disagreement, the most plausible ones being the different
functional forms of the rest mass density used in the wave extraction
method, and the different formulations (stress
formulation~(\ref{eq:I_ij_sf_3d}) versus first moment of momentum
density formulation~(\ref{eq:I_ij_fmomdf_3d})). By comparing waveforms
obtained from evolutions of oscillating neutron stars (as presented
in~\cite{shibata_03_b}), both using the quadrupole formula and by
directly reading off metric components, they find that the quadrupole
formula underestimates the wave amplitude of model
$ {\rm SCC_{SS}} $ by $ \sim 10\% $. Extrapolating these results they
arrive at the estimate that the waveforms presented
in~\cite{dimmelmeier_02_b} are accurate at best to within
$ \sim 20\% $. Shibata and Sekiguchi claim that other differences,
namely the CFC approximation versus the BSSN formulation, different
grid setups, coordinate choices and slicing conditions, or the small
discrepancy of $ \hat{A} $ in the initial model, have only negligible
impact on the waveform.

To test this conjecture, we have simulated the evolution of model SCC
with our new version of the CFC code in 2D, and extracted the wave
amplitude $ A_+^{\rm e} $ using the first moment of momentum density
formulation~(\ref{eq:I_ij_fmomdf_3d}) with $ \rho $, and also
alternatively substituting $ \rho $ by $ \rho^* $. As our results show
(see upper panel of
Fig.~\ref{fig:comparison_with_shibata_waveform_2d}), the use of
$ \rho^* $ results in a small increase of $ A_+^{\rm e} $ by about
20\% during the bounce and the ringdown phase, limiting possible
deviations due to the difference in the quadrupole formula stated
in~\cite{shibata_04_a} to about 20\%. However, the results depicted in
Fig.~\ref{fig:comparison_with_shibata_waveform_2d} exclude that
the doubling of $ A_+^{\rm e} $ observed by~\cite{shibata_04_a} for
the ringdown signal is due to the wave extraction method. On the
contrary, comparing the waveforms for model SCC and $ {\rm SCC_{SS}} $
(see lower panel of
Fig.~\ref{fig:comparison_with_shibata_waveform_2d}), both computed
with our CFC method, shows that the
strong qualitative difference found by Shibata and Sekiguchi is
clearly due to the differences in the core collapse initial model,
notably the small decrease of the differential rotation length scale
$ \hat{A} $ in model $ {\rm SCC_{SS}} $. This gives rise to an
approximately 50\% higher peak value of the amplitude during bounce,
and a strong increase of the post bounce wave amplitude, as also
observed by Shibata and Sekiguchi (compare with Fig.~13~(b)
in~\cite{shibata_04_a}).

\begin{figure}[t]
  \epsfxsize = 8.6 cm
  \centerline{\epsfbox{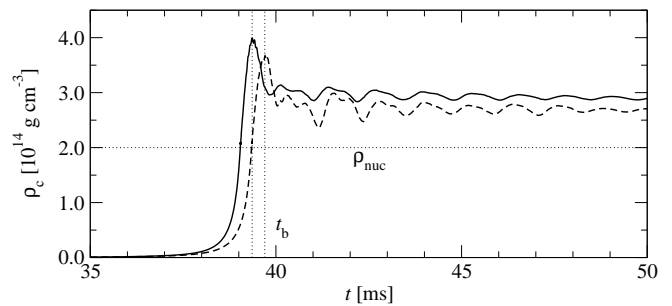}}
  \caption{Influence of differences in the initial model on the
    evolution of the central density $ \rho_{\rm c} $ for rotational
    core collapse. Changing from the collapse model SCC (solid line)
    to $ {\rm SCC_{SS}} $ (dashed line) only slightly shifts the time
    of bounce $ t_{\rm b} $ (indicated by the vertical dotted line),
    but leads to much stronger post-bounce ring down
    oscillations. Nuclear matter density
    $ \rho_{\rm nuc} $ is indicated by the horizontal dotted line.}
  \label{fig:comparison_with_shibata_central_density_2d}
\end{figure}

Furthermore, from the evolution of the central density computed with
our code (see
Fig.~\ref{fig:comparison_with_shibata_central_density_2d}), it is
evident that model $ {\rm SCC_{SS}} $ exhibits significantly stronger
ringdown oscillations than model $ {\rm SCC}$ with a somewhat
longer damping timescale, which is also in good agreement with the
results in~\cite{shibata_04_a} (see their Fig.~7~(b)). Clearly the
small difference in the rotation length parameter $ \hat{A} $ of the
initial model has a major impact on
the post-bounce dynamics of the dense core, which is in turn reflected
in the gravitational wave signal.

We have also simulated the evolution of models SCC and
$ {\rm SCC_{SS}} $ using a larger number of radial and meridional grid
points ($ n_r = 250 $ and $ n_\theta = 60 $ with a central radial
resolution $ \Delta_{r_{\rm c}} = 250 {\rm\ m} $) as compared to the
standard grid setup with $ n_r = 200 $, $ n_\theta = 30 $, and
$ \Delta_{r_{\rm c}} = 500 {\rm\ m} $ (in either case the spectral
grid resolution is $ \hat{n}_r = 33 $ and $ \hat{n}_\theta = 17 $).
Neither improving the resolution of the finite difference grid nor
discarding a significant mass fraction in the outer parts of the star
(to mimic the mass loss introduced by the regridding
method in~\cite{shibata_04_a}) have a significant impact on the
collapse dynamics or the waveform for both initial models. When
simulating the {\em same} collapse model, the observed small
differences to Shibata and Sekiguchi's results in e.g.\ the central
density or the waveform are most likely due to the use of the CFC
approximation for the spacetime metric employed in our
code. Nevertheless, for core collapse simulations, the results
obtained using either CFC or the full Einstein equations agree
remarkably well.

\subsection{Applications of the spectral solver~3 in 3D}
\label{subsection:3d_application_tests}

\subsubsection{Computation of a nonaxisymmetric spacetime metric}
\label{subsubsection:3d_metric_computation}

\begin{figure}[t]
  \epsfxsize = 8.6 cm
  \centerline{\epsfbox{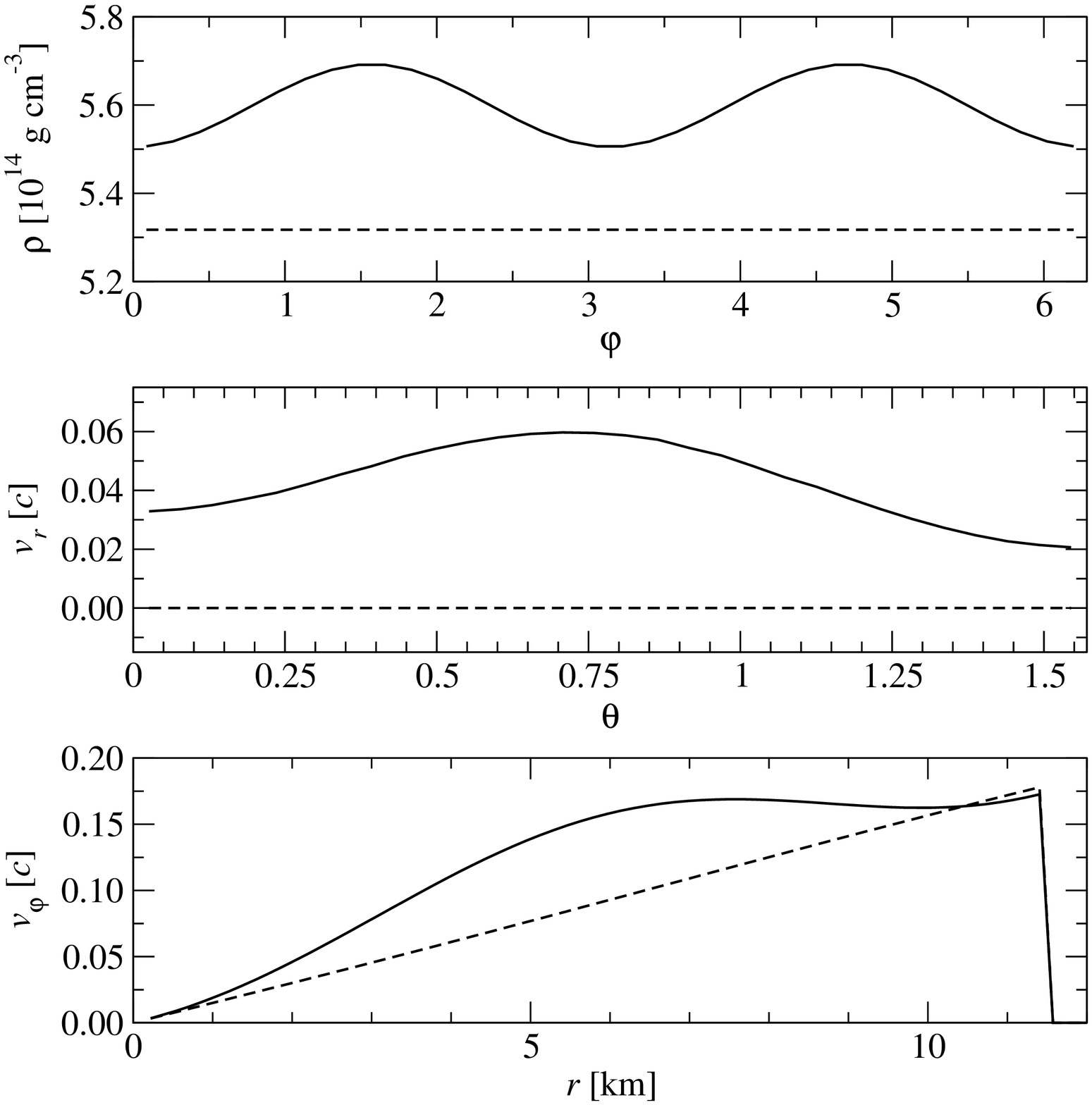}}
  \caption{Nonaxisymmetric density and velocity perturbation of the
    rapidly rotating neutron star equilibrium model RNS. By applying
    the perturbations described in the text, the original profiles
    (dashed lines) of the density $ \rho $ along the azimuthal
    direction $ \varphi $ (upper panel), the radial velocity $ v_r $
    along the meridional direction $ \theta $ (center panel), and the
    rotation velocity $ v_\varphi $ along the radial direction $ r $
    (lower panel) become strongly distorted (solid lines). The
    $ \varphi $-dependence of $ \rho $ in the upper panel shows the
    nonaxisymmetric character of the perturbation.}
  \label{fig:rns_perturbation_3d}
\end{figure}

While the previous tests were all restricted to 2D (and thus 
solvers~1 and~2 could as well be used), the genuine 3D properties of the
spectral metric solver~3 can be fully exploited and tested when
applied to the computation of the metric for a nonaxisymmetric
configuration. For this purpose we consider now the uniformly rotating
neutron star initial model RNS (see
Section~\ref{subsubsection:convergence_2d}) to which we add a
nonaxisymmetric perturbation. This is done by generalizing the
expressions in Eq.~(\ref{eq:perturbation_metric_test_2d}) through the
multiplication of a $ \varphi $-dependent term of the form
$ (1 + \sin^2 \varphi) $. The effect of such a perturbation on
representative quantities is depicted in
Fig.~\ref{fig:rns_perturbation_3d}. The metric
equations~(\ref{eq:ln_phi_alpha_metric_equation}) are then integrated
using solver~3. Convergence is reached after about 50 iterations
(threshold value $ \Delta \hat{u}^s_{\rm thr} = 10^{-6} $), and the
solution for the metric is interpolated from the spectral to the
finite difference grid.

\begin{figure}[t]
  \epsfxsize = 8.6 cm
  \centerline{\epsfbox{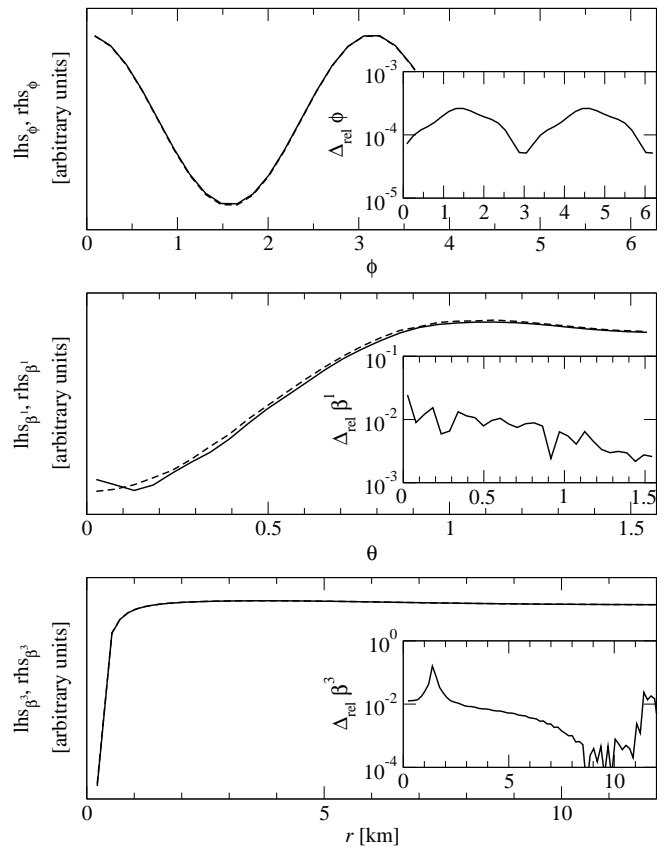}}
  \caption{Left (solid line) and right (dashed line) hand sides
    (computed on the finite difference grid) of the equation for the
    metric components $ \phi $ along the azimuthal direction $ \varphi
    $ (upper panel), $ \beta^1 $ along the meridional direction $
    \theta $ (center panel), and $ \beta^3 $ along the radial
    direction (lower panel). Even for strong nonaxisymmetric
    perturbations of the rotating neutron star model RNS, the metric
    solver~3 yields a highly accurate matching, such that the lines
    almost lie on top of one another. The insets show the relative
    difference $ \Delta_{{\rm rel}\,u} $ between the left and right
    hand sides of the equation for the same metric components. The
    relative differences are $ \lesssim 10^{-2} $, except where they
    exhibit a pole.}
  \label{fig:rns_metric_sides_3d}
\end{figure}

To exclude convergence to an incorrect solution and errors within the
interpolation routine, we compare the left and right hand sides, $
{\rm lhs}_u $ and $ {\rm rhs}_u $, of selected metric components $ u $
on the finite difference grid, in
Fig.~\ref{fig:rns_metric_sides_3d}. We note that in this figure, along
each of the profile directions, the two other coordinates are kept
fixed ($ r = r_{50} $, $ \theta = \pi / 4 $, and $ \varphi = 0 $,
respectively). The left and right hand sides of the metric
equations~(\ref{eq:ln_phi_alpha_metric_equation}) for the conformal
factor $\phi$ and the shift vector components $ \beta^1 $ and $
\beta^3 $, when evaluated on the finite difference grid, match very
accurately along all three coordinate directions. The largest
deviations are found near the rotation axis ($ \theta = 0 $) for $
\beta^1 $.

The accuracy of the metric calculation can be better quantified by
plotting the relative difference of the left and right hand sides,
$ \Delta_{{\rm rel}\,u} = |{\rm lhs}_u/{\rm rhs}_u -1| $, rather than
$ {\rm lhs}_u $ and $ {\rm rhs}_u $ alone. This is shown for the metric
quantities $ \phi $, $ \beta^1 $, and $ \beta^3 $ in the insets of
Fig.~\ref{fig:rns_metric_sides_3d}. Along any of the plotted profiles,
the spectral solver yields a solution for which the relative
difference measure is better than $ 10^{-2} $. As $ {\rm lhs}_u $ and
$ {\rm rhs}_u $ contain second spatial derivatives of the metric,
evaluated by finite differencing, this is an accurate numerical
result. We note that some of the metric components are close to zero
or change sign. Hence, the relative difference may become large or
develop a pole at some locations, as can be seen in the insets of
Fig.~\ref{fig:rns_metric_sides_3d}.

Under idealized conditions (i.e.\ without discontinuities in the
source terms of the metric equations, no artificial atmosphere, only
laminar matter flows, uniform grid spacing of the finite difference
grid, and perturbations which are regular at the grid boundaries),
such a test case also offers an opportunity to examine the order of
convergence of the metric solver~3 on the spectral and finite
difference grid, respectively. To this end we perform a metric
calculation using increasingly finer resolutions on the two grids. By
varying the number of spectral collocation points in all three spatial
directions while keeping the number of finite difference grid points
fixed (at high resolution), we observe an exponential decrease of the
relative differences $ \Delta_{{\rm rel}\,u} $ between the left and
right hand sides of the equation for the various metric components
$ u $. Correspondingly, the metric solution evaluated on the finite
difference grid exhibits second order convergence with grid resolution
for a fixed (and high) spectral grid resolution. Furthermore, the (at
least) second order accurate time integration scheme of the code in
combination with the PPM reconstruction of the Riemann solver also
guarantees second order convergence during time evolution. For fixed
time steps we actually observe this theoretical convergence order
globally and even locally (except close to the grid boundaries, where
symmetry conditions and ghost zone extrapolation spoil local
convergence).

In the three-dimensional case the computational load of the
interpolation from the spectral grid to the finite difference
grid after every metric calculation on the spectral grid becomes
significant. The time spent in the interpolation between grids can, in
fact, even surpass the computational costs of the spectral metric
solution itself (see Section~\ref{subsection:interpolation_tests}). As
a consequence, the independence of the metric execution time
$ t_{\rm m} $ on the number of finite difference grid points found in
the axisymmetric case (as shown in
Table~\ref{tab:runtime_metric_solver_2d}) cannot be
maintained. Table~\ref{tab:runtime_metric_solver_3d} reports the
summary of runtime results for a single metric computation of the
above neutron star model on an IBM RS/6000 Power4 processor. These
results indicate an (albeit sublinear) increase of $ t_{\rm m} $ with
the number of finite difference grid points. As expected, a doubling of
the spectral grid resolution e.g.\ in the $ \varphi $-direction (while
keeping $ \hat{n}_r = 33 $ and $ \hat{n}_\theta = 17 $ fixed) results
in a proportional increase of $ t_{\rm m} $. The runtime scaling
results reported in Table~\ref{tab:runtime_metric_solver_3d} also
demonstrate that the different coordinate directions contribute
equally to the computational costs.

\begin{table}[t]
  \caption{Dependence of the metric solver execution time
    $ t_{\rm m} $ on the finite difference grid resolution
    $ n_r \times n_\theta \times n_\varphi $ and the spectral grid
    azimuthal resolution $ \hat{n}_\varphi $ using the metric solver~3
    in 3D for the nonaxisymmetrically perturbed rotating neutron star
    model RNS. For typical finite difference grid point numbers, the
    ratio $ r_{n_\varphi} $ between execution times for a given
    $ n_\varphi $ and for half that resolution is smaller than 2,
    i.e.\ the increase of $ t_{\rm m} $ is less than
    linear. Furthermore, when doubling both the radial {\em and\/}
    meridional grid zones, a sublinear increase in the corresponding
    ratio $ r_{n_{r, \theta}} < 4 $ is observed. Doubling the spectral
    resolution $ \hat{n}_\varphi $ increases $ t_{\rm m} $ by
    $ r_{\hat{n}_\varphi} \sim 2 $. For comparison, the values of
    $ t_{\rm m} $ for the corresponding axisymmetric model are given
    at the bottom.}
  \label{tab:runtime_metric_solver_3d}
  \begin{ruledtabular}
    \begin{tabular}{c@{~}|rrr@{~}|rrrr@{~}}
      &
      \multicolumn{3}{c@{~}|}{$ \hat{n}_\varphi = 6 $} &
      \multicolumn{4}{c@{~}}{$ \hat{n}_\varphi = 12 $} \\
      $ n_r \times n_\theta \times n_\varphi $ &
      \multicolumn{1}{c}{$ t_{\rm m} $ [s]} &
      \multicolumn{1}{c}{$ r_{n_\varphi} $} &
      \multicolumn{1}{c@{~}|}{$ r_{n_{r, \theta}} $} &
      \multicolumn{1}{c}{$ t_{\rm m} $ [s]} &
      \multicolumn{1}{c}{$ r_{n_\varphi} $} &
      \multicolumn{1}{c}{$ r_{n_{r, \theta}} $} &
      \multicolumn{1}{c@{~}}{$ r_{\hat{n}_\varphi} $} \\
      \hline \rule{0 em}{1.0 em}%
      $ 100 \times 32 \times \wz 8 $ &  37.2 &     &     &  71.5 &     &     & 2.0 \\
      $ 100 \times 32 \times    16 $ &  39.9 & 1.1 &     &  77.8 & 1.1 &     & 2.0 \\
      $ 100 \times 32 \times    32 $ &  47.4 & 1.2 &     &  90.6 & 1.2 &     & 1.9 \\
      $ 100 \times 32 \times    64 $ &  62.3 & 1.3 &     & 116.1 & 1.3 &     & 1.9 \\ [0.7 em]
      $ 200 \times 64 \times \wz 8 $ &  48.3 &     & 1.3 &  90.7 &     & 1.3 & 1.9 \\
      $ 200 \times 64 \times    16 $ &  62.5 & 1.3 & 1.6 & 116.6 & 1.3 & 1.5 & 1.9 \\
      $ 200 \times 64 \times    32 $ &  92.0 & 1.5 & 1.9 & 166.2 & 1.4 & 1.8 & 1.8 \\
      $ 200 \times 64 \times    64 $ & 149.9 & 1.6 & 2.4 & 269.5 & 1.6 & 2.3 & 1.8 \\
      \hline
      &
      \multicolumn{3}{c@{~}|}{$ \rule{0 em}{1.5 em} \hat{n}_\varphi = 4 $} \\
      $ n_r \times n_\theta \times n_\varphi $ &
      \multicolumn{1}{c@{~}}{$ t_{\rm m} $ [s]} & & \\
      \cline{1-4} \rule{0 em}{1.0 em}%
      $ 100 \times 32 \times \wz 1  $ &  20.5 & & \\
      $ 200 \times 32 \times \wz 1  $ &  21.7 & &
    \end{tabular}
  \end{ruledtabular}
\end{table}

It is worth pointing out that the other two metric solvers we have
available in the code fail to compute the metric for the
nonaxisymmetric neutron star configuration considered in this section
due to the known limitations (excessive computing time for solver~1,
convergence problems for solver~2).

\subsubsection{Stability of symmetric configurations against perturbations}
\label{subsubsection:stability_of_symmetry}

An important requirement for any hydrodynamics code is the
preservation of the symmetry of an initially symmetric configuration
during time evolution. In a practical application this means that if a
small perturbation is added to symmetric and {\em stable\/} initial
data, the perturbation amplitude must not grow in time. Due to the
choice of spherical polar coordinates $ (r, \theta, \varphi) $, our
code is particularly well suited to test the preservation of the
symmetry of spherically symmetric and axisymmetric initial
data. Additionally, this coordinate choice implies that when
simulating axisymmetric or spherically symmetric problems, either one
or two dimensions can be trivially suppressed, respectively, which
results in considerable savings of computational time.

Next, we present results from the evolution of both a spherically
symmetric neutron star model (labeled SNS) and the axisymmetric
rapidly rotating neutron star model RNS. Model SNS has the same central
density and EoS as model RNS described in
Section~\ref{subsubsection:convergence_2d}. To each equilibrium model
SNS and RNS we respectively add an axisymmnetric $ (r, \theta) $- and
a nonaxisymmetric $ (r, \theta, \varphi) $-dependent three-velocity
perturbation of the form
\begin{equation}
  \setlength{\arraycolsep}{0.14 em}
  \begin{array}{rcl}
    v_r & = & \displaystyle
    0.02 \sin^2 \! \left( \! \pi \frac{r}{r_{\rm s}} \right)
    \left( 1 + a \sin^2 (2 \theta) \right), \\ [1.0 em]
    v_\theta & = & \displaystyle
    0.02 \sin^2 \! \left( \! \pi \frac{r}{r_{\rm s}} \right)
    a \sin^2 (2 \theta),
  \end{array}
  \label{eq:preservation_of_symmetry_perturbation_2d}
\end{equation}
and
\begin{equationarray}
  v_r & = &
  0.02 \sin^2 \! \left( \! \pi \frac{r}{r_{\rm s}} \right)
  \left( 1 + \sin^2 (2 \theta) \right)
  \left( 1 + a \sin^2 \varphi \right),
  \nonumber \\
  v_\theta & = &
  0.02 \sin^2 \! \left( \! \pi \frac{r}{r_{\rm s}} \right)
  \sin^2 (2 \theta)
  \left( 1 + a \sin^2 \varphi \right),
  \label{eq:preservation_of_symmetry_perturbation_3d}
  \\
  v_\varphi & = & v_{\varphi\,\rm ini} +
  0.02 \sin^2 \! \left( \! \pi \frac{r}{r_{\rm s}} \right) \!
  \left( 1 + \sin^2 (2 \theta) \right) \!
  \left( 1 + a \sin^2 \varphi \right),
  \nonumber
\end{equationarray}%
respectively, where $ a $ is the perturbation amplitude. Model SNS is
then evolved in time using the code in axisymmetric 2D mode, and model
RNS using the fully 3D capabilities of the code. The metric is
calculated every 100 (300) time steps in 2D (3D) and extrapolated in
between. The number of finite difference grid zones is $ n_r = 80 $,
$ n_\theta = 16 $, $ n_\varphi = 1$ in the 2D case and
$ n_r = 80 $, $ n_\theta = 16 $, $ n_\varphi = 12 $ in the 3D
case. Correspondingly, for the spectral grid we use
$ \hat{n}_r = 25 $, $ \hat{n}_\theta = 13 $, $ \hat{n}_\varphi = 4 $
in 2D, and $ \hat{n}_r = 25 $, $ \hat{n}_\theta = 13 $,
$ \hat{n}_\varphi = 6 $ in 3D.

\begin{figure}[t]
  \epsfxsize = 8.6 cm
  \centerline{\epsfbox{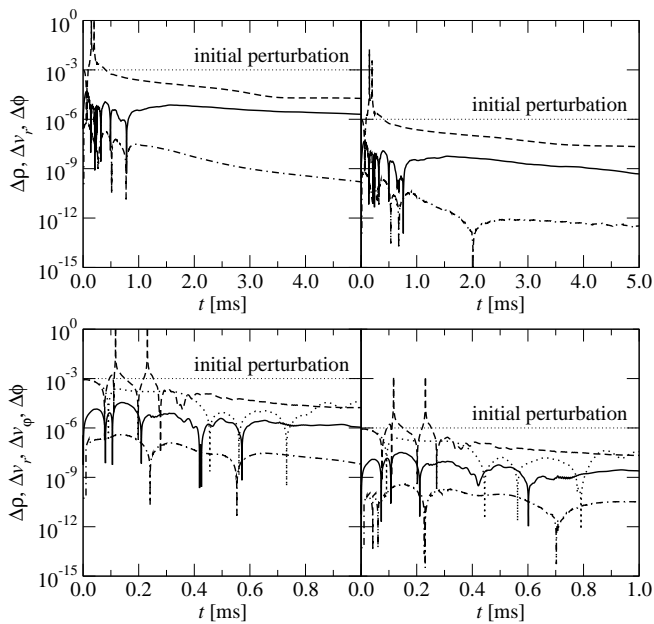}}
  \caption{Time evolution of a symmetry violating perturbation. The
    upper two panels correspond to the spherically symmetric model
    SNS, and the lower two panels to the axisymmetric model RNS. The
    relative variation in density $ \Delta \rho $ (solid line), radial
    velocity $ \Delta v_r $ (dashed line), rotational velocity
    $ \Delta v_\varphi $ (dotted line), and conformal
    factor $ \Delta \phi $ (dashed-dotted line) show a remarkable
    constancy in time (note that $ \Delta v_\varphi $ is nonzero only
    for the rotating model RNS). The symmetry violating variation of
    the different fields scale with the initial perturbation amplitude
    (horizontal dotted lines; left panels: $ a = 10^{- 3} $; right
    panels: $ a = 10^{- 6} $).}
  \label{fig:preservation_of_symmetry_ns}
\end{figure}

The results of the evolution of the symmetry violating perturbations
in both models are depicted in
Fig.~\ref{fig:preservation_of_symmetry_ns}. The upper panels
correspond to model SNS which is evolved up to 5 ms, while the bottom
panels correspond to model RNS which is only evolved up to 1 ms. The
left and right panels differ by the value of the initial amplitude
$ a $ of the velocity perturbation. We observe that the perturbation
amplitude, measured as the relative difference $ \Delta q $ of an
arbitrary matter or metric quantity $ q $ evaluated at two points of
constant $ r $ (for model SNS) and constant $ r, \theta $ (for model
RNS), remains practically unchanged for many hydrodynamic time
scales. Note that the spikes in $ \Delta q $ appearing in
Fig.~\ref{fig:preservation_of_symmetry_ns} are the poles associated
with a vanishing $ q $. Fig.~\ref{fig:preservation_of_symmetry_ns}
also shows that the amplitude of the symmetry violation $ \Delta q $
approximately scales with the amplitude $ a $ of the initial velocity
perturbation (indicated by horizontal dotted lines).

In the course of many hydrodynamic time scales, the perturbations
(which are of small amplitude, $ a \ll 1 $) will be finally damped due
to the intrinsic numerical viscosity of the schemes implemented in the
code. However, if the rotation rate $ \beta $ of a rotating neutron
star were high enough such that $ \beta \gtrsim \beta_{\rm s} $ or
even $ \beta \gtrsim \beta_{\rm d} $, perturbations of the form given
by Eq.~(\ref{eq:preservation_of_symmetry_perturbation_3d}) could
trigger the onset of {\em physically} growing modes, leading to bar
mode instabilities.

\subsubsection{Evolution of an axisymmetric uniformly rotating neutron star in 3D}
\label{subsubsection:3d_axisymmetric_neutron_star_evolution}

The ability to handle long-term evolutions of rapidly rotating
relativistic equilibrium configurations is a difficult test for any
numerical code. To demonstrate the capabilities of our code to pass
this stringent test we evolve the rotating neutron star initial model
RNS in 3D until $ t = 10 {\rm\ ms} $, which corresponds to about 10
hydrodynamic time scales and rotation periods. The simulation is
performed with a resolution for the finite difference grid of
$ n_r = 100 $, $ n_\theta = 30 $, $ n_\varphi = 8 $, and
$ \hat{n}_r = 33 $, $ \hat{n}_\theta = 17 $, $ \hat{n}_\varphi = 6 $
for the spectral grid. During the  evolution, the metric is calculated
every 100 time steps and extrapolated in between.

\begin{figure}[t]
  \epsfxsize = 8.6 cm
  \centerline{\epsfbox{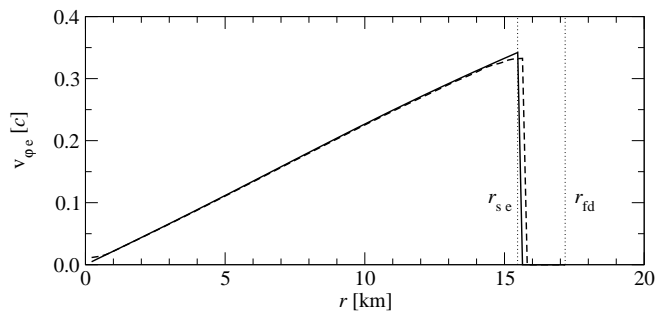}}
  \caption{Radial profile of the equatorial rotation velocity
    $ v_{\rm \varphi\,{\rm e}} $ for the unperturbed axisymmetric rapidly
    rotating neutron star model RNS evolved in 3D. The profile of
    $ v_{\rm \varphi\,{\rm e}} $ at $ t = 10 {\rm\ ms} $ (dashed line)
    closely reproduces the initial profile (solid line). The stellar
    equatorial radius $ r_{\rm s\,e} $ and the boundary of the finite
    difference grid $ r_{\rm fd} $ are indicated by vertical dotted
    lines.}
  \label{fig:axisymmetric_oscillating_rns_profile_3d}
\end{figure}

The preservation of the radial profile of the rotation velocity
$ v_{\rm \varphi\,{\rm e}} $ along the equator over a long evolution
time is shown in
Fig.~\ref{fig:axisymmetric_oscillating_rns_profile_3d}. Depicted is
the initial equilibrium solution (solid line) as a function of the
radial coordinate (in the equatorial plane) and the final
configuration (dashed line), after an evolution time of 10~ms (about
10 rotational periods). The figure shows that $ v_\varphi $ remains
close to its initial value in the interior of the star, showing the
strongest (but still small) deviations near the stellar surface (at
the interface to the artificial atmosphere). This local decrease of
$ v_\varphi $ due to interaction of stellar matter with the
atmosphere and its depencence on the order of the reconstruction
scheme has also been observed in other studies (see
e.g.~\cite{font_00_a}).

It is important to emphasize that the accurate preservation of the
rotational profile is achieved because of the use of third-order
cell-reconstruction schemes for the hydrodynamics equations, such as
PPM, as first shown by~\cite{font_00_a}. Despite the comparably coarse
resolution of the finite difference grid and the use of the CFC
approximation for the gravitational field equations, our code captures
the profile of $ v_{\rm \varphi\,{\rm e}} $ at the stellar boundary
about as accurately as codes solving the full Einstein metric
equations coupled to the hydrodynamics equations~\cite{font_02_a}, or
codes restricted to hydrodynamic evolutions in a fixed curved
spacetime (i.e.\ using the so-called Cowling
approximation)~\cite{font_00_a}.

Long-term evolutions of rotating neutron stars as the one presented
here can be effectively used for extracting the oscillation
frequencies of the various pulsation eigenmodes of the star.
This topic has been traditionally studied using perturbation theory
(see e.g.~\cite{kokkotas_99_a} and references therein). In recent
years fully nonlinear hydrodynamical codes have helped to drive
progress in the field. They have provided the quasi-radial
mode-frequencies of rapidly rotating relativistic stars, both
uniformly and differentially rotating, which is a problem still not
amenable to perturbation techniques (see e.g.~\cite{font_00_a,
  font_01_a, font_02_a, stergioulas_04_a, dimmelmeier_04_a}).

In order to test our code against existing results we show next an
example of the procedure to compute mode-frequencies using the model
RNS. The frequencies can in principle be extracted from a
Fourier transform of the time evolution of various pulsating quantities
when the oscillations are triggered by numerical truncation errors.
However, the results significantly improve when a perturbation of some
specific parity is added to the initial equilibrium model. To excite
small amplitude quasi-radial oscillations, we hence apply an $ l = 0 $
radial velocity perturbation to the equilibrium configuration of the
form
\begin{equation}
  v_r = a \sin^2 \! \left( \! \pi \frac{r}{r_{\rm s}} \right),
  \label{eq:neutron_star_oscillation_pertubation_2d_and_3d}
\end{equation}
with an amplitude $ a = -0.01 $.

\begin{figure}[t]
  \epsfxsize = 8.6 cm
  \centerline{\epsfbox{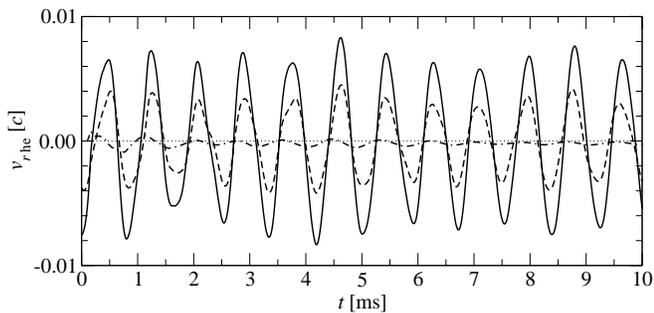}}
  \caption{Time evolution of the radial velocity at half the stellar
    equatorial radius $ v_{r\,\rm{he}} $ for the perturbed rapidly
    rotating neutron star model RNS. The radial velocity shows regular
    oscillations with neither a noticeable drift nor damping when the
    3D code is used in low resolution (solid line) as well as for the
    2D code with high resolution (dashed line). For comparison, the
    dashed-dotted line shows $ v_{r\,\rm{he}} $ when no explicit
    perturbation is added. In this case the oscillations are triggered
    by truncation errors and (mostly) by the error resulting from
    using the CFC approximation in the evolution code.}
  \label{fig:axisymmetric_oscillating_rns_2d_and_3d}
\end{figure}

Due to this perturbation, various metric and hydrodynamic quantities
exhibit very regular periodic oscillations around their equilibrium
state, as shown for the radial velocity $ v_r $ in
Fig.~\ref{fig:axisymmetric_oscillating_rns_2d_and_3d}. The pulsations,
which show no noticeable numerical damping during the entire duration
of the simulation (10 ms), are extracted at half the stellar
equatorial radius. The same oscillation pattern is obtained when
instead of using the 3D code (solid line in the figure) the model is
evolved using the code in axisymmetric mode (dashed line in
Fig.~\ref{fig:axisymmetric_oscillating_rns_2d_and_3d} with finite
difference grid size of $ n_r = 160 $, $ n_\theta = 60 $). The latter,
axisymmetric setup is currently being used in
a comprehensive parameter study of the oscillation frequencies of
rotating neutron star models~\cite{dimmelmeier_04_a}. Note that
Fig.~\ref{fig:axisymmetric_oscillating_rns_2d_and_3d} also
demonstrates that the oscillation amplitude scales linearly with the
initial perturbation amplitude $ a $ (at least if $ a \ll 1 $), which
was chosen as $ a = -0.005 $ in the 2D simulations. In the radial
velocity, neither an offset nor a noticeable drift with time can be
observed. This is in agreement with previous results using alternative
formulations and different numerical codes~\cite{font_00_a,
  font_02_a}.

\begin{table}[t]
  \caption{Comparison of the oscillation frequencies of two perturbed
    equilibrium neutron star models SNS and RNS with different axis
    ratios $ r_{\rm s\,p} / r_{\rm s\,e} $ obtained with the current
    code (both in 2D and 3D) and with the {\sc Cactus}
    code~\cite{font_02_a}. The frequencies for the fundamental mode
    $ f_{\rm F} $ and for the first harmonic $ f_{\rm H1} $ computed
    with the current code show a relative difference with respect to
    the {\sc Cactus} code (in parentheses) of at most 2\%. Due to the
    coarse spatial resolution used, the 3D code results were only
    calculated to 3 significant figures.}
  \label{tab:axisymmetric_oscillating_ns_frequencies_3d}
  \begin{ruledtabular}
    \begin{tabular}{l|@{}l@{}r@{\quad}l@{}r|@{}l@{}r@{\quad}l@{}r@{}}
      &
      \multicolumn{4}{c|}{SNS} &
      \multicolumn{4}{c}{RNS} \\
      &
      \multicolumn{4}{c|}{$ r_{\rm s\,p} / r_{\rm s\,e} $ = 1.00} &
      \multicolumn{4}{c}{$ r_{\rm s\,p} / r_{\rm s\,e} $ = 0.65} \\ [0.7 em]
      Code &
      \multicolumn{2}{l}{$ f_{\rm F} $ [kHz]} &
      \multicolumn{2}{l|}{$ f_{\rm H1} $ [kHz]} &
      \multicolumn{2}{l}{$ f_{\rm F} $ [kHz]} &
      \multicolumn{2}{l}{$ f_{\rm H1} $ [kHz]} \\
      \hline \rule{0 em}{1.0 em}%
      current (3D) &
      1.40  & (3.4) & 3.95  & (0.2) &
      1.20  & (0.4) & 3.68  & (1.0) \\
      current (2D) &
      1.463 & (0.9) & 3.951 & (0.2) &
      1.219 & (2.0) & 3.659 & (1.6) \\
      {\sc Cactus} &
      1.450 &       & 3.958 &       &
      1.195 &       & 3.717
    \end{tabular}
  \end{ruledtabular}
\end{table}

Time evolution data like the one shown in
Fig.~\ref{fig:axisymmetric_oscillating_rns_2d_and_3d} can be used to
extract the eigenmode frequencies. A Fourier transformation of
different metric and hydrodynamic quantities at various locations in
the star yields identical (discrete) frequencies.
Table~\ref{tab:axisymmetric_oscillating_ns_frequencies_3d} summarizes
the frequencies $ f_{\rm F} $ and $ f_{\rm H1} $ for the quasi-radial
fundamental mode and its first harmonic overtone, respectively. Both
frequencies obtained with the current 3D code differ only by a few
percent from those computed with the code in 2D~\cite{dimmelmeier_04_a} or
the {\sc Cactus} code, which is based on a Cartesian grid and uses the
BSSN formulation for the Einstein equations~\cite{font_02_a}.

Additionally, we have investigated the influence of grid resolution
and finite evolution time on the accuracy of the frequency
extraction. We have found that the differences in the
frequencies between the 2D and 3D simulations presented in
Table~\ref{tab:axisymmetric_oscillating_ns_frequencies_3d} can be
almost entirely attributed to the twice as long evolution time of the
2D simulation (20 ms), for which the Fourier transformation renders
more accurate frequencies. For practical evolution times of several
tens of milliseconds and for grid resolutions better than
$ n_r \sim 100 $ and $ n_\theta \sim 30 $, the extracted oscillation
frequencies are almost independent of the number of grid points used.

Note also that the mode-frequencies agree well even though we have
used different perturbation amplitudes $ a $ in the 3D and 2D
simulations (while in the {\sc Cactus} run an $ l = 0 $ rest mass
density perturbation with an amplitude $ a = 0.02 $ was used).
Table~\ref{tab:axisymmetric_oscillating_ns_frequencies_3d}
hence proves that our code is able to simulate rotating neutron stars
in a fully three-dimensional context for sufficiently long time scales
to successfully extract oscillation frequencies.

\subsubsection{Evolution of a nonaxisymmetric uniformly rotating neutron star in 3D}
\label{subsubsection:3d_nonaxisymmetric_neutron_star_evolution}

Contrary to the small amplitude nonaxisymmetric perturbations employed
in Section~\ref{subsubsection:stability_of_symmetry}, we turn now to
assess the ability of the numerical code to manage long-term stable
evolutions of strongly gravitating systems with large departures from
axisymmetry. This is an essential test for future astrophysical
applications of the code as e.g.\ the numerical investigation of bar
mode instabilities in rotating neutron stars.

For this purpose we construct a uniformly rotating neutron star model
with the same parameters as model RNS, but with only half the central
density. The finite difference grid extends out to
$ r_{\rm fd} = 80 {\rm\ km} $, with 60 equidistant radial grid points
resolving the neutron star out to $ r_{\rm s\,e} = 18.6 {\rm\ km}$.
The atmosphere is covered by 80 logarithmically spaced radial grid
points. The number of angular zones used in the finite difference grid
is $ n_\theta = 24 $ and $ n_\varphi = 32 $, respectively, while the
spectral grid has $ \hat{n}_r = 17 $, $ \hat{n}_\theta = 13 $, and
$ \hat{n}_\varphi = 12 $ grid points in 3 radial domains.

\begin{figure*}[t]
  \epsfxsize = 18.0 cm
  \centerline{\epsfbox{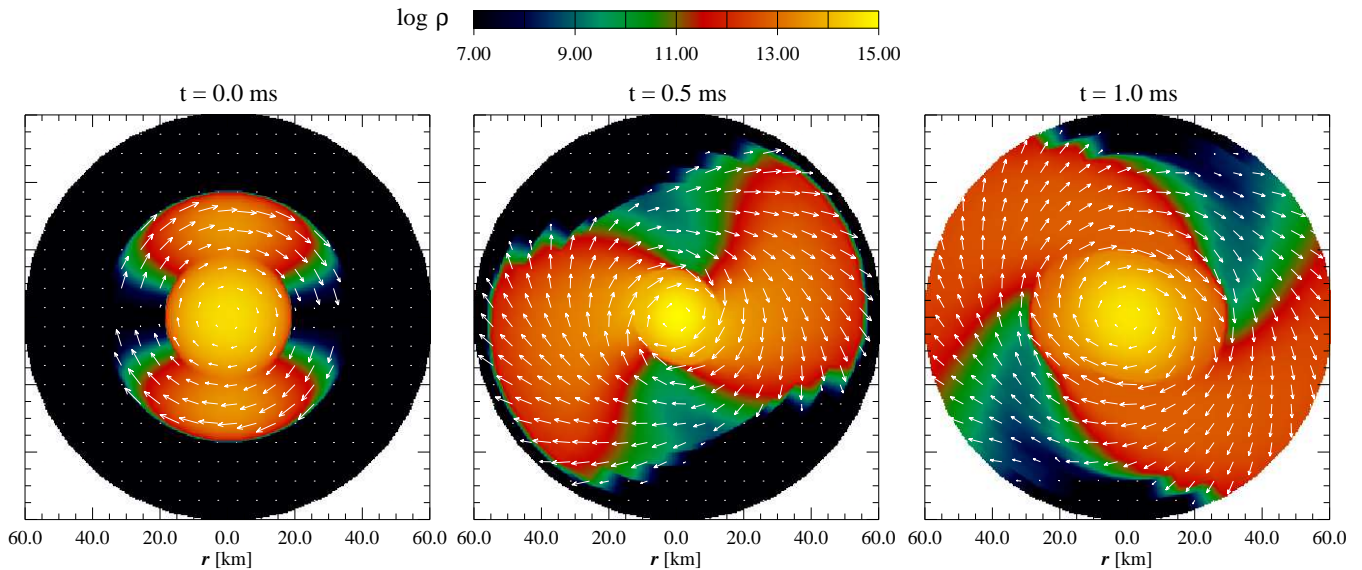}}
  \caption{Evolution of a strongly distorted nonaxisymmetric rotating
    neutron star model. The color coded distribution of $ \log \rho $
    on the equatorial plane shows how the initial perturbation (left
    panel) is partly accreted by the neutron star, and partly
    stretched into spiral arms (center panel). After about one
    rotation period of the neutron star, the trailing spiral arms have
    grown considerably in size (right panel). The rotation velocity
    $ v_\varphi $ is indicated by white arrows. Note that the
    atmosphere (color coded in black) has a density of much less than
    $ 10^7 {\rm\ g\ cm}^{-3} $, and that only the innermost 60~km of
    the computational domain are shown.}
  \label{fig:nonaxisymmatric_rns_3d_density}
\end{figure*}

On top of the equilibrium neutron star model we add a strongly
nonaxisymmetric (i.e.\ $ \varphi $-dependent) perturbation of the
rest-mass density
\begin{equation}
  \rho = \rho_{\rm ini} + a \, \rho_{\rm c} \,
  \sin^2 \! \left[ \pi \left( \frac{r}{2 r_{\rm s}} \right)^2 \right]
  \sin^{10} \varphi
  \quad
  \mbox{for } r \le 2 r_{\rm s},
  \label{eq:nonaxisymmetric_neutron_star_perturbation}
\end{equation}
with an amplitude $ a = 0.1 $, which yields an $ l = m = 2 $
bar-like structure. The rotation velocity of the uniformly rotating
unperturbed neutron star is extrapolated into the areas filled with
matter by the perturbation. The initial configuration with the
perturbation added is shown in the left panel of
Fig.~\ref{fig:nonaxisymmatric_rns_3d_density}.

\begin{figure}[t]
  \epsfxsize = 8.6 cm
  \centerline{\epsfbox{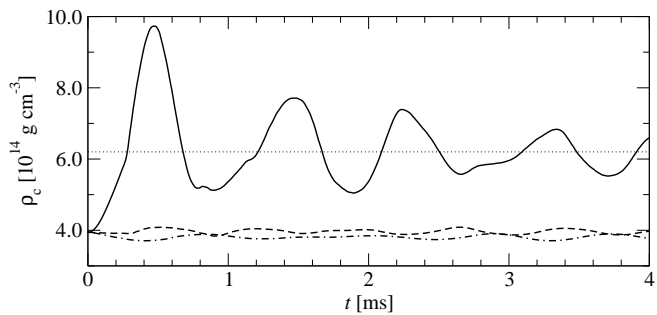}}
  \caption{Time evolution of the central density $ \rho_{\rm c} $
    for distorted nonaxisymmetric rotating neutron star models. If the
    distortion is strong ($ a = 0.1 $, solid line), matter accretion
    from the rotating bars results in a steep initial increase of
    $ \rho_{\rm c} $, which slowly settles down to a new equilibrium
    state (indicated by the horizontal dotted line). For a small
    perturbation ($ a = 0.01 $, dashed line), the evolution of
    $ \rho_{\rm c} $ follows very closely that of an unperturbed model
    (dashed-dotted line).}
  \label{fig:nonaxisymmatric_rns_3d_central_density}
\end{figure}

We have chosen this particular (albeit unphysically strong)
perturbation and velocity field in order to prevent both, an immediate
accretion of the added matter bars on to the neutron star or an
ejection. This allows us to follow the rotation of the neutron star
for a time comparable to its rotation period (which is about 1~ms for
the unperturbed neutron star). The density and rotation velocity plots
in Fig.~\ref{fig:nonaxisymmatric_rns_3d_density} after
$ t = 0.5 {\rm\ ms} $ (center panel) and $ t = 1.0 {\rm\ ms} $ (right
panel) prove this property of the chosen perturbation. These plots also
demonstrate that the corotating bar structures slowly disappear. The
innermost parts are being gradually accreted by the neutron star, which
leads to a significant initial rise in the central density, as shown in
Fig.~\ref{fig:nonaxisymmatric_rns_3d_central_density}. At later times
the more massive neutron star oscillates with a period of
$ t_{\rm osc} \sim 1.0 {\rm\ ms} $ around a new quasi-equilibrium
state, which possesses a central density of more than 50\% above
the initial equilibrium central density. Despite this strong interaction 
of the bar perturbation with the neutron star, the rotation {\em profile\/} 
inside the neutron star remains uniform throughout the evolution, although 
the rotation {\em velocity \/} nearly doubles during the oscillation maxima. 
This behavior is most likely due to the particular choice of a uniform
rotation profile for the initial bar perturbation.

For the outer parts of the initial bar, the increasing distance from
the neutron star and the sufficiently high specific angular momentum
prevents their accretion onto the neutron star. Thus the matter in this
region of the bar drifts to larger radii during the evolution. As on
the dynamical timescales considered of one rotation period there is no
efficient transport mechanism of local angular momentum by viscous
effects (which act on much longer timescales), the evolution leads to
the development of spiral arms which are clearly visible in the middle
and right panels of Fig.~\ref{fig:nonaxisymmatric_rns_3d_density}. The
outer parts of these arms are centrifugally expelled from the finite
difference grid, crossing the outer boundary at
$ t \sim 0.84 {\rm\ ms} $. By the end of the simulation, at
$ t = 4 {\rm\ ms} $, there is neither significant backscattering of
matter from the outermost boundary of the radial grid, nor there are
numerical artifacts visible at the leading or trailing edges of the
spiral arms. This proves that our numerical treatment of the radial
boundary conditions and of the artificial low density atmosphere
surrounding the star have the desired behavior.

Fig.~\ref{fig:nonaxisymmatric_rns_3d_central_density} shows that
already after an evolution time of $ \sim 1 {\rm\ ms} $, the evolution
of the spiral arms has no further significant impact in the dynamics
of the neutron star, as then the slowly decaying oscillation around
the final equilibrium state exhibits a rather regular ring-down
pattern. Plotted in this figure is also the time evolution of the central
density for a model with an amplitude $ a = 0.01 $ of the initial
perturbation given by
Eq.~(\ref{eq:nonaxisymmetric_neutron_star_perturbation}) (dashed line). In 
addition, the dashed-dotted line shows the corresponding time evolution of
$ \rho_{\rm c} $ for an unperturbed model (the small amplitude
oscillations are in this case triggered by the truncation errors of
the numerical schemes and by the use of the CFC approximation in the
evolution code). The similarity in the behavior of
$ \rho_{\rm c} $ in these cases demonstrates that for perturbations
with an amplitude $ a \lesssim 0.01 $, the dynamics of the central
neutron star is virtually unaffected by the initial bar and by the spiral
arms forming at later times. However, we observe that also for small
values of $ a $ spiral arms develop which are stable over many rotation
periods.

\begin{figure}[t]
  \epsfxsize = 8.6 cm
  \centerline{\epsfbox{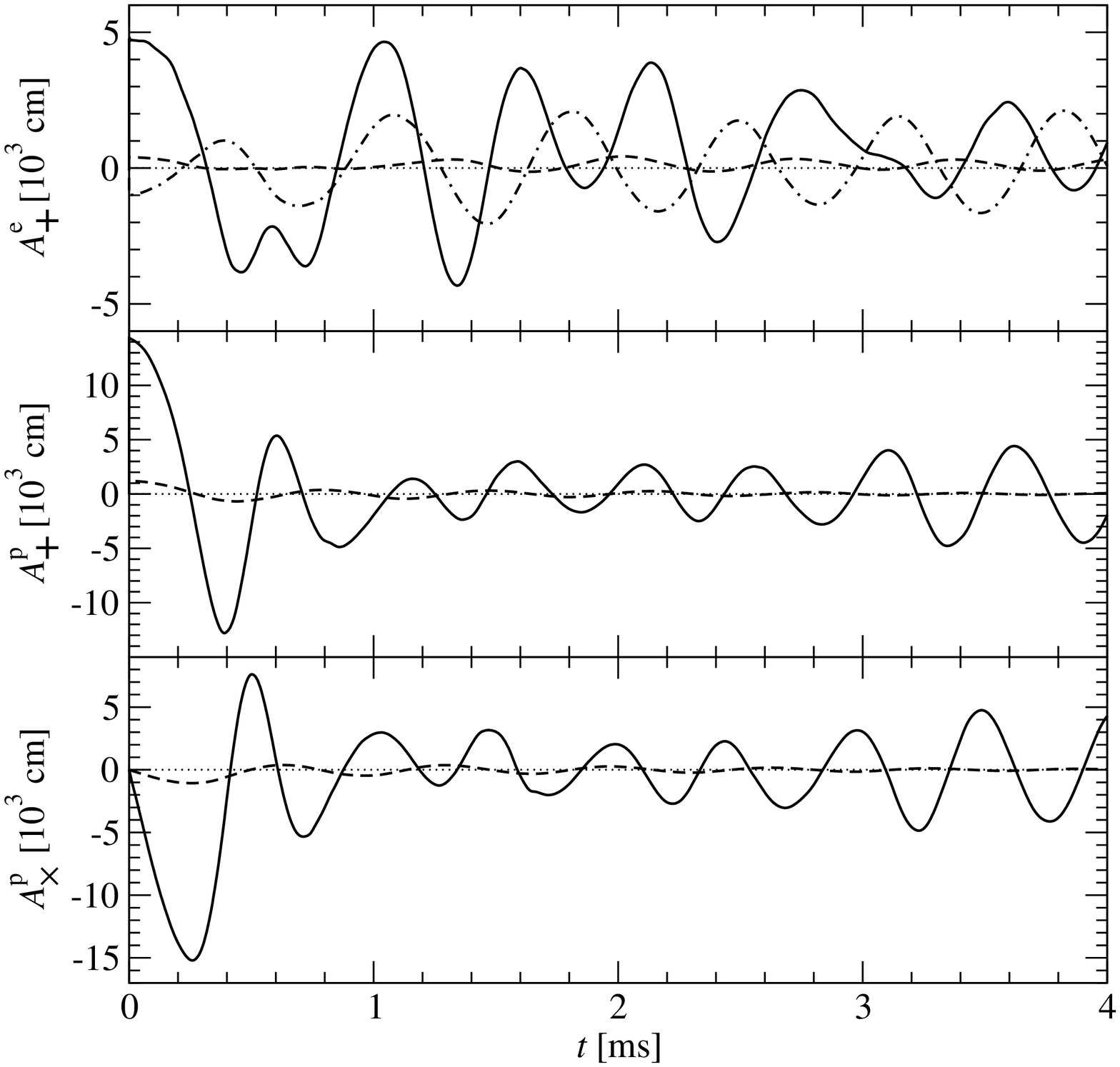}}
  \caption{Gravitational wave signal for distorted nonaxisymmetric
    rotating neutron star model. If the distortion is strong
    ($ a = 0.1 $, solid lines), the nonzero gravitational wave
    amplitudes $ A_+^{\rm e} $ (upper panel), $ A_+^{\rm p} $ (center 
    panel), and $ A_\times^{\rm p} $ (lower panel) reach peak values
    of up to $ \sim 15,000 {\rm\ cm} $. The amplitudes reduce
    significantly for $ a = 0.01 $ (dashed lines). If an axisymmetric
    perturbation with $ a = 0.1 $ is applied (dashed-dotted line),
    only the $ A_+^{\rm e} $ gravitational wave mode is present.}
  \label{fig:nonaxisymmatric_rns_3d_waveform}
\end{figure}

Apparently, strong nonaxisymmetric perturbations of the
form~(\ref{eq:nonaxisymmetric_neutron_star_perturbation}) give rise to
significant gravitational wave emission. The waveforms of the nonzero 
gravitational wave amplitudes $ A_+^{\rm e} $, $ A_+^{\rm p} $, and
$ A_\times^{\rm p} $ (as shown in the upper, center, and lower panel of
Fig.~\ref{fig:nonaxisymmatric_rns_3d_waveform}, respectively) exhibit
peak values of up to $ \sim 15 \times 10^3 {\rm\ cm} $ for the model with a
perturbation amplitude $ a = 0.1 $ (solid lines). In
Fig.~\ref{fig:nonaxisymmatric_rns_3d_waveform} we also present the
waveforms for the model with a bar perturbation of amplitude
$ a = 0.01 $ (dashed lines). Their amplitudes are roughly a factor 10
smaller than those of the corresponding waveforms of the model with
$ a = 0.1 $. Thus we can infer that the gravitational radiation
amplitude approximately scales with $ a $.

We emphasize that owing to the particular form of the
perturbation~(\ref{eq:nonaxisymmetric_neutron_star_perturbation}), the
$ \times $-mode of the gravitational radiation is zero at the equator,
$ A_\times^{\rm e} = 0 $. We also note that if instead of the
nonaxisymmetric perturbation in
Eq.~(\ref{eq:nonaxisymmetric_neutron_star_perturbation}) we use an
axisymmetric one,
\begin{equation}
  \rho = \rho_{\rm ini} + a \, \rho_{\rm c\,ini} \,
  \sin^2 \left[ \pi \left( \frac{r}{2 r_{\rm s}} \right)^2 \right]
  \quad
  \mbox{for } r \le 2 r_{\rm s},
  \label{eq:axisymmetric_neutron_star_perturbation}
\end{equation}
then the $ \times $-mode of gravitational radiation vanishes
completely, and only the $ + $-mode is present (dashed-dotted line in
the upper panel of Fig.~\ref{fig:nonaxisymmatric_rns_3d_waveform}).
Additionally, in axisymmetry the $ + $-mode on the pole is always
zero, $ A_+^{\rm p} = 0 $.

We point out that the waveform pattern for the model with the
$ a = 0.1 $ bar perturbation in
Fig.~\ref{fig:nonaxisymmatric_rns_3d_waveform} does not solely reflect
the oscillation and ring-down structure of the central neutron star,
as visible in the time evolution of $ \rho_c $ in
Fig.~\ref{fig:nonaxisymmatric_rns_3d_central_density}. For instance
the $ + $-mode at the equator (upper panel) decays on a much longer
time scale than the corresponding ring-down time of $ \rho_c $. On the
other hand, the waveforms for the two polarizations of the radiation
at the pole exhibit their peaks during the first oscillation of
$ \rho_c $ and then decay rapidly (center and lower panel). However,
after an evolution time of $ \sim 2 {\rm\ ms} $ their amplitudes
increase again. From this behavior we deduce that initially the
waveform signal is dominated by the gravitational wave emission from
the oscillating neutron star. As this contribution decays during the
ring-down, the wave emission from spiral arms becomes increasingly
important. As they expand into the atmosphere the radial weight arm in
the quadrupole formula compensates for the relatively low density of
the spiral arms, and the radiation emitted in this region becomes
visible in the signal. We cannot clearly attribute the late-time
increase in the waveform amplitude to the onset of a bar mode
instability, because the rotation parameter $ \beta $ of our model
clearly falls short of the approximate threshold for dynamical growth
of bar modes: $ \beta \sim 0.14 \ll \beta_{\rm d} $. We plan to
investigate this issue more thoroughly in the future.

The maximum amplitude $ A \sim 15 \times 10^3 {\rm\ cm} $ of the wave
signal for $ a = 0.1 $ corresponds to a dimensionless gravitational
wave amplitude $ h \sim 5 \times 10^{-19} $ at a distance of
$ r = 10 {\rm\ kpc} $ to the source. Thus, in this case of a strongly
nonaxisymmetric artificial perturbation, the typical wave amplitudes
have a value of roughly one order of magnitude above the ones of
waveforms obtained from the simplified models of rotational supernova
core collapse in axisymmetry by Dimmelmeier et
al.~\cite{dimmelmeier_02_b}. For the waveforms plotted in
Fig.~\ref{fig:nonaxisymmatric_rns_3d_waveform} we utilize the stress
formula~(\ref{eq:I_ij_sf_3d}) with $ \rho^* $ as density. The use of
this formula efficiently reduces the numerical noise in the signal as
compared with the first moment of momentum density formula and
particularly with the standard quadrupole formula. 

We consider the grid resolution used in this test simulation to be the
minimal one required for obtaining reasonably converged results. By
repeating the same model with different grid resolutions we are able
to estimate that the waveform amplitudes are correctly computed within
$ \sim 30\% $ accuracy.

\section{Conclusions}
\label{section:summary}

In this paper we have presented a new three-dimensional general
relativistic hydrodynamics code which is primarily intended for
applications of stellar core collapse to a neutron star or a black
hole, as well as for studies of rapidly rotating relativistic stars
which may oscillate in their quasi-normal modes of pulsation, emitting
gravitational radiation, or which may be subject to nonaxisymmetric
instabilities. The main novelty of this code compared to other
existing numerical relativistic codes is that it {\it combines} very accurate
state-of-the-art numerical methods specifically tailored to solve the
general relativistic hydrodynamics equations on the one hand, and the
gravitational field equations on the other hand. More precisely, the
hydrodynamic equations, formulated in conservation form, are solved
using high-resolution shock-capturing schemes based upon approximate
Riemann solvers and third-order cell-reconstruction interpolation
procedures, while the elliptic metric equations are solved using an
iterative nonlinear solver based on spectral methods. Furthermore, the
present code also departs noticeably from other three-dimensional
codes in the coordinate system used in the formulation of the
equations and in the discretization. In our approach both the metric
and the hydrodynamics equations are formulated and solved numerically
using spherical polar coordinates. In the present investigation we
have adopted the so-called conformal flatness approximation of the
Einstein equations, which reduces them to a set of five elliptic
nonlinear equations, particularly suited for the use of spectral
methods. Recently, constrained formulations of the full Einstein
equations in which elliptic equations have a preeminence over
hyperbolic equations have been reported, and appear to be amenable to
the current code.

The main purpose of the paper has been to assess the code by demonstrating
that the combination of the finite difference grid and the spectral
grid, on which the hydrodynamics and metric equations are respectively
solved, can be successfully accomplished. This approach, which we call
{\em Mariage des Maillages\/} (French for grid wedding), results in
high accuracy of the metric solver and, in practice, has allowed for
fully three-dimensional applications using computationally affordable
resources, along with ensuring long term numerical stability of the
evolution. To facilitate the {\em Mariage des Maillages\/}, i.e.\ the
combination of the finite difference grid for the hydrodynamic solver
and the spectral grid for the metric solver, a sophisticated
interpolation and grid communication scheme has been used. In
addition, we have compared our novel approach to two other, finite
difference based, methods to solve the metric equations, which we
already employed in earlier axisymmetric
investigations~\cite{dimmelmeier_02_a,dimmelmeier_02_b}.

We have presented a variety of tests in two and three dimensions,
involving neutron star spacetimes and stellar core
collapse. Axisymmetric simulations have also been performed to compare
core collapse to neutron stars using the CFC approximation and full
general relativity, for which only very recently results have become
available~\cite{shibata_04_a}. This comparison has shown the
suitability of the conformally flat approximation for such mildly
relativistic scenarios. Furthermore, the code has succeeded in
simulating the highly perturbed nonaxisymmetric configuration of a
uniformly rotating neutron star for several dynamical times. This
simulation has also been used to assess the 3D gravitational waveform
extraction capabilities of the code. In summary the numerical
experiments reported in the paper demonstrate the ability of the code
to handle spacetimes with and without symmetries in strong gravity. In
future work we plan to apply this code to simulations of stellar core
collapse to neutron stars or black holes in three dimensions, and
particularly to studies of the nonlinear development of bar mode
instabilities in rapidly rotating neutron stars.

\acknowledgments

This work is dedicated to the memory of Jean-Alain Marck who envisaged
long ago, together with S.~Bonazzola and J.M.~Ib\'a\~nez, the
potentiality of the numerical approach presented here. We gratefully
thank E.~Gourgoulhon and S.~Bonazzola for stimulating discussions and
a careful reading of the manuscript. The simulations have been
performed at the Max-Planck-Institut f\"ur Astrophysik in Garching,
Germany, and at the Laboratoire de l'Univers et de ses Th\'eories at
the Observatoire de Paris in Meudon, France. H.D.\ and E.M.\
acknowledge financial support from the SFB/Transregio~7
``Gravitations\-wellen\-astronomie'' by the DFG; J.A.F.\ and J.M.I.\
acknowledge financial support from the Spanish Ministerio de Ciencia y
Tecnolog\'{\i}a (grant AYA 2001-3490-C02-01).

\appendix

\section{Differences to previous 2D CFC simulations}
\label{section:differences_to_old_simulations}

\subsection{Compact form of the Euler equation sources}
\label{subsection:compact_hydro_sources}

In the axisymmetric CFC code presented in~\cite{dimmelmeier_02_a,
  dimmelmeier_02_b} the source terms $ Q_j $ for the hydrodynamic
momentum equations (Euler equations) were evaluated on the finite
difference grid using a formulation containing time derivatives and
explicit Christoffel symbols (see
Equation~(\ref{eq:hydro_conservation_equation_constituents})):
\begin{equation}
  Q_j = T^{\mu \nu} \left( \frac{\partial g_{\nu j}}{\partial x^\mu} - 
  {\it \Gamma}^\lambda_{\mu \nu} g_{\lambda j} \right).
  \label{eq:noncompact_sources}
\end{equation}
Using the relation between the Christoffel symbols and the derivatives
of the spacetime metric,
\begin{equation}
  {\it \Gamma}^\lambda_{\mu \nu} = \frac{1}{2} g^{\lambda \delta}
  \left( \frac{\partial g_{\delta \nu}}{\partial x^\mu} +
  \frac{\partial g_{\delta \mu}}{\partial x^\nu} -
  \frac{\partial g_{\mu \nu}}{\partial x^\delta} \right),
  \label{eq:christoffel_symbols}
\end{equation}
the sources $ Q_j $ can be written in a more compact form as
\begin{equation}
  Q_j = \frac{1}{2} T^{\mu \nu} \frac{\partial g_{\mu \nu}}{\partial x^j}.
  \label{eq:compact_sources}
\end{equation}
In this formulation, only \emph{spatial} derivatives of the metric are
needed, and the numerical evaluation of $ Q_j $ involves significantly
fewer terms, making a numerical implementation both faster and more
accurate. For these reasons, we have preferred the use of
Equation~(\ref{eq:compact_sources}) to
Equation~(\ref{eq:noncompact_sources}) in the code presented in this
paper.

\subsection{Exact numerical conservation of the hydrodynamic equations}
\label{subsection:exact_conservation}

As emphasized in Section~5.4 in~\cite{dimmelmeier_02_a}, the
conserved hydrodynamic quantity in the system of conservation
equations~(\ref{eq:hydro_conservation_equation}) is not simply the
state vector $ \mb{U} $ but rather $ \sqrt{\gamma} \mb{U} $ with
$ \sqrt{\gamma} = \phi^6 r^2 \sin \theta $. Therefore, if only the
state vector $ \mb{U} $ is evolved, this gives rise to an additional
source term $ \mb{\hat{Q}} $ which contains time derivatives of the
conformal factor $ \phi $. These generally time-dependent source terms
result in a variation of the volume-integrated state vector with time,
and thus in a violation of exact numerical rest mass and angular
momentum conservation of several percent, even though the ``physical''
sources vanish, $ \mb{Q} = 0 $ (see Figs.~9 and~10
in Ref.~\cite{dimmelmeier_02_a}).

It is not possible to evolve $ \sqrt{\gamma} \mb{U} $ in a
straightforward way and then consistently solve the elliptic metric
equations~(\ref{eq:metric_equations}) on the new time slice. This is
due to the fact that the sources for these equations contain the
pressure $ P $, which can only be extracted from $ \mb{U} $ but not
from $ \sqrt{\gamma} \mb{U} $. However, one can make use of the time 
evolution equation for the conformal factor, 
Eq.~(\ref{eq:phi_time_evolution}), to obtain an auxiliary value for 
$ \phi $ and thus for $ \sqrt{\gamma} $ on the new time slice. With 
this the state vector $ \mb{U} $ can be consistently calculated from
$ \sqrt{\gamma} \mb{U} $ after the time evolution step to the new time
slice, which in turn is used in the sources of the metric
equations~(\ref{eq:metric_equations}). These are subsequently solved
on the new time slice. With the help of this reformulation of the
hydrodynamic time evolution problem in the current code (in
combination with the compact time-independent form for the sources in
the Euler equations, Eq.~(\ref{eq:compact_sources})), we are able to 
achieve exact numerical conservation of the total rest mass and 
angular momentum up to machine roundoff errors, provided that there is 
no artificial atmosphere and no mass flow across the outer radial grid
boundary.

\subsection{Shift vector boundary conditions}
\label{subsection:shift_boundary_conditions}

The results for the evolution of the central density $ \rho_{\rm c} $
and the waveform for the core collapse model SCC (A3B2G4
in~\cite{dimmelmeier_02_b}) presented in this paper slightly differ
from those reported in the previous paper by Dimmelmeier et
al.~\cite{dimmelmeier_02_b}. This is partly due to the improvements
related to evaluating the Euler equation source terms in compact form
and using exact numerical conservation in the new code, as discussed
above. However, the main reason for the small discrepancy is that in the
simulations in~\cite{dimmelmeier_02_b} a symmetric boundary condition
for the shift vector component $ \beta^2 $ across the equatorial plane
was chosen. This leads to a nonzero value for $ \beta^2 $ at
$ \theta = \pi / 2 $ close to and after core bounce, i.e.\ when
meridional motions set in. As a consequence of this, the deviation is
stronger for models where rotation plays a significant role in the
collapse dynamics.

The physically accurate antisymmetric equatorial boundary
condition for $ \beta^2 $ which is used in the present code,
systematically yields lower post-bounce values for
$ \rho_{\rm c} $ in regular collapse type models compared to the
simulations presented in~\cite{dimmelmeier_02_b}, with a difference of
11\% on average. For models which show multiple bounce behavior,
we obtain a lower $ \rho_{\rm c} $ also at core bounce.

Accordingly, the waveform amplitudes and frequencies of the
gravitational radiation are altered by a small amount (-11\% for
$ |A^{\rm E2}_{20}|_{\rm max} $ and -18\% for $ \nu $). Despite of
these small quantitative changes, the qualitative statements related
to the influence of general relativistic effects in rotational core
collapse made by Dimmelmeier et al.~\cite{dimmelmeier_02_b} remain
unaffected, even when the antisymmetric boundary condition is used. We
particularly emphasize that the change in the boundary condition for
$ \beta^2 $ plays no role when comparing our results with the fully
general relativistic simulations by Shibata and
Sekiguchi~\cite{shibata_04_a} discussed in
Section~\ref{subsubsection:comparison_with_full_gr}.

We note that for all core collapse models presented in the
parameter study by Dimmelmeier et al.~\cite{dimmelmeier_02_b}, results
obtained with the new boundary condition for $ \beta^2 $ can be found
in the revised waveform catalogue~\cite{garching_results}.


\end{document}